\documentclass[prx,twocolumn,showpacs,superscriptaddress]{revtex4-1}
\usepackage[english]{babel}
\usepackage[dvips]{graphicx}
\usepackage{color}
\usepackage{latexsym}
\usepackage{amsmath}
\usepackage{amsthm}
\usepackage{amsfonts}
\usepackage{amssymb}

\begin{document}

\pagenumbering{arabic}

\title{Quantum phase transition with dissipative frustration}
\author{D.\ Maile}
\affiliation{Fachbereich Physik, Universit{\"a}t Konstanz, D-78457 Konstanz, Germany}
\affiliation{Institut f\"ur Theoretische Physik and Center for Quantum Science, Universit{\"a}t T{\"u}bingen, Auf der Morgenstelle 14, 72076 T{\"u}bingen, Germany}
\author{S. Andergassen}
\affiliation{Institut f\"ur Theoretische Physik and Center for Quantum Science, Universit{\"a}t T{\"u}bingen, Auf der Morgenstelle 14, 72076 T{\"u}bingen, Germany}
\author{W. Belzig}
\affiliation{Fachbereich Physik, Universit{\"a}t Konstanz, D-78457 Konstanz, Germany}
\author{G. Rastelli}
\affiliation{Fachbereich Physik, Universit{\"a}t Konstanz, D-78457 Konstanz, Germany}
\affiliation{Zukunftskolleg, Universit{\"a}t Konstanz, D-78457, Konstanz, Germany}
\begin{abstract}
We study the quantum phase transition of the one-dimensional phase model in the presence of dissipative frustration, 
provided by an interaction of the system with the environment through two non-commuting operators.
Such a model can be realized in Josephson junction chains with shunt resistances and resistances between 
the chain and the ground. 
Using a self-consistent harmonic approximation, we determine the phase diagram at zero temperature which exhibits 
a quantum phase transition between an ordered phase, corresponding to the superconducting state, 
and a disordered phase, corresponding to the insulating state with localized superconducting charge. 
Interestingly, we find that the critical line separating the two phases has a non monotonic behavior as a function of the dissipative 
coupling strength. 
This result is a consequence of the frustration between (i) one dissipative coupling that quenches the quantum phase fluctuations 
favoring the ordered phase and (ii) one that quenches the quantum momentum (charge) fluctuations leading to a vanishing phase coherence.
Moreover, within the self-consistent harmonic approximation, we analyze the dissipation induced crossover between a first and second order phase transition, showing that quantum frustration increases the range in which the phase transition is second order.
The non monotonic behavior is reflected also in the purity of the system that quantifies the degree of correlation between the system and the environment, 
and in the logarithmic negativity as entanglement measure that encodes the internal quantum correlations in the chain.
\end{abstract}
\date{\today}
\maketitle

%
%
%
\section{Introduction}
\label{sec:intro}

Owing to the recent experimental progress, the investigation of the properties of artificial quantum many-body 
systems, or  of synthetic quantum matter, has received great interest \cite{Buluta:2009ii,Cirac:2012jj,Georgescu:2014bg}.
Ultra-cold atoms in optical lattices \cite{Greiner:2002,Bloch:2012jy}, 
trapped ions \cite{Barreiro:2011jq,Blatt:2012gw} , 
exciton-polariton systems in semiconductor materials \cite{Kim2017}, 
arrays of coupled QED cavities \cite{Leib:2010fb,Hartmann:2016er},  and 
superconducting circuits made by qubits and cavities \cite{Salathe:2015bx,Barends:2015jc} 
are the most remarkable experimental platforms. 
On one side, they are considered as quantum simulators to investigate the  many-body problem in and out of thermal equilibrium. 
On the other side, they exhibit features that distinguish them from other 
strongly correlated systems in condensed matter. 
In these systems, the individual interacting units have to be considered as open or dissipative 
quantum systems as they are indeed macroscopic objects and can have relevant interactions
with the environment.
Typical examples are superconducting qubits in which energy relaxation and dephasing 
are unavoidable \cite{Makhlin:2001,Zagoskin:2011,Li:2014}. 
Generally, quantum dissipative systems or systems with quantum reservoir engineering display a variety of interesting 
phenomena \cite{Biella:2015kn,Zippilli:2015jx,HacohenGourgy:2015il,Labouvie:2016iw,Wolff:2016gn,Fitzpatrick:2017cv,Lorenzo:2017,Banchi:2017,Raghunandan:2017,FossFeig:2017}.

This intense research activity revived the study of dissipative phase transitions, 
originally initiated with Josephson junction arrays \cite{Fazio:2001}.
One-dimensional (1D) Josephson junction chains are an experimental realization 
of the 1D quantum phase  
model \cite{Wood:1982,Bradley-Doniach:1984,Jacobs:1984hs,Devillard:2011}.  
They are formed by superconducting islands with a Josephson tunneling coupling between neighboring  islands.
Here, the quantum phase transition corresponds to a superconductor-insulator transition and 
occurs due to the competition of the Josephson coupling,  which favors global phase coherence, 
and the electrostatic energy, which inhibits Cooper pair tunneling and favors the charge localization. 
The transition is activated by varying the ratio between the two energy scales, the Josephson (potential) energy $E_J$
and the characteristic charging (kinetic) energy $E_C$. 
This model - also  known as rotor model -
represents a paradigmatic statistical model to illustrate quantum phase transitions \cite{Sachdev:2007} 
and the mapping from a 1D quantum system to a 1D+1 classical one \cite{Sondhi:1997}. 
By mapping it into the XY model, theory predicts the phase transition in the 1D chain to be of Berezinsky-Kosterlitz-Thouless (BKT) type \cite{Bradley-Doniach:1984}, with the superconducting phase having quasi ordering. In the BKT scenario, the fluctuations of the local  phases diverge in the thermodynamical limit, while the fluctuations of the phase differences between neighbors are finite. In the rest of the paper we simply refer to the superconducting phase as ordered phase. 
Experiments on the scaling behavior of the resistance as a
function of the temperature in finite size chains reported the predicted superconductor-insulator transition \cite{Chow:1998bq,Kuo:2001jv}. 
The quantum phase model  also corresponds to the limit case of large average number of bosons per site 
in the lattice Bose-Hubbard model \cite{Fazio:2001,Fisher:1989ir,Bruder:1993dy}. 
Applying a gate voltage (i.e., a chemical
potential) to the chain, this system has a very
rich phase diagram \cite{Fisher:1989ir,Roddick:1993jv,Bruder:1993dy,vanOtterlo:1995ce,Freericks:1996kp,Odintsov:1996be,Glazman:1997kc,Sarkar:2007eq}.
In the limit of strong local Coulomb repulsion and when 
the gate voltage is set to a degeneracy point of two charges states of each island, the model 
maps onto the Heisenberg XXZ model  \cite{Fazio:2001,Bruder:1993dy}.
The disorder also leads to interesting effects in the phase diagram \cite{Bard:2017}, 
with glass phases that have been recently observed \cite{Duty:2017}. 
Similar complex phase diagrams have also been studied in superconducting Josephson 
circuits suitably wired up to implement the Frenkel-Kontorova model \cite{Yoshino:2010eu}, 
or in a chain formed by superinductors and small Josephson junctions  \cite{Meier:2015}.

Dissipation breaks the equivalence between the classical and the quantum case 
as the dissipation strongly affects the equilibrium phase diagram in the quantum regime, whereas 
thermodynamics and dynamics are separated in classical systems. 
In terms of the mapping to the 1D+1 classical model, 
the effect of dissipation is  to change the isotropic XY model to 
an anisotropic one leading to a dimensional 
crossover  \cite{ Panyukov:1987,Korshunov:1989,Bobbert:1990,Bobbert:1992gb}. 
Being the local superconducting phase $\hat{\varphi}$ and the electrical charge on the islands 
$\hat{Q} $ canonically conjugated operators $[ \hat{\varphi} , \hat{Q} ]= 2e i$, the transition is affected by the interplay of these degrees of freedom. 
The phases of the superconducting condensate on the islands can be regarded as rotors where 
the Josephson coupling represents a ferromagnetic interaction, whereas the charging (kinetic) energy determines the 
strength of the quantum fluctuations. 
An increase of the ratio $E_C/E_J$ leads to the transition from an ordered, classical state   
to a quantum, disordered state of the phases.

Dissipative quantum phase transitions have been intensively studied in the 1D quantum 
phase model \cite{Chakravarty:1986,Fisher:1987,Panyukov:1987,Chakravarty:1988hb,Korshunov:1989,
Zwerger:1989fo,Bobbert:1990,Bobbert:1992gb,Wagenblast:1997,Refael:2007} with 
the main result that dissipation suppresses quantum phase fluctuations thus 
favoring states with spontaneously broken symmetry and ordering of the phases. 
Experiments on linear Josephson junction chains with a tunable ratio of $E_J/E_C$ and different 
shunted resistance confirmed the predicted dissipative phase diagram \cite{Miyazaki:2002br}. 

%
%
%
\begin{figure}[t!]			
\mbox{\includegraphics[width=0.49\linewidth,angle=0.]{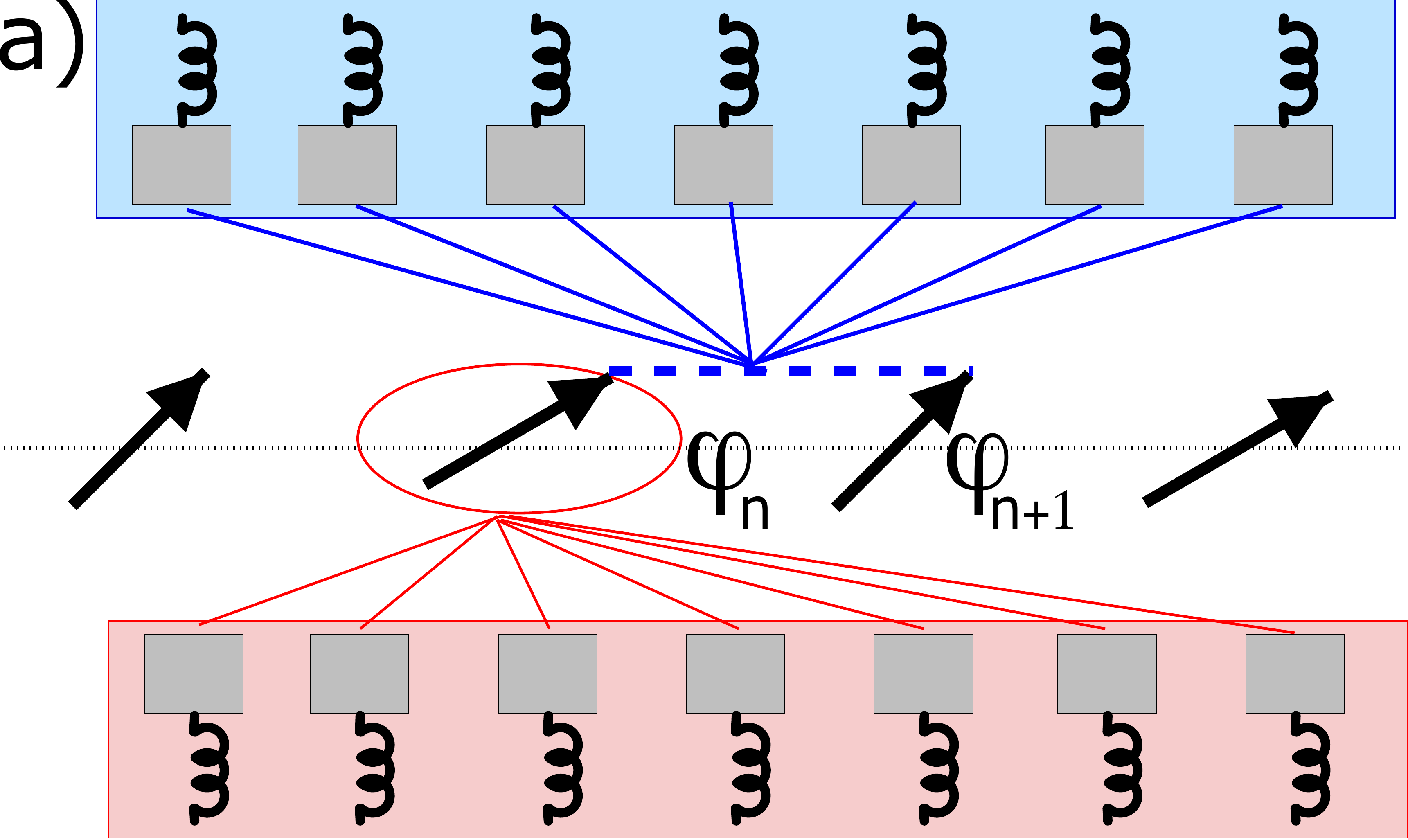}\hspace{0.1cm}\includegraphics[width=0.49\linewidth,angle=0.]{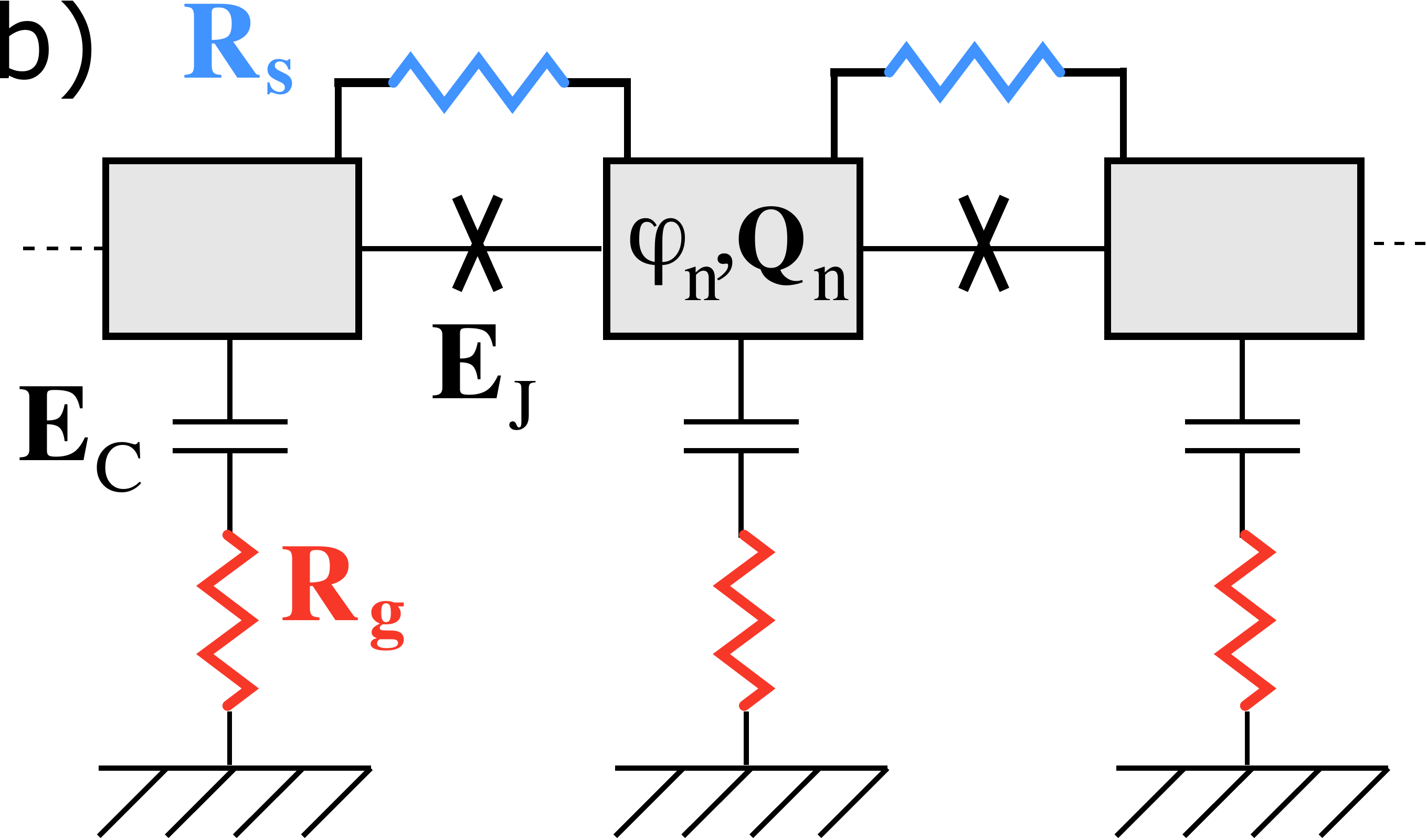}} \\	\vspace{0mm}
  	\caption{
	(a) Model for dissipative frustration in the 1D quantum phase model.	
	The phase difference $\Delta \varphi_n$ for each link $n$ is coupled linearly to a bath (blue box) and 
	the conjugated momentum of the phase is linearly coupled to a second, independent bath (red box). The external baths are represented as sets of independent harmonic oscillators. 
	(b) 1D chain of superconducting islands  with Josephson coupling of energy $E_J$ and charging energy $E_C$. 
	The shunt resistance $R_s$ corresponds to a dissipative coupling for the difference of the superconducting 
	phases $\varphi_n$, 	whereas the resistance to ground  $R_g$ yields  dissipative coupling in the charge $Q_n$. }
  	\label{fig1:sketch_model}
\end{figure}
%
%
%

A counter-example was given by a recent work in which the one-dimensional chain of Josephson 
junctions was assumed to be capacitively coupled to a proximate two-dimensional diffusive metal with
a stabilization of the insulating ground state given by increasing the dissipation strength \cite{Lobos:2011}.  
From these results, one concludes that  dissipation suppresses generally 
certain types of  fluctuations associated to one degree of freedom, favoring one or 
other phases.

Remarkably, an open quantum system coupled to two independent environments via two 
canonically conjugate operators can yield interesting effects.
This theoretical issues of ``dissipative frustration''  was analyzed for single open quantum system as 
a harmonic oscillator \cite{Kohler:2005fo,Kohler:2006ky,Cuccoli:2010dr,Kohler:2013dg,Rastelli:2016ge}, 
a single spin \cite{Neto:2003ir,Novais:2005ka,Kohler:2013ie,Bruognolo:2014jf,Zhou:2015ct}, a Y shaped Josephson network \cite{Giuliano:2008eh}, 
as well as a lattice of interacting spins \cite{Lang:2015fq}.
In other words, two environments couple to non commuting observables of a central system and 
are continuously monitoring the system leading to different and orthogonal conditioned states in 
presence of a single bath.

%
%
%
\begin{figure}[t!]			
	\includegraphics[width=0.9\linewidth,angle=0.]{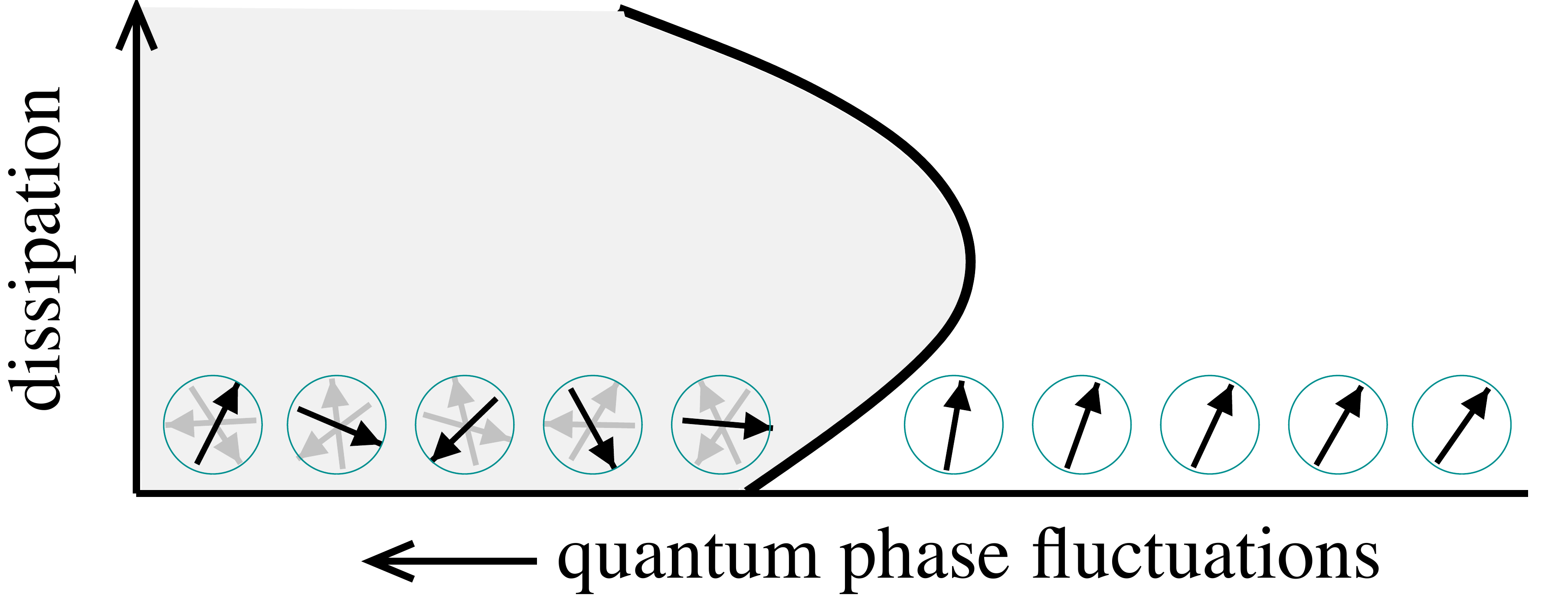}\\	
	\vspace{0mm}
  	\caption{Sketch of the phase diagram of the 1D phase model with dissipative frustration.
	The critical line between the ordered phase and the disordered phase 
	displays a non monotonic behavior.
	}  
  	\label{fig2:main_result}
\end{figure}
%
%
%

In this work we study the effect of dissipative frustration on the quantum phase transition for the one-dimensional phase model. 
The dissipative coupling through the conjugate operators is realized by 
assuming that each local phase difference is coupled to a local bath (or conventional phase dissipation) 
and each local momentum coupled to another local bath (unconventional or charge dissipation), 
see Fig.~\ref{fig1:sketch_model}(a). 
These two kinds of dissipative interaction compete since, when they are considered separately,  they suppress
different  quantum fluctuations, viz. phase or charge, whose product is bound by the uncertainty principle. \\
We show that this model can be realized by a chain of Josephson junctions with equal shunted resistance $R_s$ 
between neighboring islands - to encode the phase dissipation -  
and  resistances $R_g$ between each superconducting island and the ground -  to encode the charge dissipation - 
as shown in Fig.~\ref{fig1:sketch_model}(b). \\ 
We use a variational approach, the self-consistent harmonic approximation (SCHA)
\cite{Chakravarty:1986,Chakravarty:1988hb,Kugler:1969,Gillis:1972,Moleko:1983,Kampf:1987kt,Samathiyakanit:1973gj,Rastelli:2005bq,Rastelli:2011iw}, 
to treat the non-linear Josephson coupling between the phases.
The SCHA allows to take into account the anharmonic effects for large quantum phase fluctuations 
eventually leading to the transition.  
Within the SCHA, we construct a phase diagram for the ordered-disordered phase transition (superconductor-insulator) in terms 
of the dissipative coupling and the ratio between the two energy scales $E_J/E_C$ that measures, qualitatively speaking,
the amount of the intrinsic quantum phase fluctuations in the ordered phase of the isolated chain. 
For a given ratio between the two dissipative coupling strengths, our main result is that the critical line has a non monotic behavior 
for increasing total dissipation of the system, see Fig.~\ref{fig2:main_result}. 
On the basis of the SCHA, we discuss the order of the phase transition and the crossover from a first order to second order phase transition. \\
A non-monotonic dependence of the critical value was previously reported in a dissipative 2D Josephson array in different geometries due to non-local dissipation in Ref. \cite{Polak:2007} or due to an applied magnetic field in Ref. \cite{Polak:2005}. However,  the critical line as a function of the dissipative strength was monotonic in agreement with the expected behavior in presence of phase dissipation. \\
Since the dissipative phase transition is triggered by quantum fluctuations at zero temperature, which 
are strongly affected by the interaction of the system with the environments, we study the purity of the system that quantifies the correlation between the system and the environment.
The purity shows a  non monotonicity close to the critical point at the phase transition, pointing out that the correlation with the environment plays an important role. 
We also calculate the logarithmic negativity as entanglement measure that encodes the 
internal quantum (non classical) correlations in the system and show that this quantity can also have a non monotonic behavior 
approaching the phase transition.
From these results, we can conclude that the dissipative phase transition has a spurious nature in which 
internal (quantum) correlations as well as extrinsic (statistical) correlations have similar weight.

This paper is structured as follows.
In Sec.~\ref{sec:model} we introduce the quantum phase model with dissipative frustration 
in terms of the path integral formalism \cite{feynman-book,Kleinert:2009,Weiss:2012},  
namely we introduce the effective action in the imaginary time representation.
We also present the self-consistent harmonic approximation (SCHA) and the results of the 
phase difference fluctuations between neighboring phases, a quantity that plays a central role in the SCHA. 
In Sec.~\ref{sec:results} we discuss the results for the phase diagram in presence of dissipative frustration whose main effect is sketched in the Fig.~\ref{fig2:main_result}.
In Sec.~\ref{sec:order-phase-transition}, we classify the order of the phase transition within the SCHA 
by analyzing the behavior of the variational expansion for the free energy that represents an upper 
bound estimation of the exact free energy.
In Sec.~\ref{sec:purity-entanglement}, we present the results for the purity and  the logarithmic negativity.
Finally, in Sec.~\ref{sec:summary} we summarize our work and draw our conclusions. 
Appendixes \ref{app:derivation} and \ref{app:derivation-2} contain the derivation of the path integral action related to the unconventional (charge) dissipation.
In Appendix \ref{Sec:symplect} we recall the method to calculate the logarithmic negativity using the correlation matrix.
In Appendix \ref{Sec:config} we report further results for the entanglement measure that 
confirm the behavior discussed in the main text  for different configurations of the two 
subsystems in which the chain is bipartite.

%
%
%
\section{Model and Approximations}
\label{sec:model}

In this section we introduce the dissipative phase model and the corresponding effective action.
We then present the  SCHA and report the 
main steps of our calculations in obtaining the analytic expressions of the quantum phase fluctuations.

\subsection{Hamiltonian}

We consider a 1D chain of $N$ rotors of radius $R$ whose dynamics is described 
by the local phase operators $\hat{x}_n = R \hat{\varphi}_n$ and momenta 
$\hat{p}_n=(\hbar/i R) \partial/ \partial \hat{\varphi}_n$, 
with the commutation relation $[\hat{x}_n, \hat{p}_m]=i\hbar \delta_{nm}$.
The phases interact via a nearest-neighbor pairwise potential  
$U(\Delta\hat{\varphi}_n)= - V \cos(\Delta\hat{\varphi}_n) $, where  $\Delta\hat{\varphi}_n=\hat{\varphi}_{n+1}-\hat{\varphi}_{n}$.
We assume periodic boundary conditions $\hat{\varphi}_{N}\equiv\hat{\varphi}_0$.
The Hamiltonian of the considered system reads 
\begin{equation}
\label{eq:H_S}
\hat{H}_{S}    
=
\sum_{n=0}^{N-1} \left[ - \frac{K }{2}\left( \partial^2 /\partial\hat{\varphi}^2 \right)  - V \cos(\Delta\hat{\varphi}_n) \right]
  \, ,
\end{equation}
where $K=\hbar^2/(mR^2)$ is the energy scale associated to the kinetic energy of the rotors. 

This is the same Hamiltonian as for a chain of superconducting islands 
with a Josephson coupling $E_J$ between nearest neighbors and a
capacitance to the ground $C_0$ with charging energy $E_C= 4e^2/C_0$.
In the representation of the charge operator $\hat{N}_n =\sum_{N_n} N_n  \left| N_n \right> \left< N_n \right|$ 
with $\left| N_n \right>$  the number states and $N_n $ corresponding to the Cooper pair number in each superconducting 
island, the system Hamiltonian takes the form
\begin{equation}
\hat{H}_{S}   =   \frac{E_C}{2}  \sum_{n=0}^{N-1} {\left(\hat{N}_n - N_0^{\phantom{g}} \right)}^2  - 
E_J \left( \hat{T}_{n,n+1}^{\phantom{g}} + \hat{T}_{n,n+1}^{\dagger} \right)  \, , 
\end{equation}
with the quantum tunneling operator describing the coherent hopping of Cooper pairs given by  \cite{Zagoskin:2011,Tinkham:1996}
\begin{equation}
\hat{T}_{a,b}  =  \left|   N_{a}  ,  N_{b} + 1 \right>\left<   N_{a}  +1 ,  N_{b}  \right|  
 \, .
\end{equation}
Introducing the phase operator $\hat{\varphi}_n$ conjugate to $\hat{N}_n$,  we have the Hamiltonian  \cite{Zagoskin:2011,Tinkham:1996}
\begin{equation}
\label{eq:H_S-1DJJ}
\hat{H}_{S}    
=
\sum_{n=0}^{N-1} \left[ - \frac{E_C}{2} \left( \partial^2 /\partial\hat{\varphi}^2 \right)  - E_J \cos(\Delta\hat{\varphi}_n) \right]
  \, .
\end{equation}
The Hamiltonian (\ref{eq:H_S-1DJJ})  is based on the assumption that the quasiparticle excitations (above the gap) can be neglected, 
see Ref.~\cite{Fazio:2001}.   
At zero temperature, the behavior of the quantum phase model is fully described by the dimensionless ratio 
$g=\sqrt{V/K}=\sqrt{E_J/E_C}$.
In the limit of small phase difference fluctuations for $E_J\gg E_C$  ($g\gg 1$), one can expand the potential in Eq.~(\ref{eq:H_S-1DJJ}) to harmonic order and obtains that the average quantum 
phase difference fluctuations are controlled by the inverse of this ratio, 
viz. ${\left< \Delta \hat{\varphi}^2 \right>}_{har} = \sqrt{2} /g$.

\subsection{Effective action and dissipation}
Dissipation arises when we consider the interaction of the chain with the environment.
Then, to discuss the equilibrium properties of an open quantum system, 
the imaginary time path integral formalism allows to integrate out the degrees of freedom 
associated to the environment and focus only on the partition function associated to the degrees of freedom 
of the system, viz. the phases. 
In our case,  the effective partition function $\mathcal{Z}_{\rm eff}$ describing the phase model reads 
\begin{equation}
\label{eq:Z_eff}
\mathcal{Z}_{\rm eff}=  \prod_{n=0}^{N-1} \oint_{c}\mathcal{D}[\varphi_n(\tau)] \,\, e^{- \mathcal{S}  [\{ \varphi_n(\tau) \}]/ \hbar } \,\, ,
\end{equation}
where the symbol $\oint_{c}$  refers to the path integral over imaginary time for the interval $0< \tau < \beta$ , 
with $\beta = \hbar/(k_BT)$ and  to periodic boundary conditions for the phase variable $\varphi$, i.e. 
$\varphi(0) = \varphi(\beta)$ \cite{footnote-decompatification}\cite{Apenko:1989}. 
The effective Euclidean action for the system is given by
\begin{equation}
\label{Eq: effectiveaction}
\mathcal{S}  = 
\mathcal{S}_{diss}
-  \int_{0}^{\beta}\!\!\!\!\! d\tau E_J \cos\left(\Delta \varphi_n(\tau) \right )  \, 
\end{equation}
where the quadratic action is
\begin{align}
\label{Eq: effectiveaction_gaussian} 
\mathcal{S}_{diss} =&-
\sum_{n=0}^{N-1} 
\frac{1}{2}   \int_{0}^{\beta} \!\!\!\! \int_{0}^{\beta}\!\!d\tau d\tau' \,
F (\!\tau\!-\!\tau') \, {\left| \Delta\varphi_{n}(\tau) - \Delta\varphi_{n}(\tau') \right|}^2 \nonumber \\
&+ 
\sum_{n=0}^{N-1}  \frac{1}{2}  \int_{0}^{\beta} \!\!\!\! \int_{0}^{\beta}\!\!d\tau d\tau' \,
\widetilde{F}  (\!\tau\!-\!\tau') \, \dot{\varphi}_{n}(\tau) \, \dot{\varphi}_{n}(\tau')
\, ,
\end{align}
with $\dot{\varphi}= d\varphi/d\tau$.
Note that the action (\ref{Eq: effectiveaction_gaussian}) is locally invariant under a variation of $2\pi$ of the phase.

The first term of Eq. (\ref{Eq: effectiveaction_gaussian}) corresponds to 
the conventional or phase dissipation associated to the shunt ohmic resistance between two superconducting 
islands \cite{Fazio:2001,Chakravarty:1986,Chakravarty:1988hb,Zwerger:1989fo,Bobbert:1990,Bobbert:1992gb}. 
Using the Fourier transform in the imaginary time for the $\beta-$periodic function 
$x(\tau) = \sum_{\ell} x_{\ell} e^{i \omega_{\ell} \tau}$ with
the Matsubara frequencies $\omega_{\ell}=(2\pi/\beta) \ell$ and $\ell$ integer, 
the component of the ohmic kernel $F(\tau)$ is \cite{Chakravarty:1986,Chakravarty:1988hb,Bobbert:1990,Bobbert:1992gb,Fazio:2001,Weiss:2012}:
\begin{equation}
\label{Eq:Matsugam}
F_{\ell} =\frac{\hbar}{4\pi\beta} \left(\frac{R_q}{R_s} \right)  \left| \omega_{\ell}  \right| f_{c}(\omega_{\ell}) 
\, , 
\end{equation}
with $F_{\ell=0}=0$ and the Drude cutoff function at large frequency $\omega_c$ of the form $f_c(\omega)=0$ for $\omega/\omega_c \rightarrow \infty$. 
The parameter that quantifies  the dissipative coupling strength is associated 
to the ratio  between the quantum resistance $R_q=h/(4e^2)$ and the shunt resistance $R_s$
\begin{equation}
\label{Eq:coupling_convetional}
\alpha =  R_q/R_s =  \gamma (h/ E_C) 
\quad  \mbox{with} \,\,\, \gamma = 1/(R_s C_0) \, , 
\end{equation}
whereas the rate $\gamma$  corresponds to the friction coefficient. 
In the limit $R_s/R_q\rightarrow  \infty$, the current flowing through the shunt  resistances vanishes 
and the 1DJJ chain is not affected by the conventional dissipation.

The  second term of Eq.~(\ref{Eq: effectiveaction_gaussian}) 
includes the kinetic energy and the unconventional or charge dissipation, with the
Matsubara components given by the expression
\begin{equation}
\label{Eq:Matsutau}
\widetilde{F}_{\ell} = \frac{\hbar^2}{\beta E_C}
\left( 1 
-  \frac{R_g C_0 \left| \omega_{\ell} \right|   f_{c}(\omega_{\ell}) }{1+ R_g C_0 \left| \omega_{\ell} \right|   f_{c}(\omega_{\ell})  }
\right) \, ,
\end{equation}
whose explicit derivation is provided in the Appendices \ref{app:derivation} and \ref{app:derivation-2}. 
Here we simply observe that this expression can be derived 
by duality between the two conjugate quadratures of a harmonic oscillator 
coupled separately to two baths \cite{Kohler:2006ky,Rastelli:2016ge}.
In other words, it is possible to show that  unconventional dissipation as given by Eq.~(\ref{Eq:Matsutau}) 
yields a quenching of the momentum quantum fluctuations which is exactly equivalent to the 
quenching of the phase quantum fluctuation for an oscillator affected by ohmic damping given by 
Eq.~(\ref{Eq:Matsugam}).
In the quantum phase model, the parameter that quantifies the strength of the charge dissipative coupling 
is related to the characteristic  time scale of the impedance due to the resistance to the ground $R_g$ in 
series with the capacitance $C_0$, see Fig.~\ref{fig1:sketch_model} (b).
In contrast to the phase dissipation, the dissipative coupling  vanishes in 
the limit $R_g \rightarrow 0$.
It is useful to introduce the parameter 
\begin{equation}
\widetilde{\alpha}=  R_g/R_q  =  \tau_g (E_C / h)
\quad  \mbox{with} \,\,\,  \tau_g  = R_g C_0 \, 
\end{equation}
playing the role of the dimensionless coupling constant of the unconventional dissipation.\\
To be specific, we assume as a cutoff frequency  
$f_{c}(\omega_{\ell}) = 1/\left( 1 +  \left| \omega_{\ell} \right|/\omega_c \right)$ 
in the following. 
For $R_g=0$,  the model of the action (\ref{Eq: effectiveaction})
corresponds to the dissipative quantum rotor model discussed 
extensively in the literature \cite{Fazio:2001,Panyukov:1987,Korshunov:1989,Bobbert:1990,Bobbert:1992gb,Chakravarty:1986,Fisher:1987,Chakravarty:1988hb,Zwerger:1989fo,Wagenblast:1997,Refael:2007}.
Note that we focus on the case of homogeneous dissipation assuming
the two kernel functions $F(\!\tau\!)$ and $\widetilde{F} (\!\tau\!)$ 
to be independent of the position on the lattice (index $n$).

\subsection{The self-consistent harmonic approximation SCHA}
In the limit in which the average phase difference fluctuations are small $\sqrt{\left< \Delta \varphi^2 \right>} \ll \pi$, 
we can use the harmonic approximation and expand the potential to obtain
\begin{equation}
\label{Eq:action_harmonic}
\mathcal{S}_{\rm harm.} 
=  \mathcal{S}_{diss}  
+ 
\sum_{n=0}^{N-1} 
\int_{0}^{\beta}\!\!\!\! d\tau \, E_J  \left[ -1 +  \frac{1}{2}  \Delta \varphi_n^2(\tau) \right] \, .
\end{equation}
If the  fluctuations are strongly localized, paths 
of large fluctuations $\varphi(\tau) \sim \pi$ are extremely unlikely 
to occur.

Beyond the harmonic approximation valid at $\sqrt{\left< \Delta \varphi^2 \right>} \ll \pi$, 
the  model  of Eq.~(\ref{Eq: effectiveaction})
can not be solved exactly in general due to the presence of the interaction potential and 
we have to resort  to an approximated scheme. 
For larger values of the phase fluctuations, further anharmonic terms of the pairwise potential, have to be taken into account.
To treat this regime, we employ the self-consistent harmonic approximation (SCHA) \cite{Chakravarty:1986,Chakravarty:1988hb,Kugler:1969,Gillis:1972,Moleko:1983,Kampf:1987kt,Samathiyakanit:1973gj,Rastelli:2005bq,Rastelli:2011iw}.
Within this approach, a quadratic trial action $\mathcal{S}_{tr}$ is introduced as 
\begin{equation}
\label{Eq:trialaction}
\mathcal{S}_{tr}   = 
\mathcal{S}_{diss}   - \beta N E_J +   \sum_{n=0}^{N-1} 
\int_{0}^{\beta}\!\!\!\!\! d\tau \, \frac{1}{2} V_{tr} \, 
\Delta\varphi_n^2(\tau)  \, 
\end{equation}
which is formally equivalent to the harmonic expansion of Eq.~(\ref{Eq:action_harmonic}). 
However, one assumes $V_{tr}$ as a free variational parameter, different 
from the bare energy constant of the potential $V_{tr} \neq E_J$.
Similarly to  the harmonic expansion, 
the partition function associated to the action (\ref{Eq:trialaction}) can be   
computed by
$\mathcal{Z}_{tr} =  \prod_{n=0}^{N-1}
\oint \mathcal{D}[\varphi_n(\tau)] \,\, \exp({- \mathcal{S}_{tr} / \hbar })$, together with 
the Helmoltz free energy $\mathcal{F}_{tr}   = -(\hbar/\beta) \ln[\mathcal{Z}_{tr} ]$. 
Using the Bogoliubov inequality, an upper bound for the exact free energy
 $\mathcal{F}_{\rm eff}  = -(\hbar/\beta) \ln[\mathcal{Z}_{\rm eff}]$  
\begin{align}
\label{Eq:bogu}
\mathcal{F}_{\rm eff} \, \leq   \mathcal{F}_v, \;\;\;\;\;\;\;
 &\mathcal{F}_v=\mathcal{F}_{tr}  
+ (\hbar / \beta) {\langle  \mathcal{S} - \mathcal{S}_{tr }\rangle}_{tr} \, ,
\end{align}
where the average $  {\langle  \mathcal{S} - \mathcal{S}_{tr} \rangle}_{tr} $ is performed on the 
variational action $\mathcal{S}_{tr}$.
The minimum of the r.h.s of Eq.~(\ref{Eq:bogu}) is determined 
by taking the derivative respect to the variational parameter  $V_{tr}$ and setting it to zero.
This leads to the following self-consistent equation for the variational parameter 
\begin{equation}
\label{selfcon}
V_{sc} = E_J \,e^{-\frac{1}{2} {\langle\Delta \varphi^{2}\rangle}_{sc}} \, ,
\end{equation}
containing the fluctuations of the phase difference $ {\langle\Delta \varphi^{2}\rangle}_{sc}$
calculated on the variational action  (\ref{Eq:trialaction}) for $V_{tr} \rightarrow V_{sc}$, i.e. 
the self-consistent parameter, representing the effective spin-wave stiffness constant \cite{Chakravarty:1986}. 
This way, the SCHA captures  the anharmonic behavior of the phase fluctuations by  
an effective harmonic potential $V_{sc}$ which  approximates the actual anharmonic fluctuations.

This one-component theory of the phase transition provides a (qualitative) phase diagram in the following way:
By varying one of the parameters $g$, $\alpha$ or $\tilde{\alpha}$, 
one can determine the critical value above which there is no solution of Eq.~(\ref{selfcon}). 
This solution corresponds to a spinodal point which, in the SCHA, is associated to the transition between 
the ordered phase, characterized by (an)harmonic fluctuations of the phases, 
and the disordered phase without any long-range correlations. 
An alternative criterion to derive the critical line consists in comparing the upper bound of the exact free energy evaluated at the self-consistent solution $\mathcal{F}_{\rm eff}(V_{tr} =  V_{sc})$ with 
the value for vanishing  stiffness constant $\mathcal{F}_{\rm eff}(V_{tr} =  0 )$: 
then the critical point corresponding to the situation in which 
$\mathcal{F}_{\rm eff}(V_{sc})  \geq \mathcal{F}_{\rm eff}(0)$, identifies the transition to the disordered phase.
The latter criterion allows to distinguish between a first and second order phase transition.

In the following , we discuss both criteria to obtain the phase diagram for  
the 1D dissipative system of phases with conventional (or phase) dissipation and 
unconventional (or charge) dissipation.

\subsection{Calculation of the quantum phase fluctuations} 
\label{Sec:quantumphasefluc}
In this subsection we discuss the analytic expression for the quantum phase difference fluctuations, 
in the limit of zero temperature $\beta\rightarrow \infty$ $(T\rightarrow0)$,  calculated on the quadratic  
trial action~(\ref{Eq:trialaction}).
The Gaussian trial action can be decomposed in terms of non interacting quadratic modes  which 
 are defined by the relation $\varphi_n=(1/\sqrt{N})\sum_{k=0}^{N-1} e^{-i2\pi kn/N}\varphi_k$ 
 \footnote{Notice that, if we set $\varphi_{k}=\varphi_{k,\text{Re}}+i \varphi_{k,\text{Im}}$, we have $\varphi_n=\frac{1}{\sqrt{N}}\sum\limits_{n=0}^{N-1}\left(\varphi_{k,\text{Re}}+i \varphi_{k,\text{Im}}\right)\left(\cos\left(\frac{2\pi k n}{N}\right)-i\sin\left(\frac{2\pi k n}{N}\right)\right)$, in which we can use $\varphi_{N-k}*=\varphi_k$ (or $\varphi_{N-k,\text{Re}}=\varphi_{k,\text{Re}}$ and $\varphi_{N-k,\text{Im}}=-\varphi_{k,\text{Im}}$). With this we can find the relations $\varphi_{k,\text{Re}}=\varphi_{k,e}/\sqrt{2}$ and $\varphi_{k,\text{Im}}=\varphi_{k,o}/\sqrt{2}$, where the subscripts $e$ and $o$ stand for the $even$ and the $odd$ part of the Fourier transformation, respectively. }.
Then the average phase fluctuations are expressed as
\begin{align}
\label{eq:fluctuations}
{\langle\Delta \varphi^{2}\rangle}_{sc} & 
=  \frac{4}{N}\sum_{k=1}^{N-1}\sin^{2}\left(\frac{\pi k}{N}\right){\langle {\left|\varphi_{k}\right|}^{2}\rangle}_{\!sc}  \,
\end{align} 
in which each term corresponds to the fluctuations of a harmonic mode. 
To calculate ${\langle {\left|\varphi_{k}\right|}^{2}\rangle}_{\!sc}$, 
we express $\{\varphi_n\}$ as functions of $\{\varphi_k\}$ in the Gaussian 
action (\ref{Eq:trialaction}), and obtain the Lagrangian of $N$ independent harmonic oscillators, 
each of them affected by conventional and unconventional dissipation.

By proceeding in a similar way as in Ref.~\cite{Chakravarty:1988hb}, we arrive at the expression:
\begin{equation}
\label{eq:harmfluct} 
{\langle {\left|\varphi_{k}\right|}^{2}\rangle}_{\!sc} 
\!\! =\!\!
\sum_{l=-\infty}^{+\infty}\frac{E_C/(\hbar \beta)}
{
\left( \omega_{k}^{(sc)}  \right)^{2}
+
\frac{4\sin^{2}\left(\frac{\pi k}{N}\right) |\omega_{l}| \alpha E_C/h}{1+\frac{|\omega_{l}|}{\omega_{c}}}
+
\frac{\omega_{l}^{2}}{ 1+\frac{\tau_g |\omega_{l}|}{\left(1+\frac{|\omega_{l}|}{\omega_{c}} \right)} } 
} \, , 
\end{equation}
where the eigenfrequencies 
\begin{align}
\omega_{k}^{(sc)} & = 2 \, \frac{\sqrt{E_C V_{sc}}}{\hbar} \,  \sin\left( \pi k/ 2N \right)\, . 
\end{align}
corresponding to the frequency of the normal modes of the Josephson chain. 
Since, we are interested in the quantum regime, we
take the zero temperature limit $\beta \rightarrow \infty$,
and the sum over Matsubara frequencies transforms into an integral that can be calculated analytically. 
Thus we obtain the expression 
\begin{equation}
\label{Eq:zeroTemp}
{\langle {\left|\varphi_{k}\right|}^{2}\rangle}_{\!sc}
=
\frac{ \phi_a + \phi_b }{\pi g^2}  + 
2 \widetilde{\alpha}  \frac{\ln\left[\omega_c/ \omega_{k}^{(sc)} \right] }{1+\sigma^2_k} 
\end{equation}
where we introduced  $\sigma_{k}^{2} =  4\sin^{2}( \pi k/N ) \alpha \tilde{\alpha}$, the two phases 
\begin{eqnarray}
\phi_a  &=&\frac{ 2\pi g^2 \tilde{\alpha} }{1+\sigma_{k}^{2}}  \left[
\ln(1+\sigma_{k}^{2})+\sigma_{k}\arctan(\sigma_{k}) 
\right] \\
\phi_b  &=& \frac{1}{1+\sigma_{k}^{2}}
 \left(\!\! \frac{E_J}{\mbox{} \,\, \hbar \omega_{k}^{(sc)}}
+
2\pi g^2 \tilde{\alpha} \, 
\Gamma_{-}^{(sc)} \!\! 
\right) 
\mathcal{\bold F} \!\! \left[ \Gamma_{-}^{(sc)}  \!\!, \Gamma_{+}^{(sc)} \right] \, .
\end{eqnarray}
and the function
\begin{align}
\label{Eq:zeroTemp2}
\mathcal{\bold F} [x,y]
= 
\begin{cases} 
\frac{1}{\sqrt{1-x^{2}}}\text{arctan}\left(\frac{\sqrt{1-x^{2}}}{y}\right) & , \text{for} \, x<1\\
\frac{1}{\sqrt{x^2-1}}\text{arctanh}\left(\frac{\sqrt{x^2-1}}{y} \right) & , \text{for} \,  x>1 
\end{cases} \, 
\end{align}
with the parameter $\Gamma_{\pm}^{(sc)}   
=   \sin^{2}\left( \pi k/N \right) \alpha E_C / (\pi\hbar\omega_{k}^{(sc)}) 
\pm \pi (\hbar \omega_{k}^{(sc)}/E_J)  g^2 \tilde{\alpha}$.
We recover the previous result \cite{Rastelli:2016ge} for $V_{sc}=E_J$.
The analytical expression (\ref{Eq:zeroTemp}) for each harmonic mode was obtained in presence of
a high frequency cutoff  $\omega_c \gg \omega_{k}^{(sc)} , \gamma , 1/\tau_g$. 
Note the logarithmic dependence on $\omega_c$ in Eq.~(\ref{Eq:zeroTemp}), characteristic for the ohmic dissipation with a Drude 
cutoff \cite{Weiss:2012}.

Once the fluctuations ${\langle\Delta \varphi^{2}\rangle}_{sc}$ are expressed in terms 
of both coupling constants $\alpha$, $\widetilde{\alpha}$ and $g$, 
we use Eqs.~(\ref{eq:fluctuations})  and (\ref{Eq:zeroTemp}) to 
solve the self-consistent equation (\ref{selfcon}) numerically. 
As explained above, within the SCHA framework  
the existence of a solution of (\ref{selfcon}) corresponds 
to the ordered state of the rotors, whereas one associates its  
disappearance to a phase transition of the system 
towards a disordered state.
The SCHA approach can only be justified, a priori, for fluctuations 
$\sqrt{\langle\Delta\varphi^{2}\rangle_{sc}}\lesssim \pi$. 
Nonetheless, we use this approximation to gain a first qualitative understanding 
of the influence of the conjugate baths on the quantum phase transition.

\subsection{Absence of dissipation ($\alpha=\tilde{\alpha}=0$)}
As discussed in the introduction, in absence of dissipation, decreasing $g$ below a critical value leads to a phase transition. 
Before presenting the numerical results including dissipation, we illustrate the prediction of the SCHA equation for this case.
For $\alpha=\tilde{\alpha}=0$ and in the limit $N \gg 1$,   the self-consistent Eq. (\ref{selfcon}) simplifies to 
\begin{equation}
\label{eq:selfcon2}
V_{sc}/E_J =  e^{-\frac{1}{\pi g}\sqrt{ E_J/V_{sc}} } \, .
\end{equation}
We denote the maximum value corresponding to the critical solution of Eq.~(\ref{eq:selfcon2}) 
by  $g_{s}^{(0)}$. In correspondence of this point, the l.h.s. and r.h.s. of (\ref{eq:selfcon2}) have the same derivative with respect to the variable   $x=V_{sc}/E_J $. 
Using the latter condition together with Eq.~(\ref{eq:selfcon2}), we find  
$\sqrt{V_{sc}/E_J}=1/(2\pi g_{s}^{(0)})$  that yields  $V_{sc}/E_J = 1/e^{2}$ and
corresponds to a critical value  $g_{s}^{(0)}= e/(2\pi) \approx 0.43$
\footnote{Note that this value differs from the one 
found by Chakravarty et. al. \cite{Chakravarty:1988hb}, because they used the further approximation 
that the dispersion of the modes was purely linear.}.

%
%
%
\section{Results: solution of the self-consistent equation}
\label{sec:results}
We here present the results for the solutions of the self-consistent equation (\ref{selfcon}).
We consider a high frequency cutoff  $\hbar \omega_c=100 E_C$ 
corresponding to the regime $ \omega_c  \gg  \omega_{k}, \gamma,1/\tau_g$, 
and $N=150$ for which the phase difference fluctuations  
are converged and close to the thermodynamical limit, 
i.e. further doubling of $N$ affects the results by less than $0.07$ percent.
In this section, we set the notation $\delta \varphi_{sc} = \sqrt{\langle \Delta \hat{\varphi}_n^2\rangle_{sc}}$ for the quantum phase difference fluctuations calculated with the self consistent parameter $V_{sc}$.

\begin{figure}[t]			
	\includegraphics[width=0.49\columnwidth]{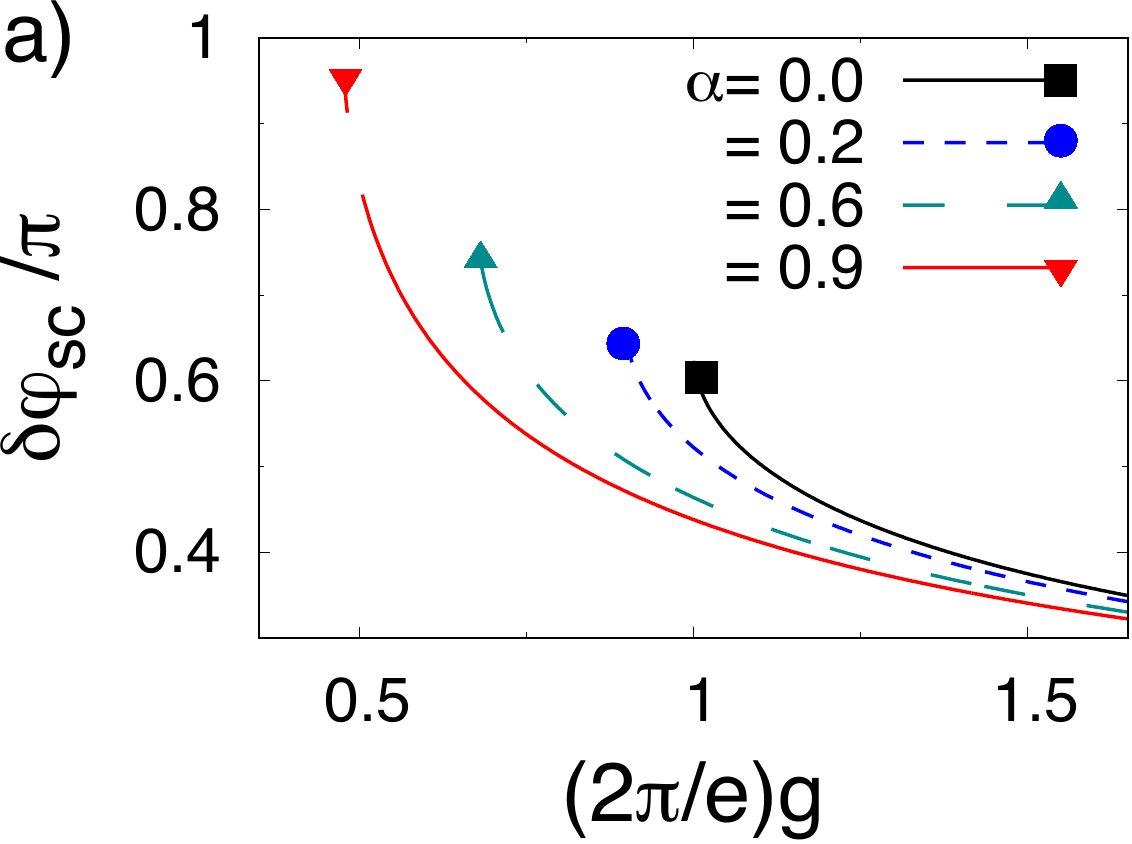}\hfill\includegraphics[width=0.49\columnwidth]{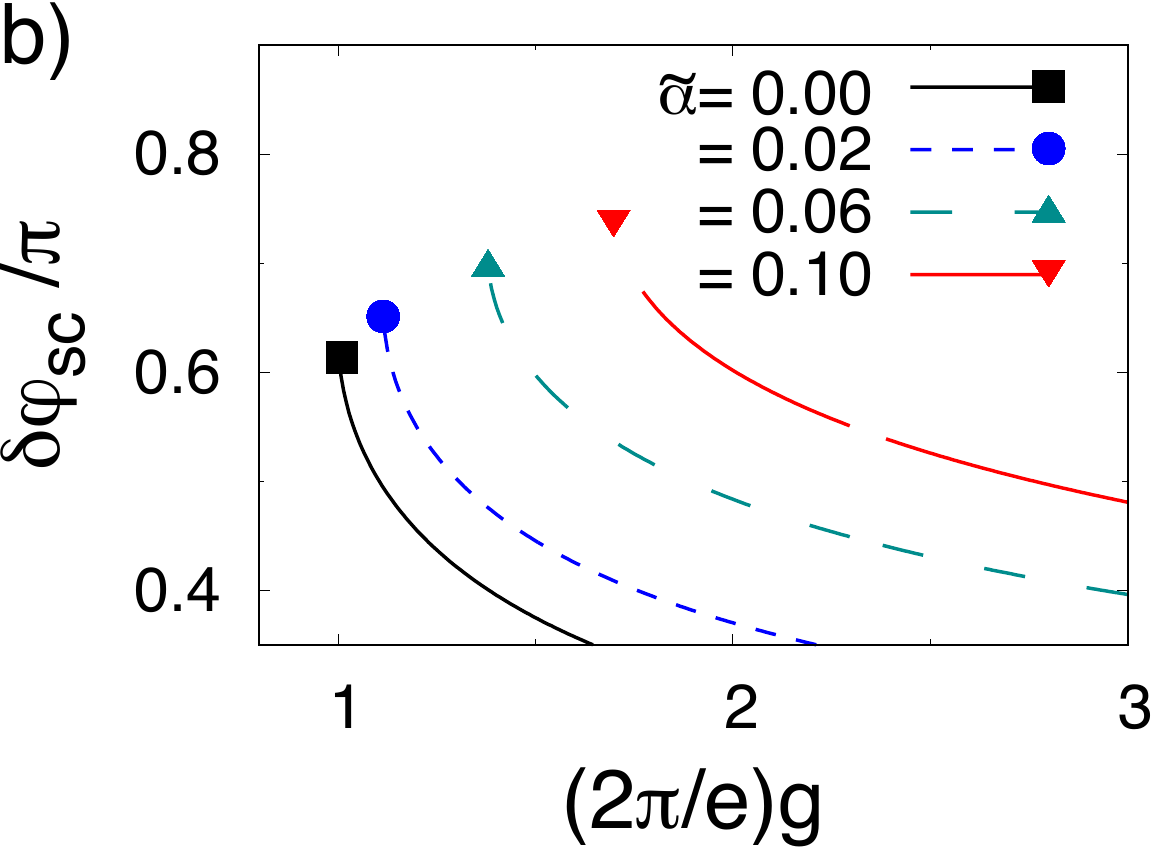}\\
	\includegraphics[width=0.49\columnwidth]{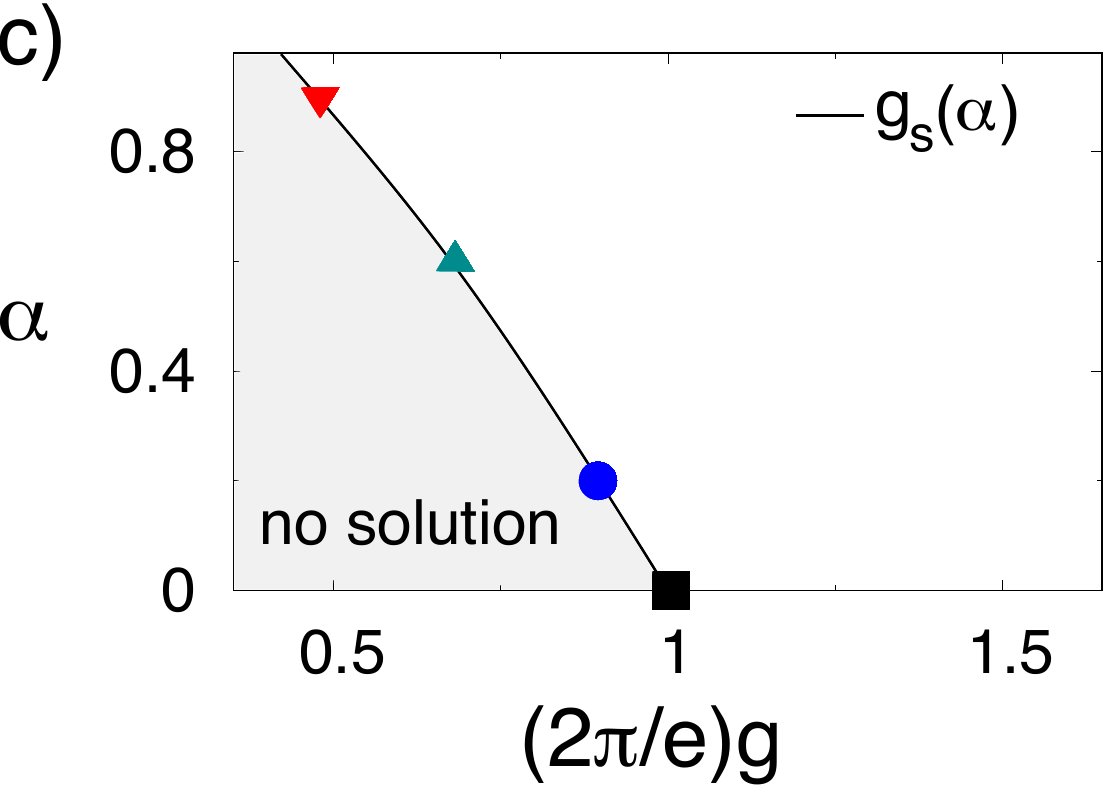}\hfill \includegraphics[width=0.49\columnwidth]{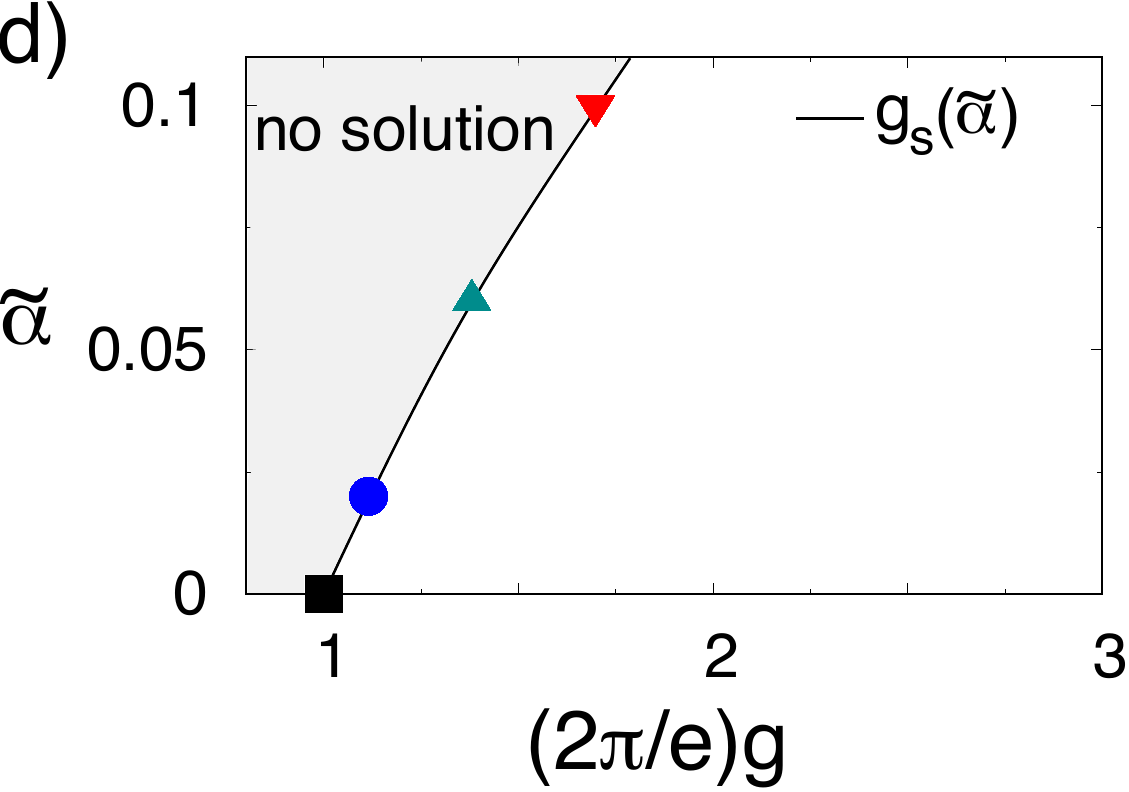}
	\caption{
	$\delta\varphi_{sc}$ as a function of the parameter $g$, for different values of the conventional damping $\alpha$ in (a), 
	and of the unconventional damping $\widetilde{\alpha}$ in (b). 
	The endpoints determine the critical values of $g_s(\alpha)$ 
	and $g_s(\widetilde{\alpha})$ below which  there is no solution of the self-consistent equation.
	In (c) and (d) the sets of the critical points $g_s(\alpha)$  and  $g_s(\widetilde{\alpha})$ are shown for the different 
	damping coefficients. The gray area denotes where the self-consistent equation has no solution. 
	}
\label{FIG:3}
\end{figure}

We first discuss the conventional dissipation  $\alpha > 0$ and $\widetilde{\alpha}=0$ 
for which we recover previous results obtained with the SCHA \cite{Chakravarty:1986,Chakravarty:1988hb}.
In Fig.~\ref{FIG:3}(a) we show $\delta\varphi$ for different 
values of $\alpha$,  by varying the system parameter $g$.
For reference, we also plot the dissipationless case $\alpha=0$ (black solid line). 
The endpoint of each line corresponds to the critical value $g_s(\alpha)$, where the SCHA solution vanishes. 

For a fixed value of $g$, the phase fluctuations decrease with increasing damping. 
As a consequence, the critical value $g_s$, determined by the critical solution of
the self-consistent equation, decreases. 
The corresponding phase diagram is shown in Fig.~\ref{FIG:3}(c), 
reporting the critical values $g_s$.
From this result, one can conclude that the dissipation stabilizes the ordered phase of the system.
Indeed, a more refined treatment beyond the SCHA yields the same qualitative behavior 
of the critical line, namely the negative slope of $\alpha$ vs $g_s$ with a shift 
towards smaller critical values \cite{Bobbert:1990,Panyukov:1987,Korshunov:1989}.

%
%
%
%
%
%
%

We now consider the opposite limit of purely unconventional dissipation affecting the system, i.e. 
$\widetilde\alpha > 0$ and $\alpha =0$. 
The behavior of the fluctuations as a function of $g$ and for different values of $\widetilde\alpha$ is shown in 
Fig.~\ref{FIG:3}(b).
Again, the black solid line corresponds to the dissipationless case $\widetilde\alpha=0$. 
Compared to the  previous results, the system displays now an opposite behavior: 
for a fixed value of $g$, the phase fluctuations increase with increasing damping. This can be  explained by  the Heisenberg uncertainty relation: 
the unconventional dissipation quenches the zero-point fluctuations of the momentum (charge) $\delta p$
leading to an increase of the phase and phase-difference fluctuations $\delta \varphi \sim \hbar/\delta p$.
As shown in Fig.~\ref{FIG:3}(d), the unconventional dissipation leads to an increasing critical value $g_s(\widetilde{\alpha})$.
A qualitatively similar result was obtained for the phase diagram of the superconductor/insulator transition 
occurring in a chain of Josephson junctions that was capacitively coupled to a metallic conducting 
film in the diffusive regime \cite{Lobos:2011}. 

%
%
%
\begin{figure}[t!]
	\includegraphics[width=0.49\columnwidth]{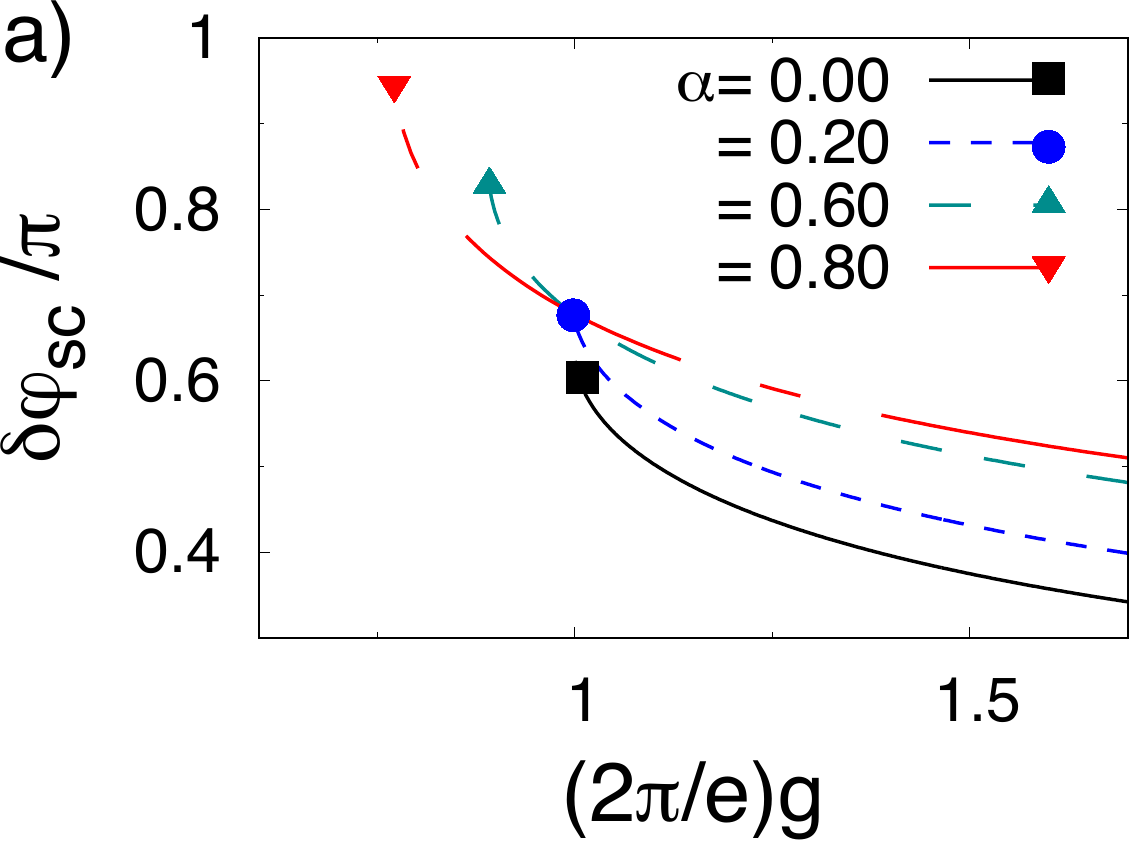}\hspace{1mm}\includegraphics[width=0.49\columnwidth]{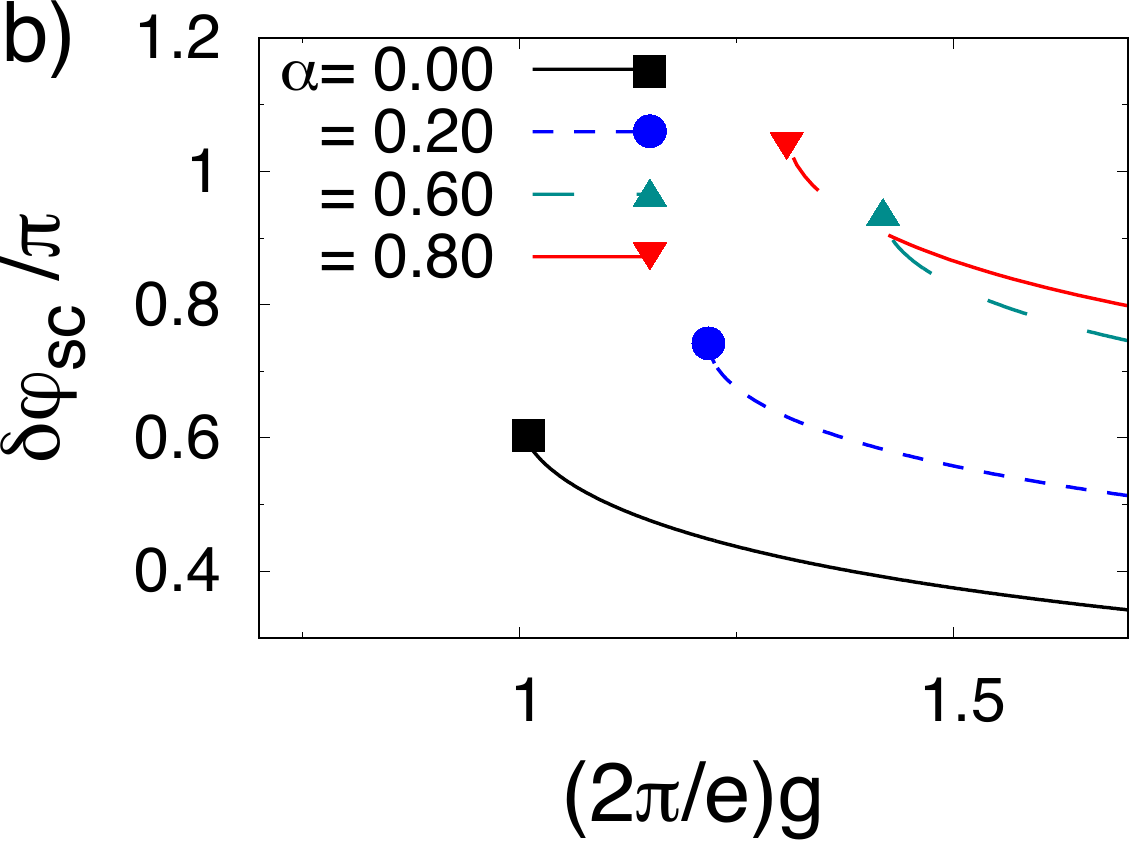}\\
	\includegraphics[width=0.49\columnwidth]{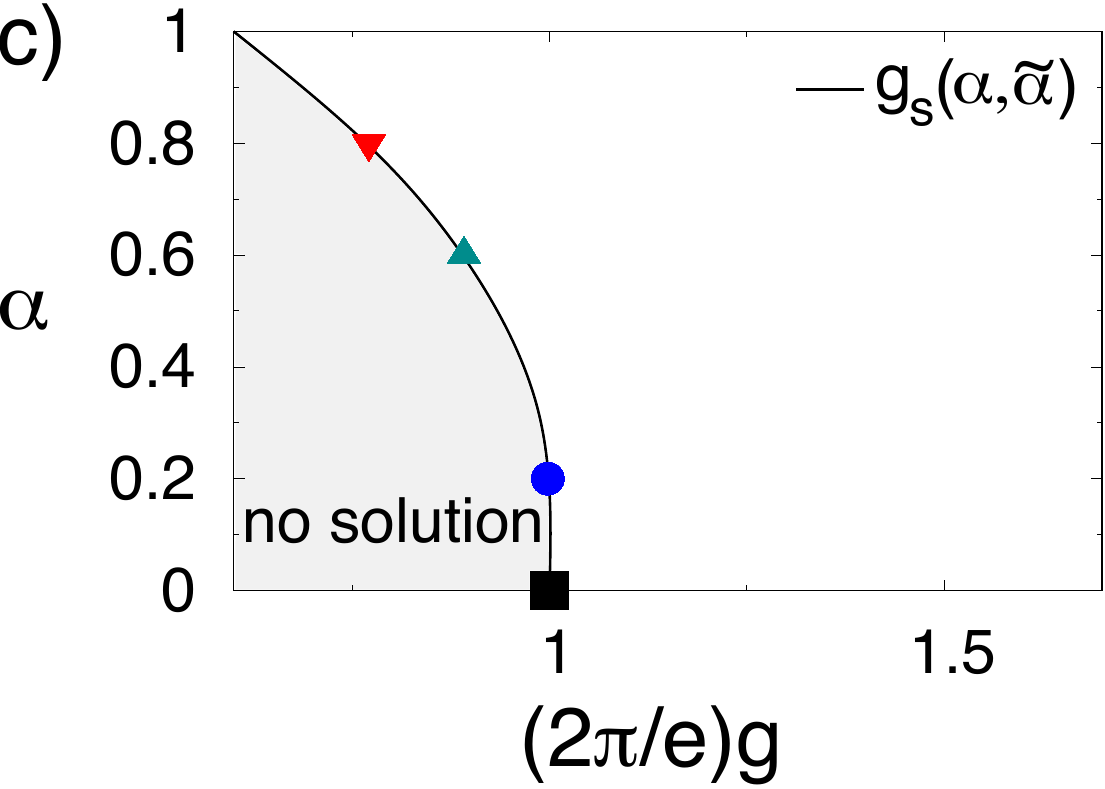} \includegraphics[width=0.49\columnwidth]{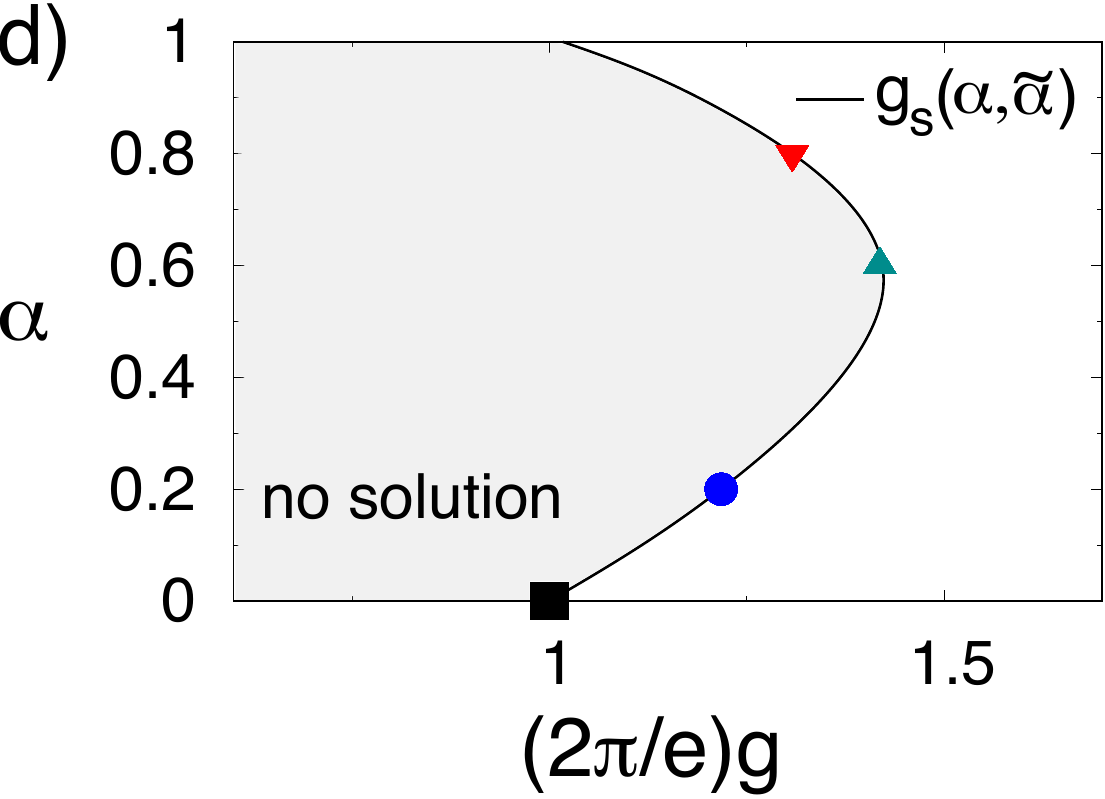}
	\caption{$\delta\varphi_{sc}$ as a function of the parameter $g$, for different values of the conventional damping $\alpha$ at a given ratio of $\widetilde{\alpha}/\alpha=0.1$  
	in (a), and $\widetilde{\alpha}/\alpha=0.3$ in (b).
	The endpoints determine the critical value of $g_s(\alpha,\widetilde{\alpha})$, 
	below which there is no solution of the self-consistent equation. These are reported in (c) for $\widetilde{\alpha}/\alpha=0.1$, and in (d) for $\widetilde{\alpha}/\alpha=0.3$. The gray area denotes where the self-consistent equation has no solution. 
	}
\label{FIG:4}
\end{figure}
%
%
%
%
%
%
%
%
%
%
\begin{figure}[b!]
\includegraphics[width=0.99\columnwidth]{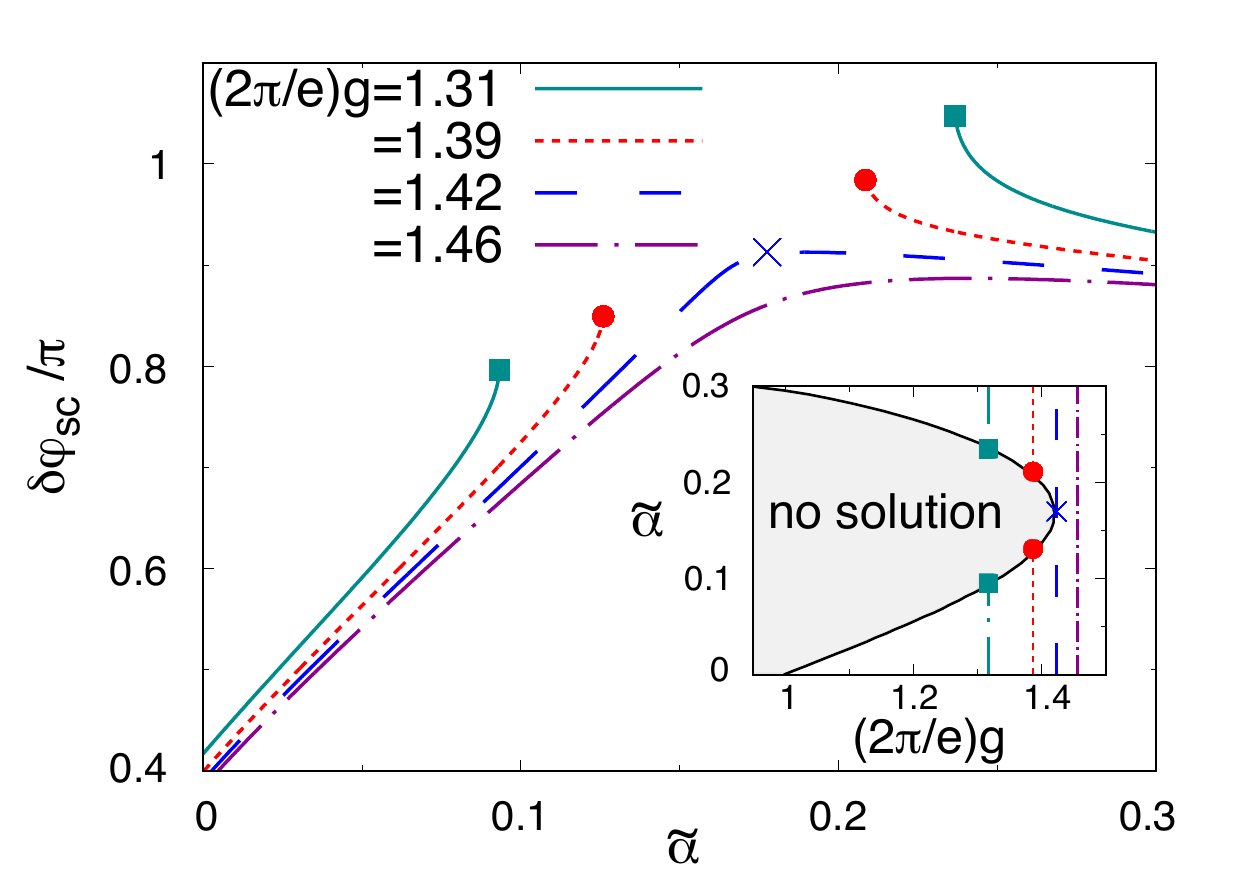}
\caption{
	$\delta\varphi_{sc}$ as a function of the parameter $\widetilde{\alpha}$ (with the ratio $\widetilde{\alpha}/\alpha=0.3$), 
	for different values of $g$.
	The end points determine the critical value(s), 
	above which there is no solution of the self-consistent equation 
	(see shaded area in the inset displaying the phase diagram in $\widetilde{\alpha}$ and $g$).
}
\label{FIG:5}
\end{figure}
%
%
%
%

We now analyze the general case when both types of 
dissipation are present: $\alpha > 0 $ and $\widetilde\alpha > 0$. 

Since conventional (or phase) dissipation quenches the phase fluctuations whereas 
unconventional (or momentum) dissipation yields a quenching of the momentum fluctuations, we expect a competition of the two types of dissipative interactions  
as they affect two canonically conjugate operators.  
In Figs.~\ref{FIG:4}(a) and (c) we show the results for a given ratio $\widetilde\alpha/\alpha=0.1$, 
for which we obtain a qualitatively similar behavior 
to the case of $\widetilde\alpha=0$.
A different behavior occurs in the regime when momentum dissipation has a stronger influence. As an example of this regime, we show in Figs.~\ref{FIG:4}(b) and \ref{FIG:4}(d) the results for the ratio $\widetilde\alpha/\alpha=0.3$. 
In this case, the trend appears to be inverted:  increasing the overall dissipative coupling strength,
the values $g_s(\alpha,\widetilde{\alpha})$ exhibit a non monotonic behavior.
Starting from a small value of the dissipation ($\alpha =0.2 $ or  $\widetilde{\alpha} =0.06$),
the critical value increases, as in the case of purely unconventional dissipation. 
However, for larger values of dissipation ($\alpha > 0.6 $ or  $\widetilde{\alpha}  > 0.18$)
$g_s(\alpha,\widetilde{\alpha})$ decreases, as in the case of a purely conventional dissipation.

In order to gain a better understanding of this regime, we also report the phase fluctuations as a function on the damping 
coefficient at a fixed ratio $\widetilde\alpha/\alpha$, and for different values of $g$ (see Fig. 5).
For large values $(2\pi/e)g \geq 1.42 $, we always obtain a solution for the self-consistent equation 
for all values of the damping coefficient.
As long as $(2\pi/e)g \leq 1.42$, there is a solution for both small and large values of the dissipative strength interaction, 
whereas there is a region of no solution  at intermediate values. 
This result stems from the behavior of the quantum fluctuations for the position or momentum 
of a harmonic oscillator with two non commuting dissipative interactions.
In this case, the fluctuations show  a non monotonic behavior as a function of the dissipative coupling 
strength \cite{Rastelli:2016ge}.
For instance, at  $(2\pi/e)g=1.42$ , in Fig.~\ref{FIG:5}, it is possible to observe a weak non monotonic behavior. 
However, in contrast to a pure harmonic oscillator for which strong fluctuations are always allowed at any scale, 
the solution for the self-consistent equation vanishes at large phase fluctuations and this produces  
a cut of the lines for values $(2\pi/e)g < 1.42$, shown in Fig.~\ref{FIG:5}.

%
%
%
\begin{figure}[b]
		\includegraphics[width=0.99\columnwidth]{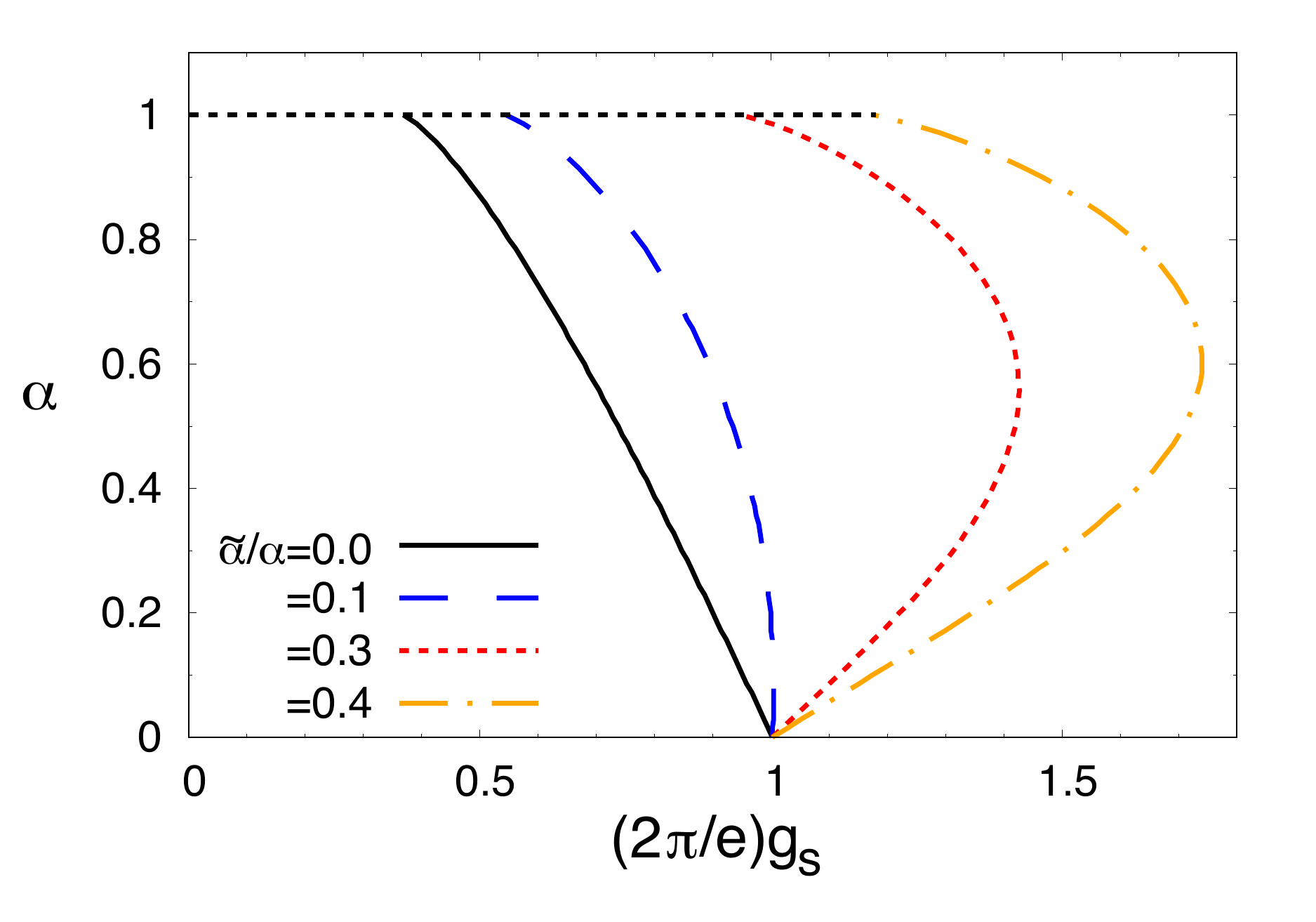}
		\caption{Critical line $g_s(\alpha,\widetilde{\alpha})$ with dissipative frustration.
		Right to the transition line is the ordered phase, on the left the rotors are randomly orientated (see Fig.~\ref{fig2:main_result}). By increasing the ratio $\widetilde\alpha /\alpha$ the non monotonic behavior is more pronounced.  
		}
\label{fig:phase}
\end{figure}
%
%
%
%

Finally, we analyze the evolution of the phase diagram between the two regimes
of Fig.~\ref{FIG:4}(c) and (d), 
and plot the critical line $g_s(\alpha,\widetilde{\alpha})$  and for different ratios $\widetilde{\alpha} / \alpha$ in Fig.~\ref{fig:phase}.
The region to the right of the transition line presents a solution of the self-consistent equation and is associated to
the phase with phase ordering, whereas in the region to the left there is no self consistent solution.
%
%
Further, as previously reported (see Ref.~\cite{Chakravarty:1988hb} for example), above the critical damping $\alpha_C=1$ the system is always in the ordered phase.

As discussed above, at small dissipative coupling strengths we observe an evolution from the regime of negative to positive slope of the critical line.
Moreover, the critical line exhibits a change in the global behavior.
In the regime $\widetilde\alpha/\alpha < \xi$, with the critical threshold $\xi \approx0.1$, 
the critical value $g_s$ decreases with the dissipative coupling.
In the opposite regime $\widetilde{\alpha}/\alpha > \xi$,  the critical value $g_s$
first increases with the dissipative coupling and then decreases again at larger damping.
Such non monotonic behavior is more pronounced for larger values of the ratio $\widetilde{\alpha}/\alpha$. 
The phase diagram reported in Fig.~\ref{fig:phase} implies the interesting possibility that, 
for a given ratio of the parameter $g$, the system exhibits two phase transitions by increasing 
the dissipation: the first one from the ordered phase to the disordered phase and then, 
by further increasing the damping,  one drives the system back to the ordered phase, see Fig.~\ref{FIG:5}.

%
%
%
\section{ORDER of the PHASE TRANSITION}
\label{sec:order-phase-transition}

In this section we consider the phase diagram of the system by using a criterion of Eq.~(\ref{Eq:bogu}) based on 
the upper bound $\mathcal{F}_{v}(V_{tr} =  V_{sc})$ of the free energy $\mathcal{F}_{\rm eff}$, 
where $V_{sc}$ is the self-consistent solution of Eq.~(\ref{eq:selfcon2}).
When $\mathcal{F}_{v}(V_{sc})  \geq \mathcal{F}_{v}(0)$, 
the real local minimum in the SCHA corresponds to the solution $V_{tr}=0$ 
which represents the real upper bound estimation of the exact free energy.
Within the SCHA, this point corresponds to the phase transition in which the spin-wave stiffness vanishes 
and the system is in the disordered state.

\begin{figure}[t!]
	\includegraphics[width=0.49\columnwidth]{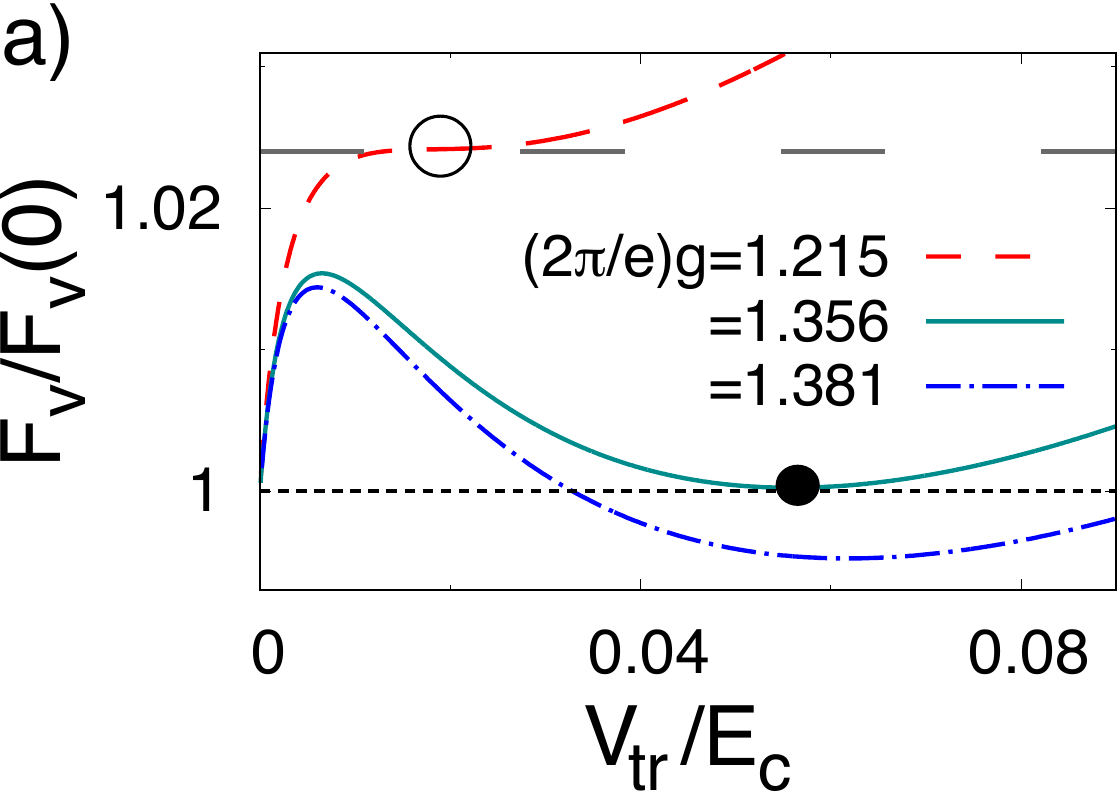}\hfill\includegraphics[width=0.49\columnwidth]{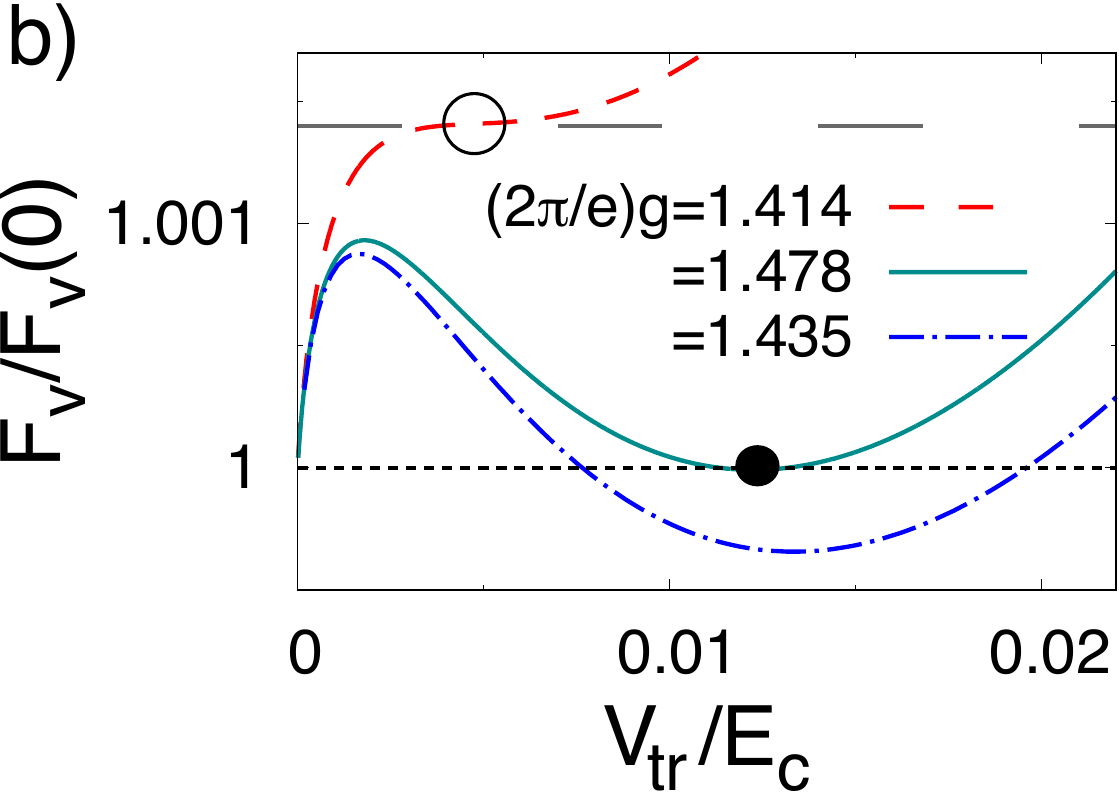}\\
	\includegraphics[width=0.49\columnwidth]{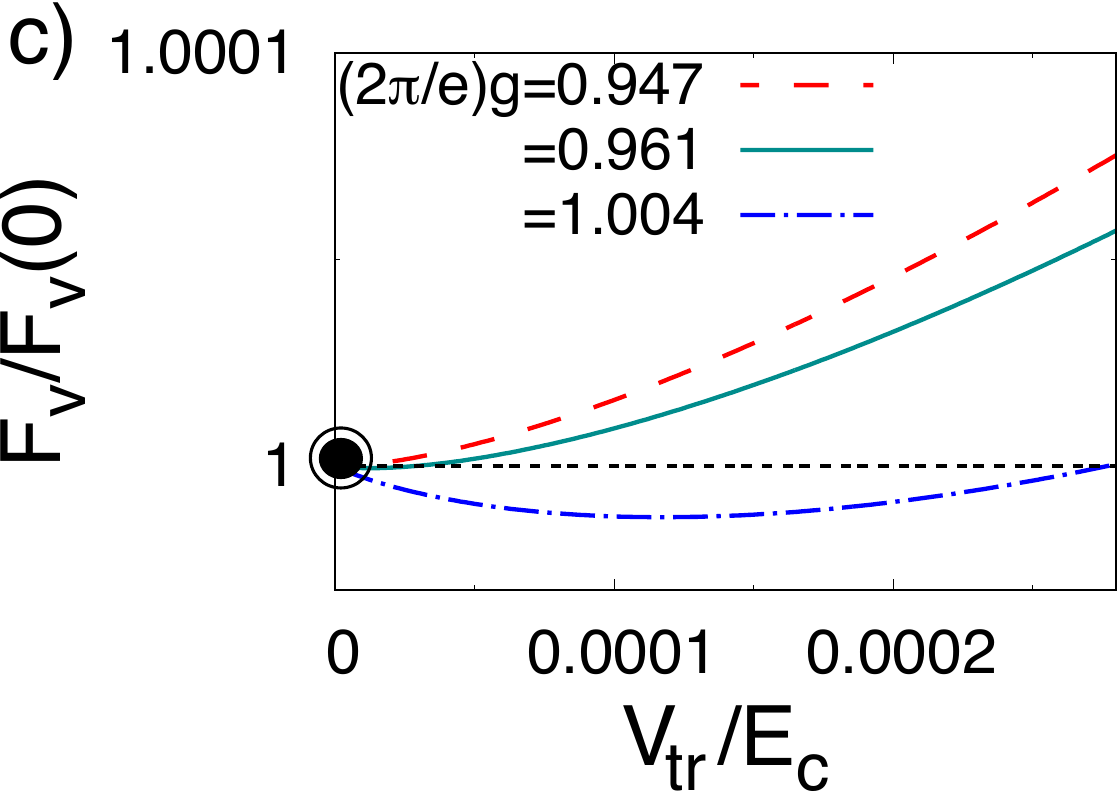}\hfill \includegraphics[width=0.49\columnwidth]{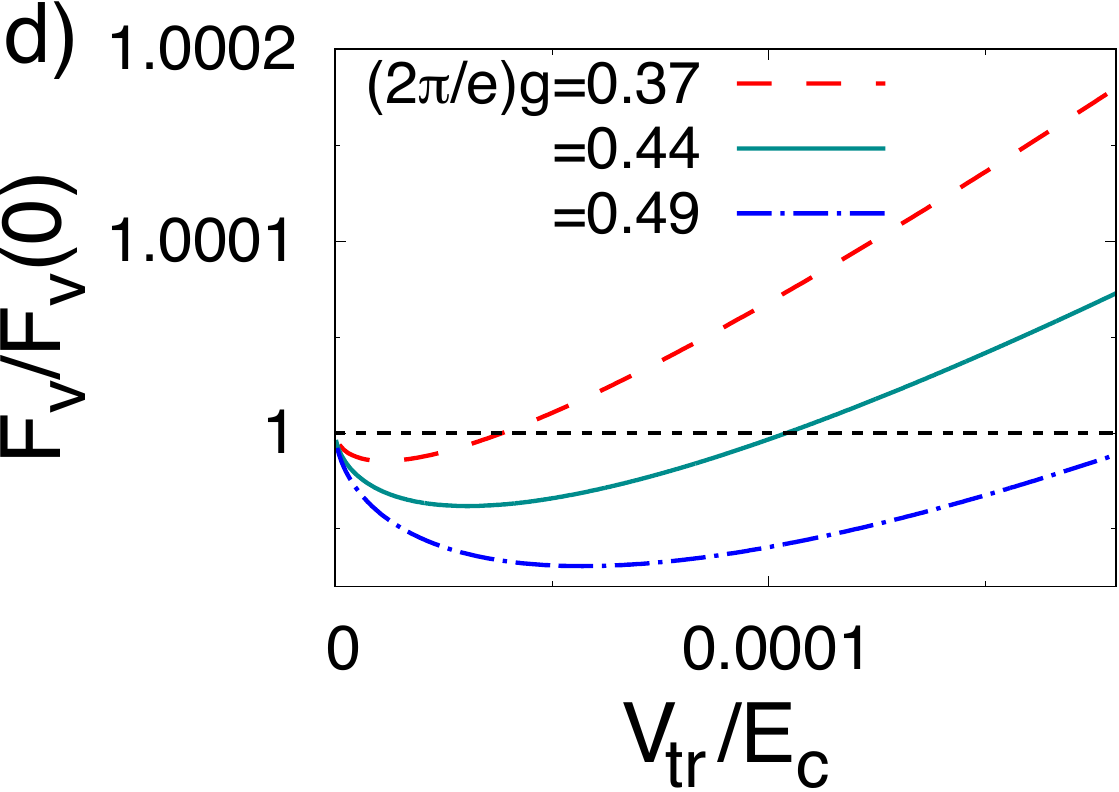}

	\caption{Upper bound $\mathcal{F}_{v}(V_{tr}) $ for free energy in Eq. (\ref{Eq:bogu}) as a function of $V_{tr}$ in presence of dissipative frustration, for the ratio $\widetilde{\alpha}/\alpha=0.3$ and a) $\alpha = 0.2$, b) $\alpha =0.6$, c) $\alpha =1.0$ and d) $\alpha =1.5$. The empty circles correspond to the spinodal point, i.e. the disappearance of the finite solution $V_{sc}$ in the self consistent equation (\ref{eq:selfcon2}). The solid dots correspond to the phase transition according to the criterion $\mathcal{F}_{v}(V_{sc})=\mathcal{F}_{v}(0)$.  }
	\label{FIG:7}
\end{figure}

An example of the behavior of  $\mathcal{F}_{v}(V_{tr}) $ is reported in 
Fig.~\ref{FIG:7}, showing the free energy for different values $\alpha$ with frustrated dissipation.
In this figure, the circle corresponds to the spinodal point of the self-consistent equation, while
the black dots to the condition $\mathcal{F}_{v}(V_{sc})  =  \mathcal{F}_{v}(0)$.
The latter condition occurs at a value of $g_c$ which is generally larger than the $g_s$ found by the self-consistent Eq.~(\ref{eq:selfcon2}).
Hence, the phase transition shifts to larger values of $g$.  
This jump, from a finite value of $V_{sc}$ to zero, corresponds to a first order phase transition.

However, by increasing the coupling strength $\alpha$ we see that the 
difference between the spinodal point and the global minimum disappears.  
In particular, for values  $\alpha > 1$  the system is always in the ordered state and the point 
at $V_{tr}=0$ is a maximum for all values of $g>0$. At the point $\alpha_C = 1$  the transition turns to be of second order. 

%
%
%
%
%
%
%

\begin{figure}[t!]
	\includegraphics[width=0.99\columnwidth]{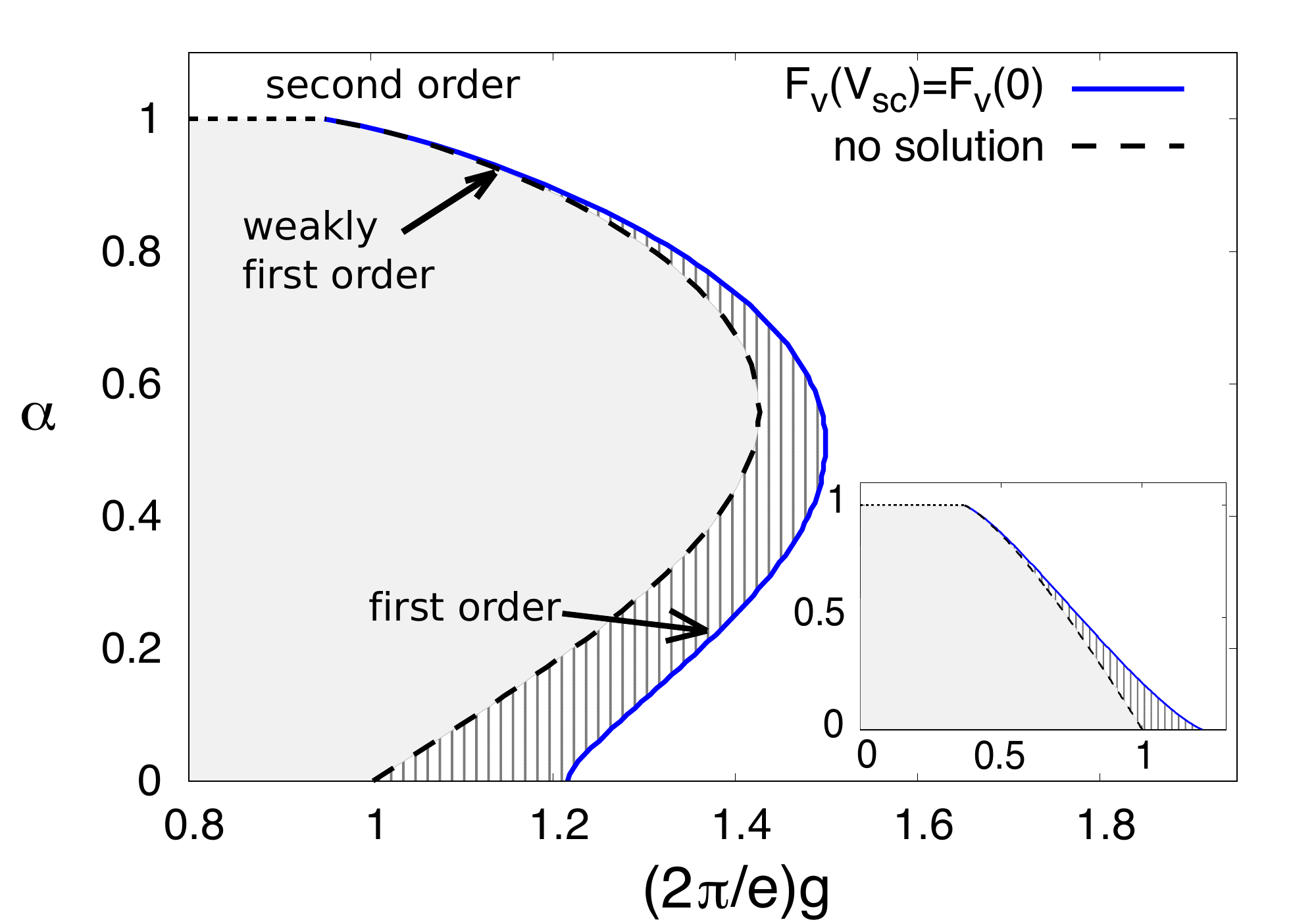}
	\caption{Quantum phase diagram with dissipative frustration for the ratio $\widetilde{\alpha}/\alpha = 0.3$. The solid line and the dashed line correspond, respectively, to the phase transition according to the criterion $\mathcal{F}_{v}(V_{sc})=\mathcal{F}_{v}(0)$ and the vanishing of the solution in the self consistent equation. The inset shows the case with conventional dissipation ($\widetilde{\alpha}=0$). }
	\label{FIG:8}
\end{figure}

To summarize we can identify three regimes. 
In the low damping regime we have a first order phase transition, see Fig.~\ref{FIG:7}(a,b).
Increasing the damping the jump in the parameter gets smaller and we call the transition "weakly" first order, Fig.~\ref{FIG:7}(c).
Further increasing the damping, $V_{sc}$ tends continuously to zero and we have a second order phase transition, Fig.~\ref{FIG:7}(d).
We illustrate those three regimes by plotting the phase diagrams originating from the self-consistent equation discussed in the previous section with the one obtained from the free energy minimum. Fig.~\ref{FIG:8} shows the phase diagrams of the two different criteria (dashed black line for the spinodal points,  blue solid line for the free energy approach).

%
%
%
%
%
%
%

%
%
%
\section{Purity and Entanglement}
\label{sec:purity-entanglement}
A natural question is whether the two ordered phases at weak and strong dissipative coupling can be characterized 
by another intrinsic property beyond the (classical)  ordering of the phases.
We discuss this issue in the next section by studying the purity and the logarithmic negativity.

In the SCHA, the system is described by an effective density matrix $\hat{\rho}_{sc}$ which is formally Gaussian. Using the representation with the amplitudes of the harmonic modes, the  elements of $\hat{\rho}_{sc}$ read
\begin{equation}
\langle\{\varphi_{k}\}|\hat{\rho}_{sc}|\{\varphi_{k}'\}\rangle=\prod_{\nu=\text{Re},\text{Im}}\prod_{k=1}^{N-1}\frac{1}{\sqrt{\pi\langle {\left|\varphi_k\right|}^2 \rangle_{sc}}}e^{-\frac{\mathcal{S}_{k,\nu}}{\hbar}},
\label{Eq:density_matrix}
\end{equation}
with $\{\varphi_k\}=\{\varphi_{k=1},\varphi_{k=2},..\}$ and $\varphi_k=\left(\varphi_{k,\text{Re}}+i\varphi_{k,\text{Im} }\right)$. 
The exponent reads
\begin{equation}
\frac{\mathcal{S}_{k,\nu}}{\hbar}= \frac{\left(\varphi_{k,\nu}+\varphi_{k,\nu}'\right)^2}{4\langle {\left|\varphi_k\right|}^2 \rangle_{sc}}+\frac{\left(\varphi_{k,\nu}-\varphi_{k,\nu}'\right)^2}{4}\langle {\left|\dot{\varphi}_k\right|}^2 \rangle_{sc},
\label{Eq:action_density_matrix}
\end{equation}
where we used $\langle \varphi^2_{k,\text{Re}}\rangle = \langle \varphi^2_{k,\text{Im}}\rangle$.\\
However, even if $\hat{\rho}_{sc}$ is a Gaussian functional of the fluctuations, we calculate the quantities in this section by solving the self-consistent equation, which takes the anharmonicity of the cosine potential into account. 

\subsection{Purity}
As a measure of the correlations between the system and the environment, 
we calculate the purity of the system which is defined as 
\begin{equation}
\mathcal{P} = \mbox{Tr}\left[ \hat{\rho}_{sc}^2 \right],
\end{equation}
where $\hat{\rho}_{sc}$ is  the reduced density matrix describing our system, the one dimensional superconducting chain.
For pure quantum states, one has $\mathcal{P} = 1$ (isolated system) whereas $\mathcal{P}  < 1$ for statistical mixture 
of states \cite{Nielsen:2010}. 

Due to the fact that our system is described by an effective ensemble of independent harmonic modes,  
the purity is simply related to the inverse product of the phase difference $\langle {\left| \varphi_k\right|}^2 \rangle $ and 
momentum (charge) fluctuations $\langle \left|\dot{\varphi}_k\right|^2 \rangle$ (we drop the subscript ${sc}$ for the fluctuations from now on).
For the isolated system, without dissipation ($\alpha=\tilde{\alpha}=0$), 
increasing the parameter $g$, the phase fluctuations decrease while the charge  
fluctuations  increase. 
Anyway,  the product of the two fluctuations is invariant 
and the purity remains  $\mathcal{P}=1$, viz. 
the system is in a pure quantum state.
Hence, we express the purity of the general case as 
\begin{equation}
\label{Eq:purity}
\mathcal{P} = \prod_k \mathcal{P}_k
= \prod_k \
\sqrt{ \frac{ {\langle {\left|\varphi_k\right|}^2 \rangle}_0 {\langle {\left|\dot{\varphi}_k\right|}^2 \rangle}_0 }
{  \langle {\left|\varphi_k\right|}^2 \rangle   \langle {\left|\dot{\varphi}_k\right|}^2 \rangle } }
 \, , 
\end{equation}
where $\langle {\left|\varphi_k\right|}^2 \rangle_0$ and $\langle {\left|\dot{\varphi}_k\right|}^2 \rangle_0$ denote the fluctuations without dissipation and the expression for the velocity fluctuations is given by 
\begin{equation}
\label{Eq:zeroTempv}
 \langle {\left|\dot{\varphi}_k\right|}^2 \rangle
=
g^2\frac{ ( \phi_c+  \phi_d )}{\pi }  + 
\frac{\alpha\sin\left(\frac{\pi k}{N}\right)}{\pi^2}  \frac{\ln\left[\omega_c/ \omega_{k}^{(sc)} \right] }{1+\sigma^2_k} \, ,
\end{equation}
with 
\begin{eqnarray}
\phi_c  &=&\frac{\alpha}{\pi g^2}\frac{ \sin\left(\frac{\pi k}{N}\right) }{1+\sigma_{k}^{2}}  \left[
\ln(1+\sigma_{k}^{2})+\sigma_{k}\arctan(\sigma_{k}) 
\right] \\
\phi_d  &=& \frac{1}{1+\sigma_{k}^{2}}
\left(\!\! \frac{\mbox{} \,\, \hbar \omega_{k}^{(sc)} }{E_J}
-
\frac{\alpha \sin\left(\frac{\pi k}{N}\right)}{\pi g^2}  \, 
\Gamma_{-}^{(sc)} \!\! 
\right) 
\mathcal{\bold F} \!\! \left[ \Gamma_{-}^{(sc)}  \!\!, \Gamma_{+}^{(sc)} \right] , 
\nonumber\\
\end{eqnarray}
where $\sigma_{k},\;\mathcal{\bold F} $ and $\Gamma_{\pm}^{(sc)}$ have been introduced in  Sec.~\ref{sec:model}.
Inserting the expressions (\ref{Eq:zeroTemp}), (\ref{Eq:zeroTempv}) in (\ref{Eq:purity}), we  
calculate the purity and discuss the influences of the baths.

The interaction with the external environment always leads to a mixing of the quantum states with a purity lower than one.
This occurs for each single harmonic mode $ \mathcal{P}_k  < 1$.
Then, the purity of the whole system is given by the product of all $\{\mathcal{P}_k\}$ 
corresponding to a small number in the limit of large $N \gg 1$.
Therefore,  it is more convenient to analyze the behavior of the geometric mean of the purity 
defined as $\mathcal{P}^{1/N}$.

\begin{figure}[t!]
	\includegraphics[width=0.49\columnwidth]{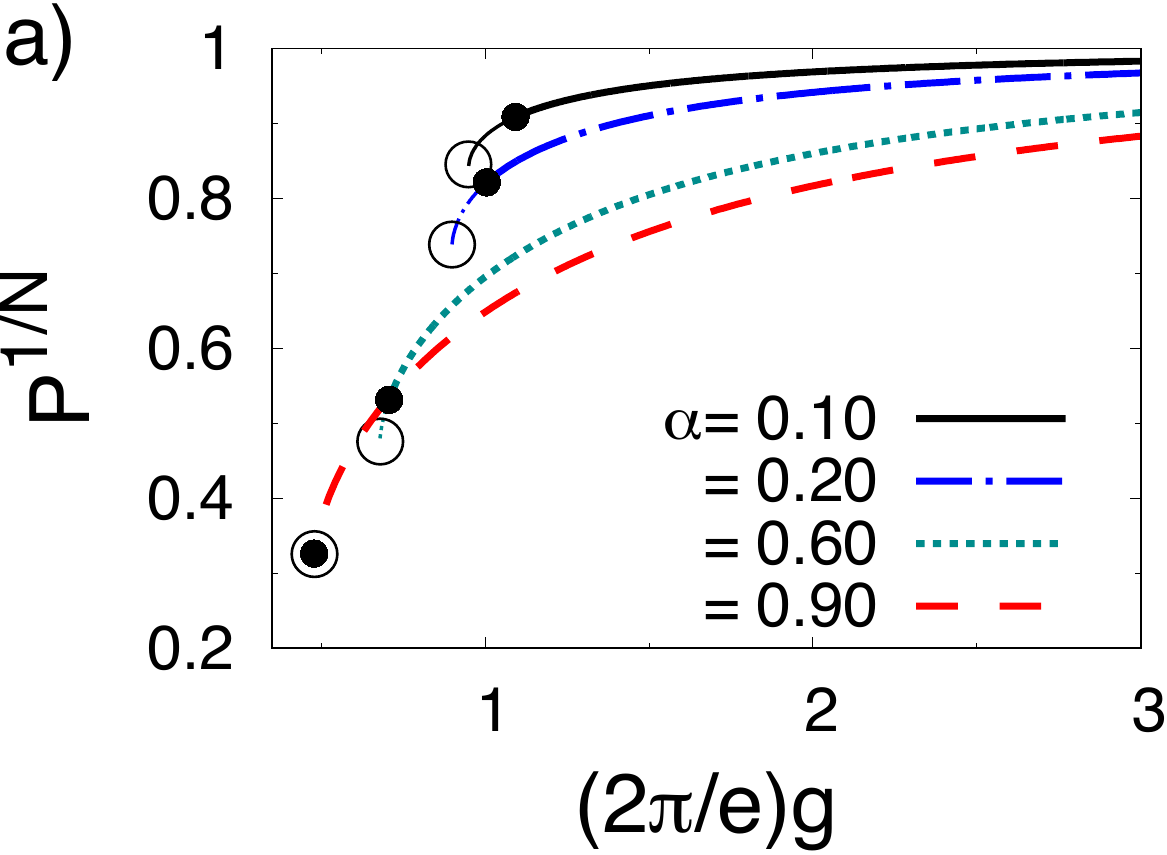}\hfill\includegraphics[width=0.49\columnwidth]{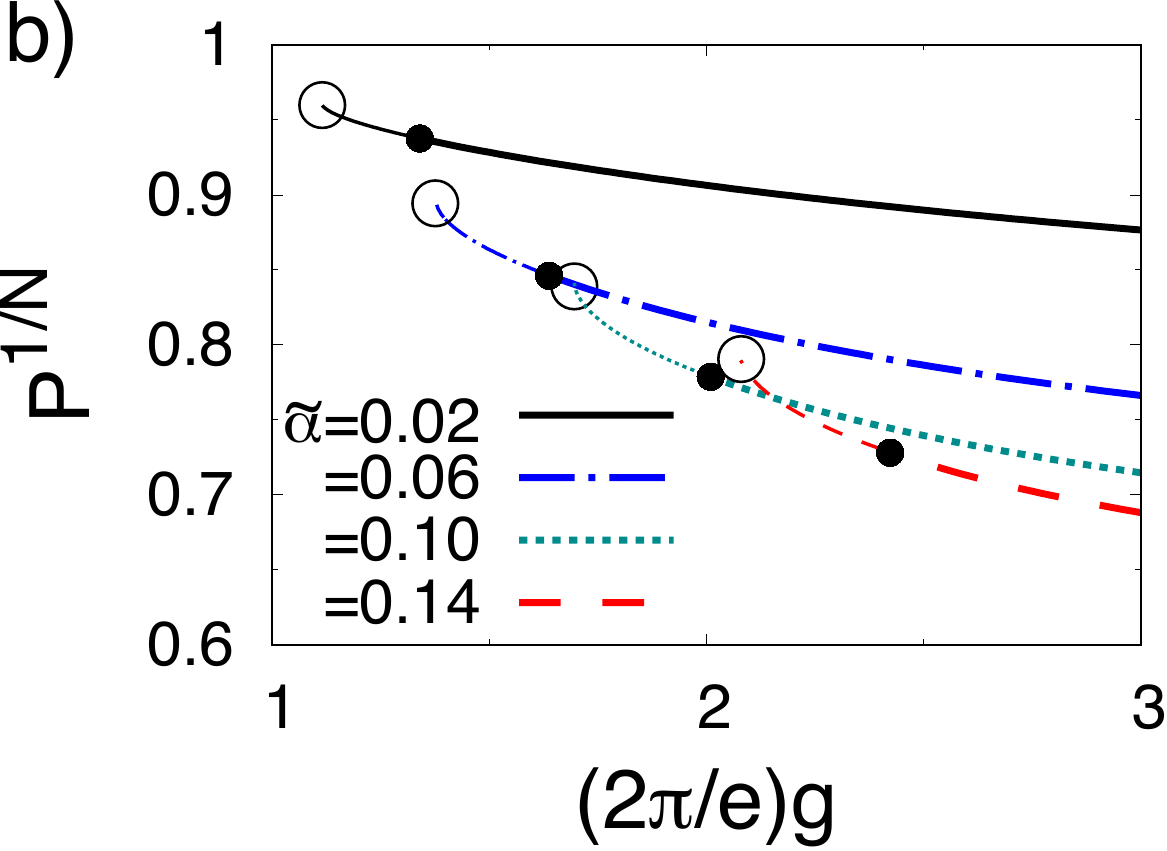}\\
	\vspace{0.5cm}
	\includegraphics[width=0.98\columnwidth]{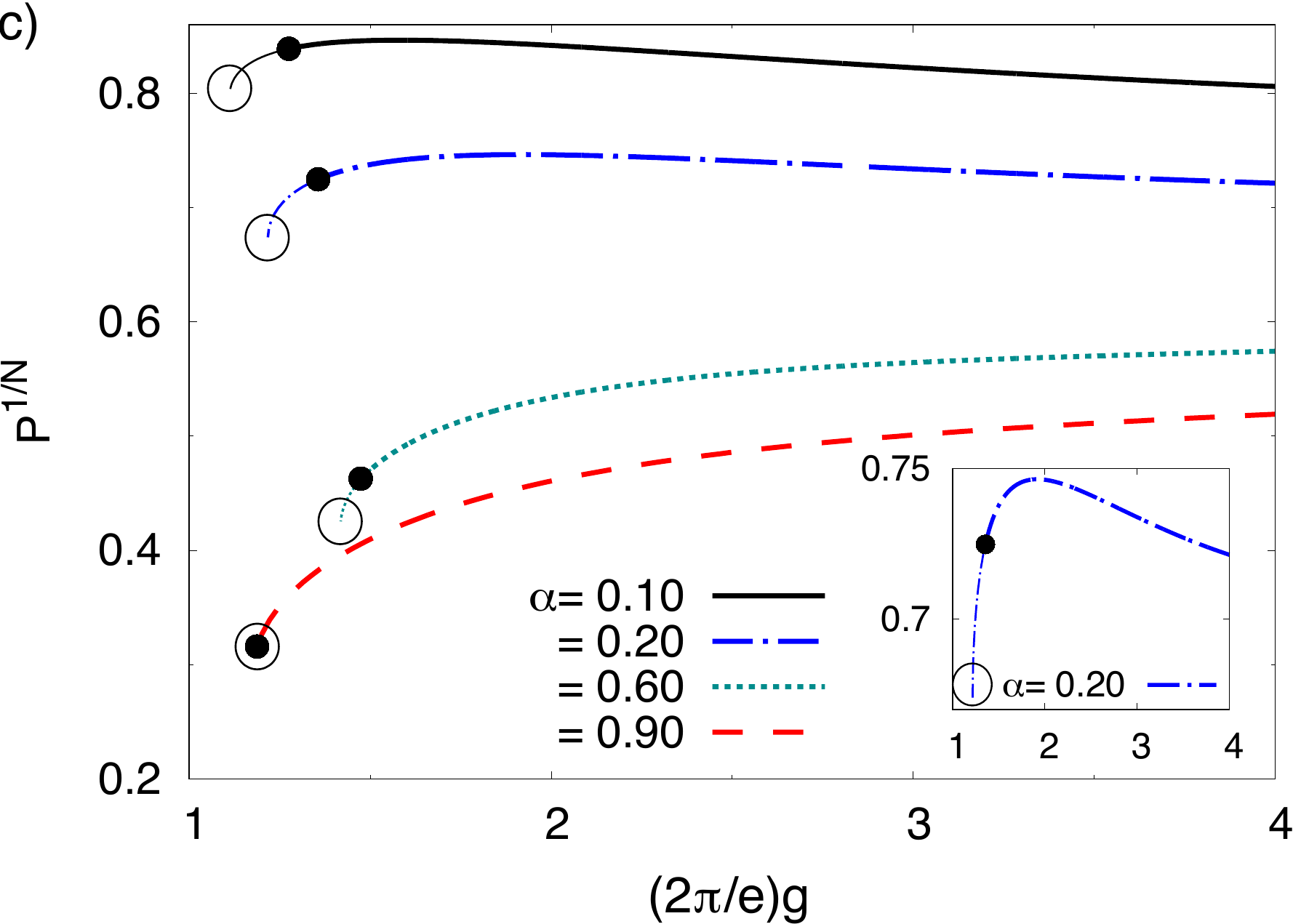}
	\caption{Geometric mean  purity  $\mathcal{P}^{1/N}$, for $N=150$ ,  as a function of $g$.  
	The solid dot corresponds to the phase transition $\mathcal{F}_v(V_{sc})=\mathcal{F}_v(0)$ while the open circles mark the disappearance of the self-consistent solution.  
	a) Conventional dissipation for different values of $\alpha$.   
	b) Unconventional dissipation for different values of $\widetilde{\alpha}$.
	c) Frustrated case for the ratio $\widetilde{\alpha}/\alpha = 0.3$. 
	The inset is a zoom for the case $\alpha=0.2.$}
	\label{FIG:9}
\end{figure}

Fig.~\ref{FIG:9}(a) shows the mean purity  as a function of $g$ for different values of $\alpha$ 
in the case of conventional dissipation $(\widetilde{\alpha}=0)$ whereas Fig.~\ref{FIG:9}(b) reports the 
mean purity  in the case of pure unconventional dissipation $(\alpha=0)$.
The black solid dots in Fig.~\ref{FIG:9} correspond to the critical value $\mathcal{F}_v(V_{sc})=\mathcal{F}_v(0)$ 
and the end points (open circles) correspond to the vanishing of the solution in the self-consistent equation.
In both cases, as expected, the purity of the system decreases by increasing the dissipative coupling with the bath, 
with conventional dissipation $\alpha$ or the unconventional $\widetilde{\alpha}$. 
However, the purity shows the opposite behavior by varying $g$, in particular close to the critical point: 
it  decreases for the conventional dissipation and increases for the unconventional one.

The mean purity in the case of frustrated dissipation ($\alpha>0$ and $\widetilde{\alpha}>0$) is shown 
in Fig.~\ref{FIG:9}(c).
Remarkably, for $\alpha = 0.1$ and $\alpha = 0.2$, the mean purity has a non-monotonic behavior as a function of $g$.
This non-monotonicity is a characteristic feature of the dissipative frustration acting on the system 
since it combines the two opposite trends on the purity in the presence of a single type of dissipative interaction 
(phase or charge) affecting the system, see Fig.~\ref{FIG:9}(a) and (b).

%
%
%
%
%
%
%

\subsection{Entanglement}
In this section we analyze an entanglement measure to describe the non-classical correlations present in the quantum phase model 
with dissipative frustration.
Specifically, we discuss the behavior of the logarithmic negativity $E_{\mathcal{N}}$, a suitable entanglement measure to characterize 
Gaussian states \cite{Adesso2007,Werner2002}.\\
Before we discuss the logarithmic negativity and the results for  our system, a remark is needed.
To compute $E_{\mathcal{N}}$ we use a Gaussian density matrix $\hat{\rho}_{sc}$, see Eqs.~(\ref{Eq:density_matrix}) and (\ref{Eq:action_density_matrix}). The results we obtain in this way naturally can be different from the exact measure of  the quantum phase model with the cosine interaction having non-Gaussian correlations.
Since $E_{\mathcal{N}}$ fails to be superadditive, the results obtained with the Gaussian state do not represent a lower bound and can overestimate the amount of entanglement \cite{Cirac:2006}.
However, $E_{\mathcal{N}}$ is a simple and straightforward quantity to compute and it can provide a first, rough estimate of the possible behavior of the entanglement in our problem. \\
The logarithmic negativity is based on the Peres-Horodecki criterion (or Positive Partial Transpose, PPT) \cite{Peres:1996,HORODECKI1996} 
which states that if the global density matrix $\hat{\rho}$ for two combined subsystems $A$ and $B$ is separable 
(= no entanglement but only classical correlations), then the partial transpose density matrix respect to one of the two subsystems, for instance  
$\hat{\rho}^{T_A}$, has still positive eigenvalues. 
Hence, the amount of negativeness of the eigenvalues of $\hat{\rho}^{T_A}$ can be considered as a measure of the non-separability 
between $A$ and $B$, viz. entangled states are present.
Following this criterion, one defines the logarithmic negativity in our case as 
\begin{equation}
E_{\mathcal{N}} \left[ \hat{\rho} \right]= \log_2\left( ||\hat{\rho}^{T_A}||_1 \right ) = 
 \log_2 (  1 - 2 \sum_{\lambda_k< 0 } \lambda_k[\hat{\rho}^{T_A}] )
\, ,
\end{equation}
where  $\lambda_k[\hat{\rho}^{T_A}] $ are the eigenvalues 
of $\hat{\rho}^{T_A}$  and  $||M||_1$ denotes the trace norm of a matrix $M$,  $||M||_1=\text{tr}(\sqrt{M^\dagger M})$ and 
corresponds to the sum of the absolute values of its eigenvalues
\footnote{Since $\hat{\rho}_{sc}^{T_A}$  is still Hermitian with $\text{tr} \hat{\rho}_{sc}^{T_A}=\sum_k \lambda_k =1$ then we have 
$ ||\hat{\rho}^{T_A}||_1 = \sum_k | \lambda_k|  = 1 - 2  \sum_{\lambda_k < 0} | \lambda_k|$.}.
The PPT criterion is a sufficient condition, implying that even for $E_{\mathcal{N}} = 0$, the two subsystems can still have some 
entanglement  \cite{Fazio:2007}.

We calculate the logarithmic negativity $E_{\mathcal{N}}[\hat{\rho}_{sc}]$ for our system using the correlation covariance matrix \cite{Simon:2000,Werner2002,Adesso2007}.
A more detailed discussion of the formalism is given in Appendix \ref{Sec:symplect}, illustrating the case of two coupled oscillators.

We introduce the canonical conjugated variables $\hat{q}_n=\hat{Q}_n/(2e)$, i.e. the scaled charge operators on each superconducting island forming the chain, 
with the commutation relation $[\hat{\varphi}_n, \hat{q}_m] = i \delta_{nm}$.
We also define the vector
\begin{equation}
\hat{R}=
\left( \hat{R}_1,  \hat{R}_2, \dots, \hat{R}_N  \right)^T  \, , 
\end{equation}
with $\hat{R}_n = ( \hat{\varphi}_n , \hat{q}_n )$.  
The full symmetric covariance matrix $\hat{\sigma}[\hat{\rho}_{sc}]$ 
of size $(N \times N)$ is formed by the block elements $(2 \times 2)$ that read 
\begin{equation}
\hat{\sigma}_{nm}[\hat{\rho}_{sc}]= \langle\hat{R}_{l}\hat{R}_{m}+\hat{R}_{m}\hat{R}_{l}\rangle  /2 \, . 
\end{equation}
The correlation functions are given by
\begin{align}
\langle \varphi_n^2 \rangle
&= \langle \varphi^2 \rangle
=
\frac{1}{N} \sum_{k=1}^{N-1} 
 \langle {\left|  \varphi _k\right|}^2 \rangle 
\label{Eq:phase} \\
\langle \varphi_l\varphi_m\rangle &=
\frac{1}{N} \sum_{k=1}^{N-1} \cos\left(\frac{2\pi}{N}k(m-l)\right)  \langle {\left|  \varphi _k\right|}^2 \rangle , 
\label{Eq:phaselm}
\end{align}
and similar expressions for $\langle q_n^2 \rangle = \langle q ^2 \rangle  \propto \langle \dot{\varphi}^2 \rangle$ 
and $\langle q_n  q_m \rangle \propto   \langle \dot{\varphi}_n \dot{\varphi}_m \rangle$, whereas 
we have $\langle  \hat{\varphi}_n \,  \hat{q}_m \rangle = 0$.

After having partitioned the superconducting Josephson chain in two subsystems formed by the local variables $n_A=1,\dots,N_A$ and 
$n_B=1,\dots,N_B$, it is possible to show that the covariance matrix $\hat{\sigma} [\hat{\rho}_{sc}^{T_A}]$
associated to $\hat{\rho}^{T_A}_{sc}$ is  easily obtained by time reversal symmetry operations \cite{Simon:2000}, viz. by inverting 
all momenta of subsystem $A$, namely 
\begin{equation}
\hat{\sigma}[\hat{\rho}_{sc}] \rightarrow  \hat{\sigma}[\hat{\rho}_{sc}^{T_A}]
\quad \mbox{with}    \quad
\langle q_{n_A}  q_{m_B} \rangle \rightarrow  - \langle q_{n_A}  q_{m_B} \rangle \, ,
\end{equation}
and leaving  unmodified the products in each subsystem $\langle q_{n_A}  q_{m_A} \rangle$ and 
 $\langle q_{n_B}  q_{m_B} \rangle$. 
Finally, the connection between the  logarithmic negativity and the covariance matrix of the partial transpose matrix 
$\hat{\sigma}[\hat{\rho}_{sc}^{T_A}]$ is given by the formula  \cite{Serafini:2007} 
\begin{equation}
E_{\mathcal{N}}[ \hat{\rho}_{sc} ] = 
-\sum_k \log_2 \left( \min[1 , (2 c_k)]  \right) \, ,
\label{logneg}
\end{equation}
where the quantities $\{c_1,c_2,...,c_N\} $ are the {\sl symplectic} eigenvalues (spectrum) 
associated to the covariance matrix $\hat{\sigma}[\hat{\rho}_{sc}^{T_A}]$.
The symplectic spectrum is computed by finding the real  eigenvalues of 
the real symmetric matrix  $\hat{S} = - i \mathcal{O} \hat{\sigma}[\hat{\rho}_{sc}^{T_A}]$, 
namely  the product of the covariance matrix with the symplectic block diagonal matrix
\begin{equation}
\mathcal{O}=\bigoplus_{n=1}^{N}
\left(\begin{array}{cc}
0& 1\\ 
-1& 0
\end{array} \right) \, . \label{Eq:block}
\end{equation}
In the diagonal form, the matrix $\hat{S}$ reads $\text{diag}\{\pm c_1,\pm  c_2,...,\pm  c_N\}$ \cite{Serafini:2007}
(see Appendix \ref{Sec:symplect} for more details). 

The symplectic eigenvalues $\{ c_k \}$ are continuous functions of the correlation functions of the system. 

We find that the symplectic spectrum and hence the logarithmic negativity does not vary with $g$ 
without coupling with the environment. 
This results can be understood by scaling analysis of the symplectic spectrum as a function of the normal modes.
In other words, for Gaussian states, the degree of quantum correlations between  coupled harmonic oscillators (viz. the local phases) 
does not depend on the amount of the phase-difference quantum fluctuations 
$\sim {\langle \Delta \hat{\varphi}^2  \rangle}_0 \sim 1/\sqrt{g} $.
By contrast, when the chain is coupled to the environment, we find that a such dissipation interaction 
always yields a decreasing of the entanglement with  respect to the value of the isolated system.

\begin{figure}[btp]
\includegraphics[width=0.49\columnwidth]{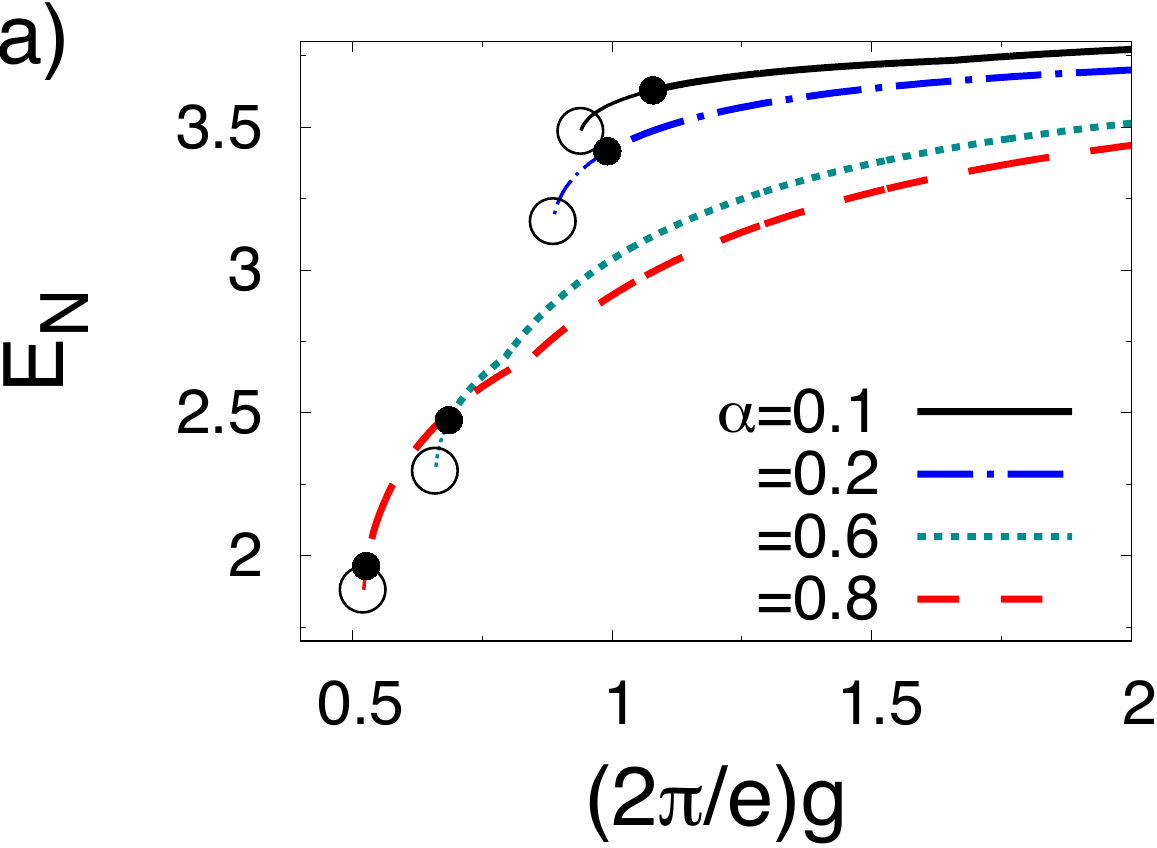}\hfill\includegraphics[width=0.49\columnwidth]{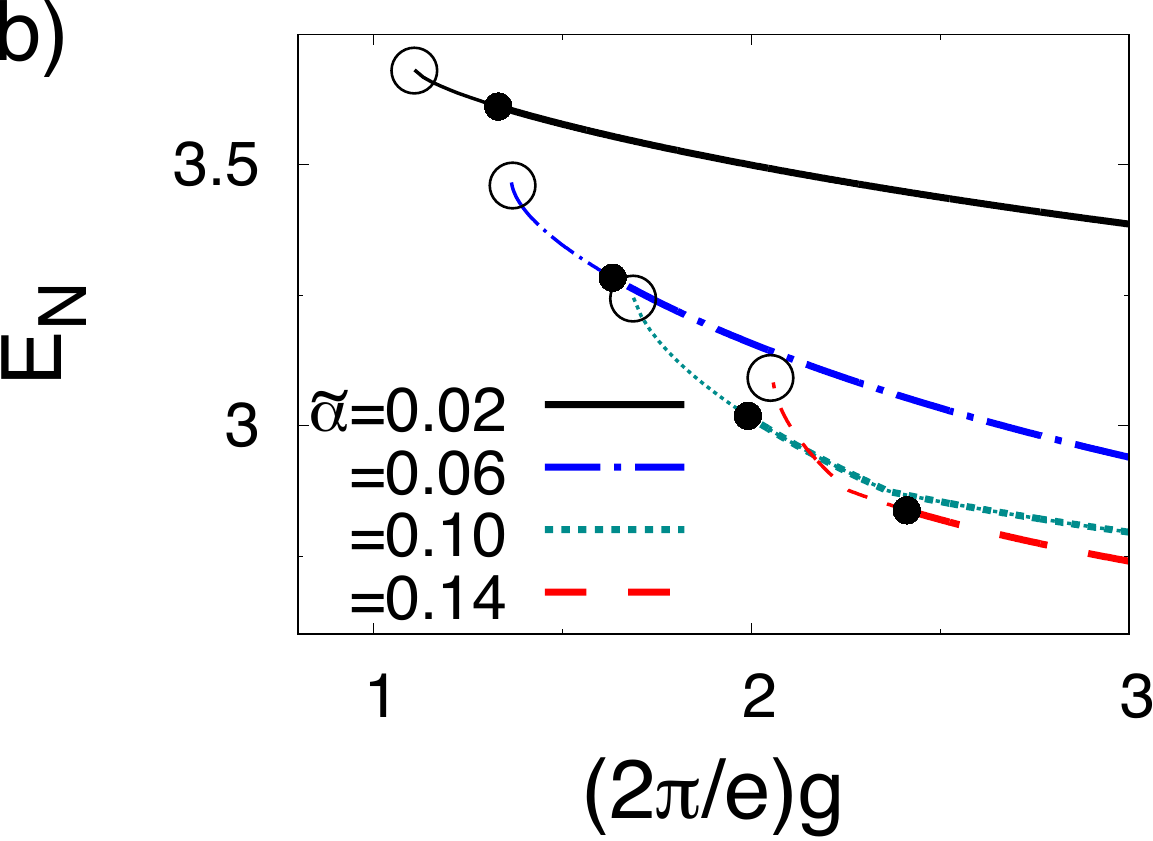}\\
\vspace{0.5cm}
\includegraphics[width=0.98\columnwidth]{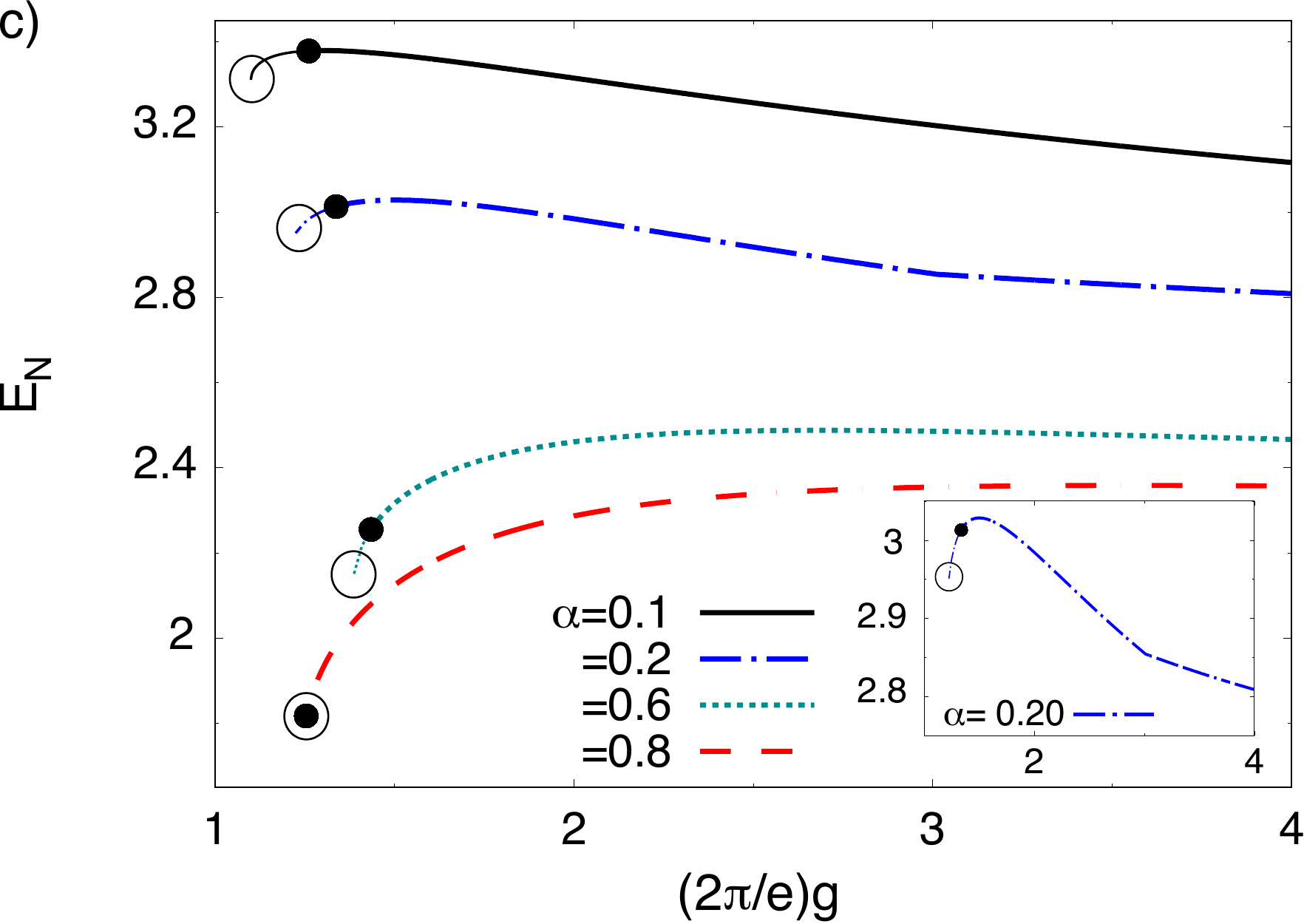}
\caption{Logarithmic negativity for  $N=9$ sites and subsystem sizes $N_A=4,N_B=5$ 
as a function of  $g$. 
The solid dot corresponds to $\mathcal{F}_v(V_{sc})=\mathcal{F}_v(0)$ 
while the open circles mark the disappearance of the self-consistent solution Eq.~(\ref{eq:selfcon2}).  
a) Conventional dissipation for different coupling of $\alpha$. 
b) Unconventional dissipation for different couplings $\widetilde{\alpha}$. 
c) Frustrated dissipation with the ratio $\widetilde{\alpha}/\alpha = 0.3$.	The inset is a zoom for the case $\alpha=0.2.$}
\label{Fig:10}
\end{figure}

Generally, the logarithmic negativity depends on the specific configuration for the partition of the system 
in two subsystems $A$ and $B$ since the correlation functions between different sites are long-ranged, 
see Eq.~(\ref{Eq:phaselm}).
Here, as example of results, in Fig.~\ref{Fig:10}, we present the case for the logarithmic negativity by dividing the periodic lattice (ring) 
formed by $N=9$ sites partitioned in two compact subsystems of size $N_A=4$ and $N_B=5$.
Our qualitative results and conclusions do not depend on this specific choice and 
further configurations are discussed in Appendix \ref{Sec:config}. 
%
%
%
%
%
%
%
%
%

Fig.~\ref{Fig:10}(a) and (b) show the logarithmic negativity as a function of $g$ when only one type of 
dissipative interaction is affecting the system, $\widetilde{\alpha} = 0$ or $\alpha=0$ respectively.
The global behavior is very similar to the results obtained for the purity. 
However, the logarithmic negativity shows a non-continuous behavior of the derivative, with 
kinks appearing for lower values of $g$.
This can be explained by considering the formal definition of $E_{\mathcal{N}}$: 
decreasing $g$, the kinks correspond to the point where a symplectic eigenvalue 
becomes less than the fixed threshold $(c_k<1/2)$,  yielding an additional term in the sum of Eq.~(\ref{logneg}). 

Finally, we report the most interesting case of dissipative frustration with 
$\widetilde{\alpha}/\alpha=0.3$  in Fig.~\ref{Fig:10}(c). 
As for the purity,  for a given ratio $\widetilde{\alpha}/\alpha < 1$, 
the logarithmic negativity can display a non-monotonic behavior  for not too large values of the dissipative interaction.

\section{Summary}
\label{sec:summary}
To summarize, we studied a 1D quantum phase model with dissipative frustration defined as a system 
coupled to the environment through two non-commuting observables, namely the phase and its 
conjugated operator, Fig.~\ref{fig1:sketch_model}(a). 
We showed that this system can be readily implemented using one dimensional Josephson junction chains 
formed by superconducting islands connected by Josephson coupling.
In these systems, the local phases and charges are the canonically conjugated variables. 
The conventional (phase) dissipation arises from the shunt resistances in parallel between two neighboring islands 
whereas the unconventional (charge) is related to the resistance connecting the local capacitance to the ground, 
Fig.~\ref{fig1:sketch_model}(b). 
When the two dissipative interactions affect separately the system, they yield quenching of, respectively, the 
quantum phase fluctuations or quantum charge fluctuations. 
When both two dissipative interaction are present, frustration emerges due to the uncertainty relation that 
sets a lower bound to the product of the two fluctuations.  

Quantum fluctuations play a crucial role in the quantum phase transition occurring the  1D quantum phase model.
This corresponds to the superconductor
vs insulator phase transition in the Josephson chain, associated to the presence of
 phase ordering or not.
Using the Self Consistent Harmonic Approximation (SCHA), we derive the qualitative phase diagram of the system 
under the influence of the dissipation. 
The dissipative frustration operating in the system leads to a non-monotonic behavior 
of the quantum fluctuations \cite{Kohler:2006ky,Rastelli:2016ge}  
which translates into the non-monotonic behavior of the critical line in the 
phase diagram at fixed ratio of the two dissipative coupling strengths.

The dissipative frustration has a genuine quantum origin since it is related to the non-commutativity of quantum operators.
Hence, we analyzed the effects of the dissipative frustration in the average quantities characterizing 
the state of the system.
In particular, we discussed two quantum thermodynamical quantities, the purity and the 
entanglement measure of the logarithmic negativity, which have no analog in classical systems. 
We found that, within the SCHA approach, both quantities show a non-monotonic  behavior approaching the critical point associated to the 
dissipative phase transition.

In conclusion, our results for a specific system demonstrate that dissipative frustration 
can lead to interesting effects and novel features in the physics of   
open quantum many body systems.

%
%
%
%
\section{Acknowledgements}
The authors thank Luigi Amico, Denis Basko, Daniel Braun, Rosario Fazio and Nils Schopohl for useful discussions. Sabine  Andergassen acknowledges the Wolfgang Pauli Institute for the kind hospitality.
We acknowledge financial support from the Deutsche Forschungsgemeinschaft
(DFG) through ZUK 63, SFB 767 and the Zukunftskolleg, and by the
MWK-RiSC program. 
%

%
%
%

\appendix

%
%
%
\section{Unconventional or charge  dissipation in the equations of motions}
\label{app:derivation}

\begin{figure}[tbp]
	\includegraphics[width=0.9\columnwidth]{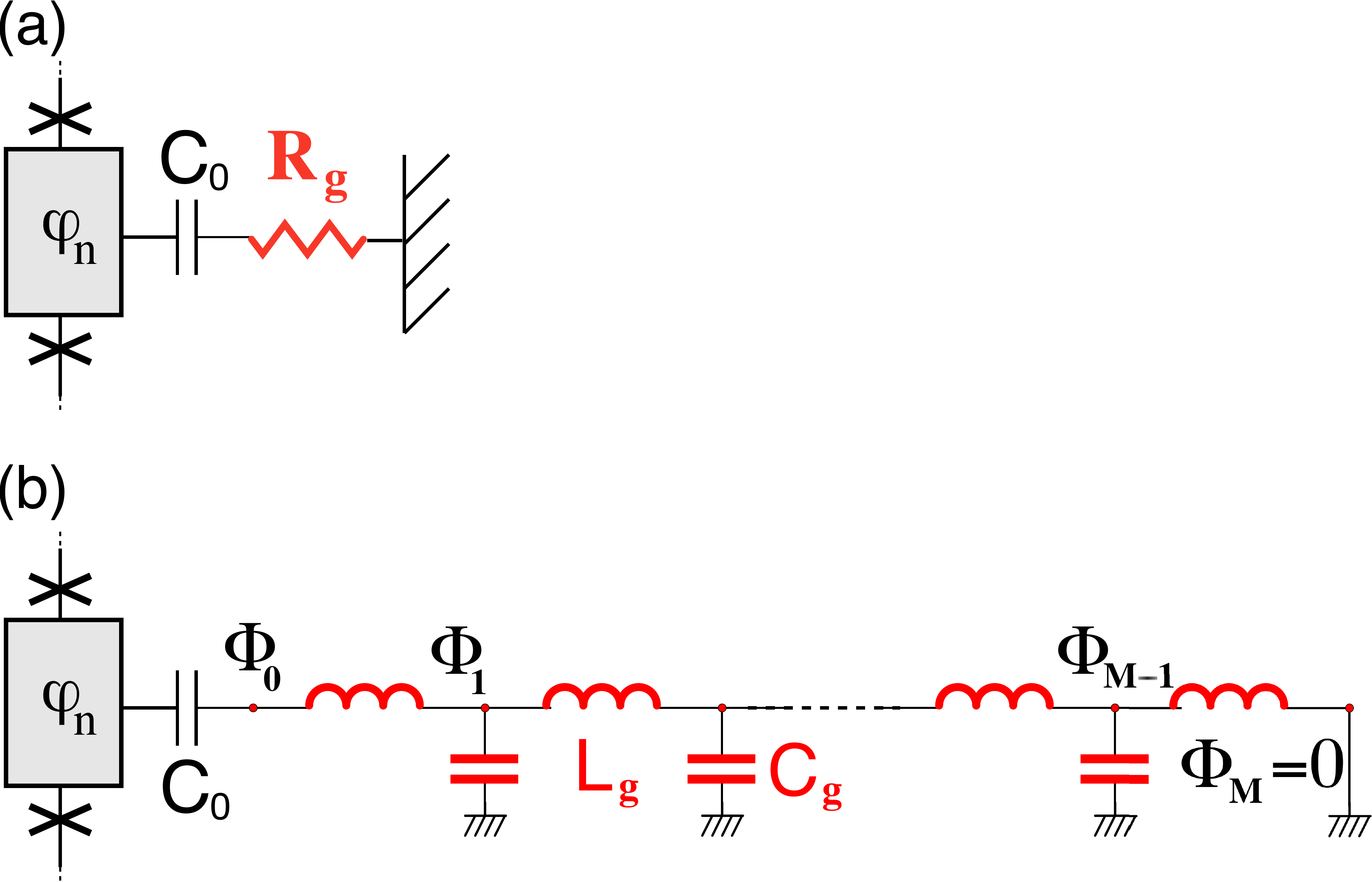}
	\caption{(a) One superconducting island of the Josephson junction chain 
	with local phase $\varphi$  connected to the ground through the series 
	formed by the capacitance $C_0$ and the resistance $R_g$. (b) Equivalent circuit simulating the dissipation by using the Caldeira-Leggett model, 
	with a transmission line formed by discrete elements, that contains the inductance $L_g$ and the capacitance $C_g$, with the characteristic impedance $R_g=\sqrt{L_g/C_g}$.}
	\label{FIG:appendix}
\end{figure}

In this appendix we discuss the dissipation obtained by coupling a superconducting 
island to the ground via an impedance formed by the series of capacitances.
We directly include capacitances in the equation whereas the resistive elements are taken into account by a standard Caldeira-Leggett approach, i.e. introducing 
a discrete line formed by capacitances $C_g$ and inductances $L_g$,
as shown in Fig.~\ref{FIG:appendix}.
This line is formed by $M$ elements. We will consider the limit $M\rightarrow \infty$ 
to recover the full dissipative ohmic behavior.
To simplify the notation, we set the local superconducting phase in the island of the Josephson 
junction chain $\varphi_n \rightarrow \varphi$.

%
%
%
%
%
%
%

Referring to Fig.~\ref{FIG:appendix}(b), we discuss the circuit using 
the equations of motion  for a lumped number of circuit elements \cite{Devoret:2004-LesHouches}.
We  use the phase nodes variables $\phi_{m}$, with $m=0,\dots,M$,  
with the boundary condition $\phi_M=0$.
We start by  the Kirchoff's equation for the energy conservation 
at each node  $m=1,\dots,M-1$ of the circuit Fig.~\ref{FIG:appendix}(b), 
\begin{equation}
C_g \frac{d^2  \phi_{m} }{dt^2} =
-\frac{1}{L_g} \left(  2\phi_m -  \phi_{m-1} -  \phi_{m+1} \right) \, .
\end{equation}
Introducing the vector $\vec{\phi}' = (\phi_1, \phi_2, \dots, \phi_m, \dots, \phi_{M-1})$ and 
the frequency $\omega_g^2 = 1 /(L_gC_g)$, 
the previous equation can be cast in the matrix form
\begin{equation}
\label{eq:first_form}
\frac{d^2    \vec{\phi}'   }{dt^2} 
\!\!=\! 
-\omega_g^2  
\left(
\begin{array}{ccccc}
   2		 		&     - 1 			&        0				&    \dots		&  \dots			\\
  -1				&        2			&      - 1				&     0		&  \dots			\\
    0				&       - 1  			&        2				&   -1		&  \dots			\\
    \dots			&      \dots			&    \dots				&    \dots		&  -1				\\
   \dots 			&      \dots			&       0				&  -1			&   2 
\end{array}
\right)
  \vec{\phi}' 
+
\omega_g^2
\left[
\begin{array}{c}
\phi_0 \\
0 \\
\dots \\
0
\end{array}
\right] \, .
\end{equation}
The eigenvectors, $e_{k}(m)= \sqrt{2/M} \sin\left[ \pi k m / M \right]$ of the matrix $\bar{\bar{M}} $, with $m=1,\dots,M-1$, span the matrix $\bar{\bar{M}} = \bar{\bar{U}} \bar{\bar{D}} {\bar{\bar{U}}}^{-1} $,
where $ \bar{\bar{D}} $ is the diagonal form 
and $ \bar{\bar{U}}$ ($\bar{\bar{U}}^{-1}$) contains the (normalized) eigenvectors.  
This corresponds to the unitary transformation  $\phi_m = \sum_{k=1}^{M-1}e_{k}(m) \Phi_k$ for $m=1,\dots,M-1$. Eq.~(\ref{eq:first_form}) reads then in terms of the normal modes $\Phi_k$ 
\begin{equation}
\label{eq:second_form}
\frac{d^2  \Phi_{k} }{dt^2} =  -\Omega_k^2    \Phi_{k}  + \omega_g^2 \sum_{k=1}^{M-1} \epsilon_k \phi_0 \, ,
\end{equation}
with the spectrum $\Omega_k = 2\omega_g \sin\left[  \pi k  / (2M) \right] $ 
and $\epsilon_k = e_{k}(1)= \sqrt{2/M} \sin\left[ \pi k  / M \right] $. 
The solution of  Eq.~(\ref{eq:second_form}) is given by
\begin{eqnarray}
\Phi_{k}(t) &=& \Phi_{k}^{(0)}(t) +  \frac{\omega_g^2}{\Omega_k}  \epsilon_k 
\int^{t}_{t_0} \!\!\! dt' \sin\left[ \Omega_k (t-t') \right] \phi_0(t')  \nonumber \\
&=&
 \Phi_{k}^{(0)}(t) +   \frac{\omega_g^2}{\Omega_k^2}    \epsilon_k \left[ \phi_0(t)   -   \cos\left[ \Omega_k (t-t_0) \right]   \phi_0(t_0) \right] \nonumber \\
&-&  
\frac{\omega_g^2}{\Omega_k^2}   \epsilon_k  
\int^{t}_{t_0} \!\!\! dt'  \cos\left[ \Omega_k (t-t') \right]   \frac{d \phi_0(t') }{dt'} \, , 
\label{eq:solution_phi_k_t}
\end{eqnarray}
with $ \Phi_{k}^{(0)}(t) $ being the solution of the associated homogeneous Eq.~(\ref{eq:second_form}). 
Then, we write the dynamics equation for the node $m=0$ 
\begin{eqnarray}
C_0 \frac{d^2 \left(  \phi_0 - \varphi \right)}{dt^2} &=&   -\frac{1}{L_g} \left(  \phi_0 -  \phi_{1} \right)  \nonumber \\ 
&=&  -\frac{\phi_0}{L_g}  + \frac{1}{L_g } \sum_{k=1}^{M-1}  \epsilon_k \Phi_k(t)  \, .
\label{eq:case_M}
\end{eqnarray}
Inserting the solution (\ref{eq:solution_phi_k_t}) for $\Phi_k(t) $  into Eq.~(\ref{eq:case_M}), 
after some algebra, 
we obtain
\begin{equation}
\label{eq:case_M_2}
C_0 \frac{d^2 \left(  \phi_0 - \varphi \right)}{dt^2} =
-\frac{\phi_0}{M L_g} + \delta I_0(t) 
- \int^{+\infty}_{-\infty}\!\!\!\!\!\!\! dt' Y(t-t')  \frac{d \phi_0(t') }{dt'}  \, ,
\end{equation}
where we set the initial time $t_0 \rightarrow -\infty$ and $\delta I_0(t) $ is 
a time function depending on the initial conditions.
This function corresponds to the noise and we can ignore it for the rest of the discussion.
The relevant quantity appearing in Eq.~(\ref{eq:case_M_2}) is the response function given by
\begin{equation}
\label{eq:admittance_caldeira-leggett}
Y(t)
= 
\theta(t) 
\frac{2  }{M L_g} \sum_{k=1}^{M-1}  \left(1 - \frac{\Omega_k^2}{4\omega_g^2}\right)   \cos\left[ \Omega_k t \right]   \, . 
\end{equation}
Finally, we take the thermodynamic limit for the number of the elements in the line $M\rightarrow\infty$ 
such that the real part of the Fourier transform of the response function $Y(t)$, associated to the dissipation, 
becomes finite and reads   
\begin{equation}
\lim_{M\rightarrow\infty} \mbox{Re}\left[ Y(\omega) \right] =(1/R_g) f_c(\omega ) \, ,
\end{equation}
with the high frequency cutoff $\sim \omega_g$ that we neglect hereafter to simplify the notation.
Omitting the noise and using the Markovian approximation (viz. the decay rate $\sim 1/\omega_g$ of the function $Y(t)$  much larger than the time scale of the evolution of the phases), we have 
\begin{equation}
\frac{d^2   \phi_0 }{dt^2} = - \frac{1}{\tau_g} \frac{d  \phi_0 (t) }{dt}  +   \frac{d^2  \varphi  }{dt^2}  \, ,
\end{equation}
with $\tau_g = C_0 R_g$. 
We then consider the particular solution $  \phi_0 (t) = \int_{-\infty}^{t} \!\!\! dt' \exp\left[ - (t-t') / \tau_g\right] d\varphi(t')/dt'$
with the property 
\begin{equation}
\frac{d \phi_0 (t)}{dt} =  \frac{d\varphi(t)}{dt}  - \frac{1}{\tau_g} \int_{-\infty}^{t} \!\!\!\!\!\! dt' \exp\left[ - (t-t') / \tau_g\right] \frac{d\varphi(t')}{dt'} \, .
\end{equation}
In this way we can show that 
\begin{equation}
\frac{d^2   \left( \phi_0  - \varphi  \right)}{dt^2} 
=  
-\frac{1}{\tau_g} \frac{ d\varphi(t)}{dt} 
+
\frac{1}{\tau_g^2}
\int_{-\infty}^{t} \!\!\!\!\!\!\! dt' e^{- \frac{t-t'}{\tau_g}}  \frac{d\varphi(t')}{dt'}  
\label{eq:main_result_appendixA} \, . 
\end{equation}
In the final step, we recover the index for each element  $\varphi \rightarrow \varphi_n$ and 
write the equation for the phase $\varphi_n$ 
(in the Markovian limit) of the 1DJJ shown Fig.~\ref{fig1:sketch_model} as
\begin{equation}
\label{eq:varphi_n}
\frac{d^2   \left( \varphi_n - \phi_0  \right)}{dt^2}   =  -\omega_0^2 \left(2 \varphi_n - \varphi_{n-1} -\varphi_{n+1} \right) - \gamma \frac{d\varphi_n}{dt} 
\, , 
\end{equation}
with $\omega_0 = \sqrt{E_C E_J}/\hbar$ and $\gamma=1/(R_s C_0)$. 
Using the main result  Eq.~(\ref{eq:main_result_appendixA}), the Fourier transform of Eq.~(\ref{eq:varphi_n})
reads
\begin{equation}
\label{eq:varphi_n_2}
- \frac{\omega^2  \varphi_n(\omega)  }{1+i\tau_g\omega}= 
  -\omega_0^2  \left[ 2\varphi_n - \varphi_{n-1}  - \varphi_{n+1}   \right]_{\omega}  - i \gamma \omega \varphi_n(\omega) 
\end{equation}
in which we are interested only to the l.h.s. describing the effect of the unconventional (charge) dissipation 
in frequency space.
Thus, we conclude that the unconventional dissipation corresponds to a damped (imaginary) mass 
in the equation of motion of the local phases $\varphi_n$.

We finally observe that, using the Wick's rotation from real frequency $\omega$ to  Matsubara frequency $-i \omega_{\ell}$
and restoring the capacitance (mass) in the l.h.s. of Eq.~(\ref{eq:varphi_n_2}),  we get 
\begin{equation} 
 \frac{C_0 \omega^2_{\ell} }{1+ \tau_g \omega_{\ell} }  
\sim  
\widetilde{F}_{\ell}  \, \, \omega_{\ell}^2    \, ,
\end{equation}
where the propagator $\widetilde{F}_{\ell}$ is given by Eq.~(\ref{Eq:Matsutau}) with the cutoff function $f_c =1$ and $\omega_{\ell}>0$.
A   rigorous derivation will be given in the following Appendix B.

%
%
%
\section{Unconventional or charge dissipation with the path integral in the imaginary time}
\label{app:derivation-2}

In this appendix we derive the unconventional or charge dissipation 
introduced in the main text, in the path integral formalism in imaginary time

As first step, we recall the Lagrangian in the imaginary time of the Josephson junction chains 
with each junction shunted by the resistance $R_s$, 
\begin{eqnarray}
\mathcal{S}_{JJ}
&=& 
\sum_{n=0}^{N-1} 
\frac{1}{2}   \int_{0}^{\beta} \!\!\!\! \int_{0}^{\beta}\!\!d\tau d\tau' \,
F (\!\tau\!-\!\tau') \, {\left| \Delta\varphi_{n}(\tau) - \Delta\varphi_{n}(\tau') \right|}^2   \nonumber \\
&-& 
\sum_{n=0}^{N-1}  E_J  \cos \left(  \Delta\varphi_{n}(\tau)  \right)   \, ,
\end{eqnarray}
where $E_J$ is the Josephson energy and the function $F (\tau) $ encoding the ohmic dissipation of $R_s$ refers to Eq.~(\ref{Eq:Matsugam})
discussed in the main text.
Then we assume that each local superconducting island $n$ is coupled to an external bath (external impedance) 
leading to the general form of the Lagrangian 
\begin{equation}
\mathcal{S} =  \mathcal{S}_{JJ} +  \int^{\beta}_0\!\!\!\! d\tau \sum_{n=0}^{N-1}\mathcal{L}_0^{(n)} \, . 
\end{equation}
The external impedance  is  formed by the capacitance $C_0$ and a resistance $R_g$,
as depicted in Fig.~\ref{FIG:appendix}(a). 
The dissipative element $R_g$ is described by the Caldeira-Leggett model, viz.  
as an  ensemble of $M$ discrete  elements forming a transmission line, 
as depicted in the Fig.~\ref{FIG:appendix}(b). 
In the thermodynamic limit $M \rightarrow \infty$, such a line is equivalent to the resistance $R_g$,
as we show in the following. 

To construct the Lagrangian we have to consider the electrostatic energy associated to each link containing a capacitance and the associated inductive energy \cite{Devoret:2004-LesHouches}.
The result is  
\begin{eqnarray}
\frac{\mathcal{L}_0^{(n)}}{ \mu_0 } 
&=& 
\frac{C_0}{2} {\left( \dot{\varphi}_n - \dot{\phi}_0^{(n)}  \right)}^2   
+ 
\frac{{\left(  \phi_{0}^{(n)}- \phi_{1}^{(n)}   \right)}^2}{2 L_g} \nonumber \\
&+& \!\! 
\sum_{m=1}^{M-1} 
\!\! 
\left[  
\frac{C_g}{2}  {\left( \dot{\phi}_m^{(n)}  \right)}^2  
\!\!\!+\!
\frac{{\left(  \phi_{m+1}^{(n)}- \phi_{m}^{(n)}  \right)}^2}{2 L_g}   
\! \right]  , 
\label{eq:L_0_n}
\end{eqnarray}
with $\mu_0= \Phi_0^2 /(4\pi^2 )  $ and  $\Phi_0 = h /(2e)$ the flux  quantum. 
Then we express the partition function of the whole system in the imaginary time path integral formalism \cite{Weiss:2012}
\begin{eqnarray}
Z 
&=&  \!\!
\prod_{n,m} \! \oint \! D[\varphi_n(\tau)]
e^{ - \frac{\mathcal{S}_{JJ} }{\hbar}  }
 \!\! \oint \! D[\phi_{m}^{(n)}(\tau)] 
e^{ 
-\frac{1}{\hbar}\sum_{n=0}^{N-1}  \int\limits_{0}^{\beta} d\tau  \mathcal{L}_0^{(n)} 
} \nonumber \\
&\equiv&
\prod_n \oint D[\varphi_n(\tau)] e^{ - \frac{\mathcal{S}_{JJ} }{\hbar}  } \,\,\,  \mbox{F}_{ch}[\varphi]  \, .
\end{eqnarray}
Hereafter, we  focus only on one superconducting island $n$ described by the phase $\varphi_n$, and 
to simplify the notation we drop the index $n$.
Hence, we consider the Lagrangian $\mathcal{L}_0$ (without index n) in Eq.~(\ref{eq:L_0_n}). 
Now can diagonalize the part containing the transmission line for the 
phases $m=1,\dots,M-1$ via the unitary transformation introduced in the previous Appendix A. 
Then the ensemble of the harmonic modes $\Phi_k$ represents the effective bath affecting the phase $\phi_0$ and that 
eventually becomes equivalent to a dissipative resistance.
Only the phase $\phi_0$ is directly coupled  capacitively to the superconducting phase  $\varphi$ 
of the local island forming the 1DJJ. 
Thus we obtain
\begin{eqnarray}
\mbox{F}_{ch}[\varphi] &=&  
\oint D[\phi_{0}(\tau)]   
e^{-\frac{1}{\hbar} \int\limits_{0}^{\beta} d\tau \left[ \frac{\mu_0 C_0}{2} {\left( \dot{\varphi} - \dot{\phi}_0  \right)}^2   \right] }  \nonumber \\
& & \prod_{k=1}^{M-1} \oint D[\Phi_{k}(\tau)] 
e^{-\frac{1}{\hbar} \int\limits_{0}^{\beta} d\tau \mathcal{L}_g } 
\label{eq:path-integral-to-calculate}
\, , 
\end{eqnarray}
where $\mathcal{L}_g$ is given by 
\begin{equation}
\frac{\mathcal{L}_{g}}{\mu_0}
=
\frac{\phi_0^2}{2L_g} - \frac{\phi_0}{L_g} \sum_{k} \epsilon_k \Phi_k 
+ 
\frac{C_g}{2}  \sum_{k=1}^{M-1}  \left[    \dot{\Phi}_k^2   + \frac{\Omega_k^2}{2}  \Phi_k^2  \right] 
\end{equation}
and the spectrum $\Omega_k^2$ and $\epsilon_k$ are defined above. 
In order to derive the final effective functional for the phase $\varphi$, we have first to integrate out the harmonic modes $\Phi_k$ and then the phase variable $\phi_0$  directly coupled 
to the superconducting phase  $\varphi$ via the capacitance $C_0$. 
Using the Matsubara Fourier transformation $\Phi_k(\tau) = \sum_k \Phi_{\ell}^{(k)} \exp(i \omega_{\ell} \tau)$ 
and $\phi_0(\tau) = \sum_k \phi_{\ell}^{(0)} \exp(i \omega_{\ell} \tau)$, with $\omega_{\ell}=2\pi \ell / \beta$ ($\ell$ integer), 
we  express the action $\mathcal{L}_g $ as  
\begin{eqnarray}
& & \frac{1}{\hbar}  \int_{0}^{\beta} \!\!\!\! d\tau \mathcal{L}_g   =  
\frac{\beta\mu_0}{\hbar L_g} \sum_{\ell=0}^{\infty}  \left(1-\frac{\delta_{\ell,0}}{2} \right) {\left| \phi_{\ell}^{(0)} \right|}^2    
 \nonumber \\
&-&
\frac{\beta\mu_0}{\hbar L_g}   \sum_{k}^{M-1} \epsilon_k 
\left[ \phi_{0}^{(0)} \Phi_{0}^{(k)} + 
 \sum_{\ell=0}^{\infty} \left( \phi_{\ell}^{(0)}  \Phi_{\ell}^{(k) *} + \phi_{\ell}^{(0)*}  \Phi_{\ell}^{(k)} \right)
\right] \nonumber \\
&+&
\frac{\beta C_g\mu_0}{\hbar}  
\sum_{k=1}^{M-1} \sum_{\ell=0}^{\infty} \left[ \omega_{\ell}^2 + \Omega_k^2 \left(1-\frac{\delta_{\ell,0}}{2} \right) \right] {\left| \Phi_{\ell}^{(k)} \right|}^2 
\end{eqnarray}
to be inserted in the path integral Eq.~(\ref{eq:path-integral-to-calculate}) with the metric \cite{Kleinert:2009,Weiss:2012}
\begin{equation}
\! \oint \!\! D[\Phi_{k}(\tau)]   
\! \rightarrow \! 
\! \int\!\! \frac{d\Phi_{0}^{(k)} }{\sqrt{2\pi\hbar\beta/(\mu_0C_g )}}
\prod_{\ell=1}^{\infty} \! \int \!\!\!\  \frac{d \Phi_{\ell}^{(k),Re} d  \Phi_{\ell}^{(k),Im} }{\pi \hbar  /(\beta \mu_0 C_g \omega_{\ell}^2)} ,
\end{equation}
with $\Phi_{\ell}^{(k),Re}$ and $\Phi_{\ell}^{(k),Im}$ the real and imaginary part of $\Phi_{\ell}^{(k)}$ $(\ell \neq 0)$, respectively. 
After performing the Gaussian integral, we derive the effective action for the phase $\Phi_M$
\begin{eqnarray}
\label{eq:F_phi_1}
\mbox{F}(\phi) &=&  
\oint D[\phi_{0}(\tau)]   
e^{-\frac{1}{\hbar} \int\limits_{0}^{\beta} d\tau \left[ \frac{\mu_0 C_0}{2} {\left( \dot{\varphi} - \dot{\phi}_0  \right)}^2   \right] }  \nonumber \\
& & \left[ \prod_{k=1}^{M-1} Z_h(\Omega_k) \right] 
e^{-\frac{1}{\hbar} \Delta S_{eff}[\phi_0] },
\end{eqnarray}
where $Z_h(\Omega)$ is the partition function of a harmonic oscillator of frequency $\Omega$ that we omit hereafter, and
 $\Delta S_{eff}[\phi_0]$ the effective action for the phase $\phi_0$ which reads in Matsubara space  
\begin{equation}
\label{eq:Delta_S_eff}
\Delta S_{eff}[\phi_0]
\! = \!  
\beta\mu_0 \!
\sum_{\ell=0}^{\infty}
\left[ 
\frac{1}{L_g M}  \! \left(\! 1\!-\! \frac{\delta_{\ell,0}}{2} \! \right)  
\!+ \omega_{\ell} \mbox{Y}_{\ell} \right]   {\left| \phi_{\ell}^{(0)} \right|}^2  \, .
\end{equation}
In Eq.~(\ref{eq:Delta_S_eff}),  the first term represents an effective inductance for the phase $\phi_0$ that
vanishes in the limit $M\rightarrow \infty$, whereas 
the relevant term is the second one with the function
\begin{equation}
\label{eq:Y_ell}
\mbox{Y}_{\ell} 
=
\frac{2 \omega_{\ell}}{M L_g} \sum_{k=1}^{M-1} 
 \left( 1 - \frac{\Omega_k^2}{4\omega_g^2} \right) \frac{1}{\omega_{\ell}^2+\Omega_k^2}. \\
\end{equation}
Note the similarity of $\mbox{Y}_{\ell} $ in Eq.~(\ref{eq:Y_ell}) 
with Eq.~(\ref{eq:admittance_caldeira-leggett}) for the response 
function (admittance) of the transmission line.
With some algebra,  setting $R_g=\sqrt{L_g/C_g}$, we cast $\mbox{Y}_{\ell}$ in the following form  
\begin{eqnarray}
\label{eq:Y_ell_2}
\mbox{Y}_{\ell} 
&=& 
\left(1/R_g\right)  \frac{2}{\pi} 
{\left[
x_{\ell} \,  \frac{\pi}{2M} 
\sum_{k=1}^{M-1}  \frac{1-\sin^2\left( \frac{\pi k}{2M}\right)}{x^2_{\ell} +\sin^2\left( \frac{\pi k}{2M}\right) }
\right]}_{x_{\ell}=\frac{\omega_{\ell}}{2\omega_g}}
\nonumber \\
_{(M \rightarrow \infty)} &=& 
\left(1/R_g\right) 
 \frac{2}{\pi} 
{\left[  x_{\ell} \int_{0}^{\frac{\pi}{2}} \!\!\! d\theta 
\frac{1-\sin^2(\theta)}{x^2_{\ell} +\sin^2(\theta) }
\right]}_{x_{\ell}=\frac{\omega_{\ell}}{2\omega_g}}
\nonumber \\
&=& 
\left(1/R_g\right)  f_c\left( \omega_{\ell} \right ) \label{eq:dissipation},
\end{eqnarray}
where in the second line we have taken the  limit $M\rightarrow \infty$ replacing the sum 
with the continuous integral. 
$x_{\ell}= \omega_{\ell}/(2\omega_g)$ corresponds to the cutoff function with high frequency  $\omega_c=2\omega_g$. 
For the specific choice of the circuit discussed here leading to Eq.~(\ref{eq:dissipation}), 
we get $f_c(\omega_{\ell}) = \sqrt{1+x^2_{\ell}}-x_{\ell}$.
However, details of the specific form of the cutoff functions are irrelevant  for the results analyzed in the main text.
In the limit in which $\omega_c$  represents the high frequency involved in the problem, 
we expect only logarithmic corrections to the average phase difference fluctuations,
see Eq.~(\ref{Eq:zeroTemp}).

Summarizing we have shown that
\begin{equation}
\label{eq:Delta_S_eff_2}
\Delta S_{eff}[\phi_0]
=
\beta   \frac{\hbar}{2\pi}  \frac{R_q}{R_g} \sum_{\ell=1}^{\infty} \omega_{\ell}  f_c(\omega_{\ell})      {\left| \phi_{\ell}^{(0)} \right|}^2  \, ,
\end{equation}
where $\mu_0 =\Phi_0/(2\pi) = R_q \hbar/(2\pi)$.
Indeed, this is exactly the same form as for the dissipative function describing a shunt resistance for the Josephson junction 
phase difference, see Eq.~(\ref{Eq:Matsugam}) for which we have given, en passant, a demonstration.

In the last part, we have to perform the integral in Eq.~(\ref{eq:F_phi_1}) with the action Eq.~(\ref{eq:Delta_S_eff_2}),
with the use of the metric
\begin{equation}
\! \oint \!\! D[\phi_{0}(\tau)]   
\! \rightarrow \! 
\! \int\!\! \frac{d\phi_{0}^{(0)} }{\sqrt{2\pi\hbar\beta /(\mu_0 C_0)}}
\prod_{\ell=1}^{\infty} \! \int \!\!\!\  \frac{d \phi_{\ell}^{(0),Re} d  \phi_{\ell}^{(0),Im} }{\pi \hbar /(\beta \mu_0 C_0 \omega^2_{\ell})} .
\end{equation}
The Gaussian integral is then carried out using the Matsubara frequency representation,  which yields
\begin{align}
\mbox{F}(\phi) &\sim 
\exp
\left[ \frac{\hbar \beta}{E_C} \sum_{\ell=1}^{\infty}  \frac{\hbar}{2\pi}
\left( \frac{\omega_{\ell}^2 }{1+\omega_{\ell} \tau_g f_c(\omega_{\ell})}  \right)
{\left| \varphi_{\ell} \right|}^2 
\right] \, .
\end{align}
The latter expression corresponds to the part containing the unconventional or charge damping kernel $\tilde{F}(\tau-\tau')$ 
in the total action of the system for each local phase $\varphi_n$ in Eq.~(\ref{Eq: effectiveaction_gaussian}),
with the propagator given by $\tilde{F}_{\ell}$ in Eq.~(\ref{Eq:Matsutau}).

%
%
%
%
\section{Covariance matrix and logarithmic negativity}
\label{Sec:symplect}
To illustrate the method used in Sec.~\ref{sec:purity-entanglement} to compute the logarithmic negativity
from the covariance matrix, we  discuss in this appendix the simple example of two coupled oscillators.
In particular, we calculate the symplectic eigenvalues and show how it is related  to the Heisenberg uncertainty principle.
We refer to the works Refs.~\cite{Adesso2007,Werner2002,Simon:2000,Serafini:2007} 
for extended discussions.

We consider two harmonic oscillators described by the two position and momentum operators which define the vector 
\begin{equation}
\hat{R}= (\hat{R}_0, \hat{R}_1) =  (\hat{x}_0,\hat{p}_0,\hat{x}_1,\hat{p}_1)^T \, ,
\end{equation}
with the Hamiltonian 
\begin{equation}
\hat{H}=\frac{1}{2m}(\hat{p}_0^2+\hat{p}_1^2)+\frac{k_0}{2}(\hat{x}_0^2+\hat{x}_1^2)+\frac{k}{2}(\hat{x}_0-\hat{x}_1)^2.
\end{equation}
The corresponding commutator relation reads
\begin{equation}
[\hat{R}_i,\hat{R}_j]=i \hbar \delta_{ij} 
\left(\begin{array}{cc}
0& 1\\ 
-1& 0
\end{array}
\right).
\end{equation}
The $(4 \times 4)$ matrix
\begin{equation}
\mathcal{O}=\bigoplus_{n=0}^{1}
\left(\begin{array}{cc}
0& 1\\ 
-1& 0
\end{array} \right) \,  \label{Eq:block}
\end{equation}
is the symplectic  matrix, which is invariant under symplectic transformations 
$S^T\mathcal{O}S=\mathcal{O}$, with $S\in \text{Sp}(4,\mathbb{R})$ denoting the symplectic group.
The covariance matrix  reads
\begin{equation}
\sigma=\left(\begin{array}{cccc}
\langle \hat{x}_0^2 \rangle	& 0 & \langle \hat{x}_0\hat{x}_1\rangle  & 0  \\ 
0	& \langle \hat{p}_0^2\rangle & 0 & \langle \hat{p}_0\hat{p}_1\rangle \\ 
 \langle \hat{x}_0\hat{x}_1\rangle 	& 0 &  \langle \hat{x}_1^2\rangle & 0 \\ 
0	&\langle \hat{p}_0\hat{p}_1\rangle  & 0  & \langle \hat{p}_1^2\rangle 
\end{array} \right).
\end{equation}
The  Heisenberg uncertainty principle is equivalent 
to the condition that the eigenvalues of the matrix given by the sum of $\sigma$ 
and $(i\hbar/2) \mathcal{O}$ are always positive or zero, namely
\begin{equation}
\left[ \sigma +i\frac{\hbar}{2}\mathcal{O} \right] \ge 0 \, . 
\label{Eq:uncertainty}
\end{equation}
In other words, the l.h.s. of Eq.~(\ref{Eq:uncertainty}) 
has to be positive semi-definite such that the matrix $\sigma$ has a physical meaning.
As the covariance matrix is positive and symmetric, according to the Williamson's theorem, 
it is always possible to cast it in a diagonal form using a symplectic transformation
\begin{align}
S^T\sigma S= \mathcal{B} ,\;\;\;\; \text{with}\;\;\;S\in \text{Sp}(4,\mathbb{R}) \, , 
\end{align}
where 
\begin{equation}
\mathcal{B}=\left(\begin{array}{cccc}
b_0	& 0 & 0 & 0  \\ 
0	& b_0 & 0 & 0 \\ 
0	& 0 &  b_1 & 0 \\ 
0	& 0 & 0  & b_1
\end{array} \right) \, .
\end{equation}
The quantities \{$b_0,b_1$\} are called symplectic eigenvalues and build the symplectic spectrum of the covariance matrix. 
Hence via the symplectic transformation of the l.h.s of Eq.~(\ref{Eq:uncertainty}) we get
\begin{alignat}{4}
S^T(\sigma +i\frac{\hbar}{2}\mathcal{O})S&\ge 0\\
\Leftrightarrow \;\;\;\;\; \mathcal{B} + i\frac{\hbar}{2}\mathcal{O}& \ge 0.
\end{alignat}
Because of the positive semi definiteness all eigenvalues $\lambda_{k}$ with $k=1,\dots,4$
of the l.h.s. have to satisfy $\lambda_{k} \ge0$. 
This leads to $b_0 \ge \hbar/2 $ and  $b_1 \ge\hbar/2$.
For instance, for a single harmonic oscillator, one can obtain 
$b_0^2 =  \langle \hat{x}_0^2\rangle\langle \hat{p}_0^2\rangle $.

We now find the symplectic eigenvalues associated to the correlation matrix $\sigma$ 
by computing the orthogonal eigenvalues 
of the matrix ($- i \mathcal{O} \sigma$) with  $\{\pm  b_1,\pm  b_2\}$  \cite{Serafini:2007}. 
After some algebra, one obtains 
\begin{align}
b_0  & = \sqrt{(\langle \hat{x}^2 \rangle+\langle \hat{x}_0\hat{x}_1\rangle)(\langle \hat{p}^2 \rangle+\langle \hat{p}_0\hat{p}_1\rangle)	}\\
b_1 & = \sqrt{(\langle \hat{x}^2 \rangle-\langle \hat{x}_0\hat{x}_1\rangle)(\langle \hat{p}^2 \rangle-\langle \hat{p}_0\hat{p}_1\rangle)	}.
\end{align}
With the center of mass position $\hat{X}$ and momentum $\hat{P}$ as well as the corresponding relative coordinates $\hat{r}$ and $\hat{p}$ we perform the canonical transformation
\begin{align}
\hat{x}_0&=\hat{X}+(1/2)\hat{r} &\hat{x}_1=\hat{X}-(1/2)\hat{r}\\ \hat{p}_0&=(1/2)\hat{P}+\hat{p}  &\hat{p}_1=(1/2)\hat{P}-\hat{p}. 
\end{align} 
With this we can rewrite the terms for the position
\begin{align}
\langle \hat{x}^2 \rangle+\langle \hat{x}_0\hat{x}_1\rangle&= 2 \langle \hat{X}^2\rangle \label{Eq:trafox}\\
\langle \hat{x}^2 \rangle-\langle \hat{x}_0\hat{x}_1\rangle&= \frac{1}{2} \langle \hat{r}^2\rangle
\end{align}
and for the momentum
\begin{align}
\langle \hat{p}^2 \rangle+\langle \hat{p}_0\hat{p}_1\rangle&= 2 \langle \hat{P}^2\rangle \\
\langle \hat{p}^2 \rangle-\langle \hat{p}_0\hat{p}_1\rangle&= \frac{1}{2} \langle \hat{p}^2\rangle.\label{Eq:trafop}
\end{align}
and we obtain that the inequality for the symplectic eigenvalues corresponds to the Heisenberg's uncertainty principle
\begin{align}
b_0  & = \sqrt{\langle \hat{X}^2 \rangle\langle \hat{P}^2\rangle}\ge \frac{\hbar}{2}\\
b_1 & = \sqrt{\langle \hat{r}^2 \rangle\langle \hat{p}^2\rangle	}\ge \frac{\hbar}{2}.
\end{align}
In the ground state of the system we know that $\langle \hat{X}^2 \rangle = \hbar/(4m\omega_0)$ and $\langle \hat{P}^2 \rangle = \hbar2m\omega_0/2$ yielding $b_0= \hbar /2$. The relative coordinates are described by the same relations but oscillate with the frequency $\omega_r$ which also leads to  $b_1=\hbar/2$. \\
In order to calculate the logarithmic negativity,  one has to repeat the same procedure for 
the covariance matrix $\sigma[\hat{\rho}^{T_A}]$ associated to the partially transposed system $\rho^{T_A}$. 
Since the partial transpose operation corresponds to $\langle \hat{p}_0\hat{p}_1\rangle \rightarrow 
- \langle \hat{p}_0\hat{p}_1\rangle$, we obtain directly
\begin{align}
\tilde{b}_0  
& = \sqrt{(\langle \hat{x}^2 \rangle+\langle \hat{x}_0\hat{x}_1\rangle)(\langle \hat{p}^2 \rangle -\langle \hat{p}_0\hat{p}_1\rangle)	}\\
\tilde{b}_1 & 
= \sqrt{(\langle \hat{x}^2 \rangle-\langle \hat{x}_0\hat{x}_1\rangle)(\langle \hat{p}^2 \rangle+\langle \hat{p}_0\hat{p}_1\rangle)	}.
\end{align}
and with the Eqs. (\ref{Eq:trafox})-(\ref{Eq:trafop}) 
\begin{align}
\tilde{b}_0  & = 2\sqrt{\langle \hat{X}^2 \rangle\langle \hat{p}^2\rangle}\\
\tilde{b}_1 & = \frac{1}{2}\sqrt{\langle \hat{r}^2 \rangle\langle \hat{P}^2\rangle	}.
\end{align}
Note that the symplectic eigenvalues $\tilde{b}_{0}, \tilde{b}_{1}$ 
of $\sigma[\hat{\rho}^{T_A}]$ contain products of variables which are not conjugate. 
The explicit expression reads
\begin{align}
(2/\hbar) \tilde{b}_0 &= \sqrt{\frac{\omega_r}{\omega_0}} = \left(1+\frac{2k}{k_0}\right)^{\frac{1}{4}}>1\\
(2/\hbar) \tilde{b}_1 &= \sqrt{\frac{\omega_0}{\omega_r}} = \frac{1}{\left(1+\frac{2k}{k_0}\right)^{\frac{1}{4}}}<1. 
\end{align}

Recalling that the logarithmic negativity is defined by $E_{\mathcal{N}}[ \hat{\rho} ] = 
-\sum_{k=0}^{1} \log_2 \left( \min[1 , (2/\hbar) \tilde{b}_k)]  \right)$, 
the symplectic eigenvalue $\tilde{b}_1<1$ will contribute to the logarithmic negativity  
from which one concludes that the two oscillators are  entangled.

%
%
%
%
\section{Logarithmic negativity  for different partitions}
\label{Sec:config}
%
%
%
%
%
%
%
\begin{figure}[b!]
	\includegraphics[width=0.33\columnwidth]{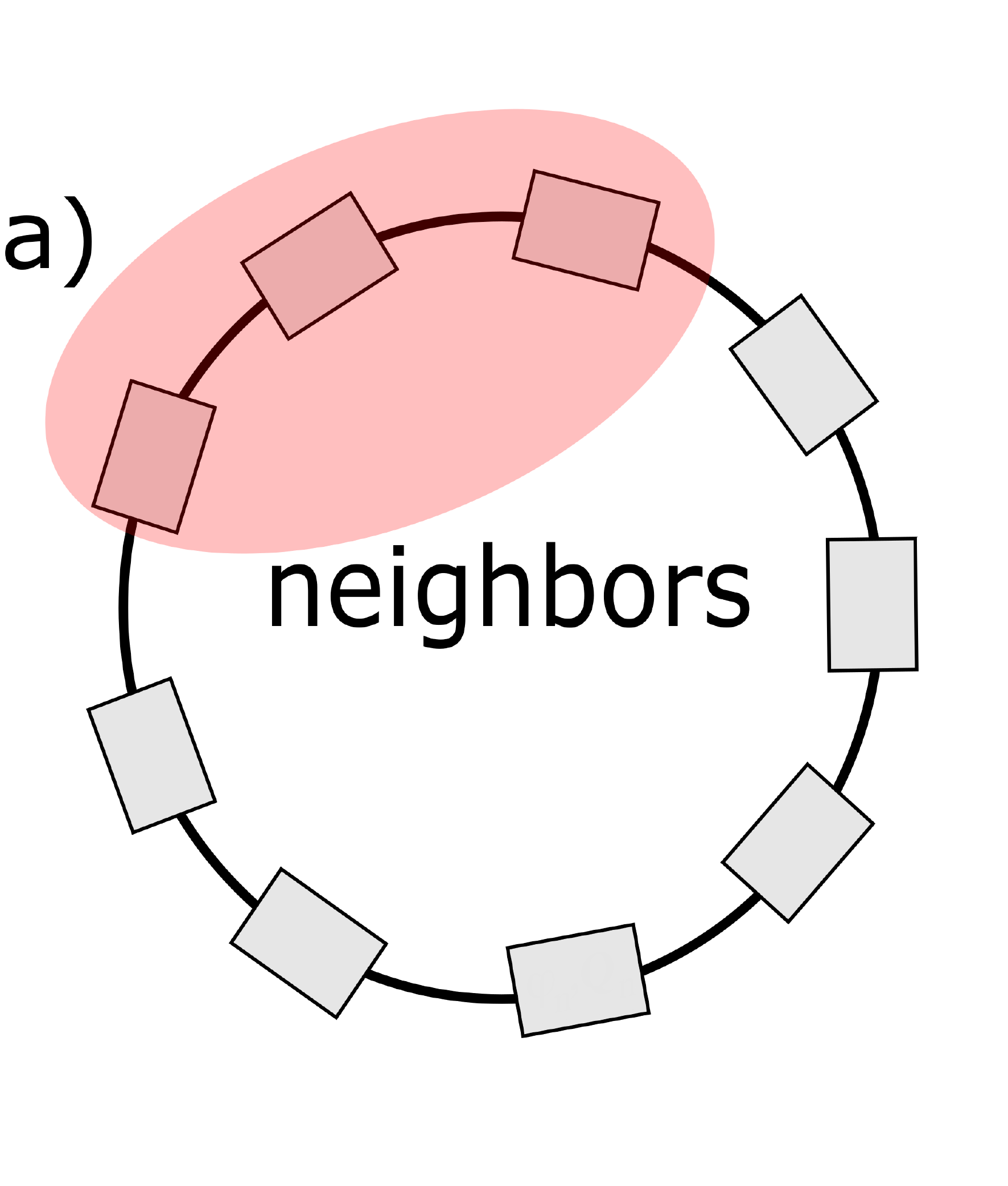}\hfill\includegraphics[width=0.33\columnwidth]{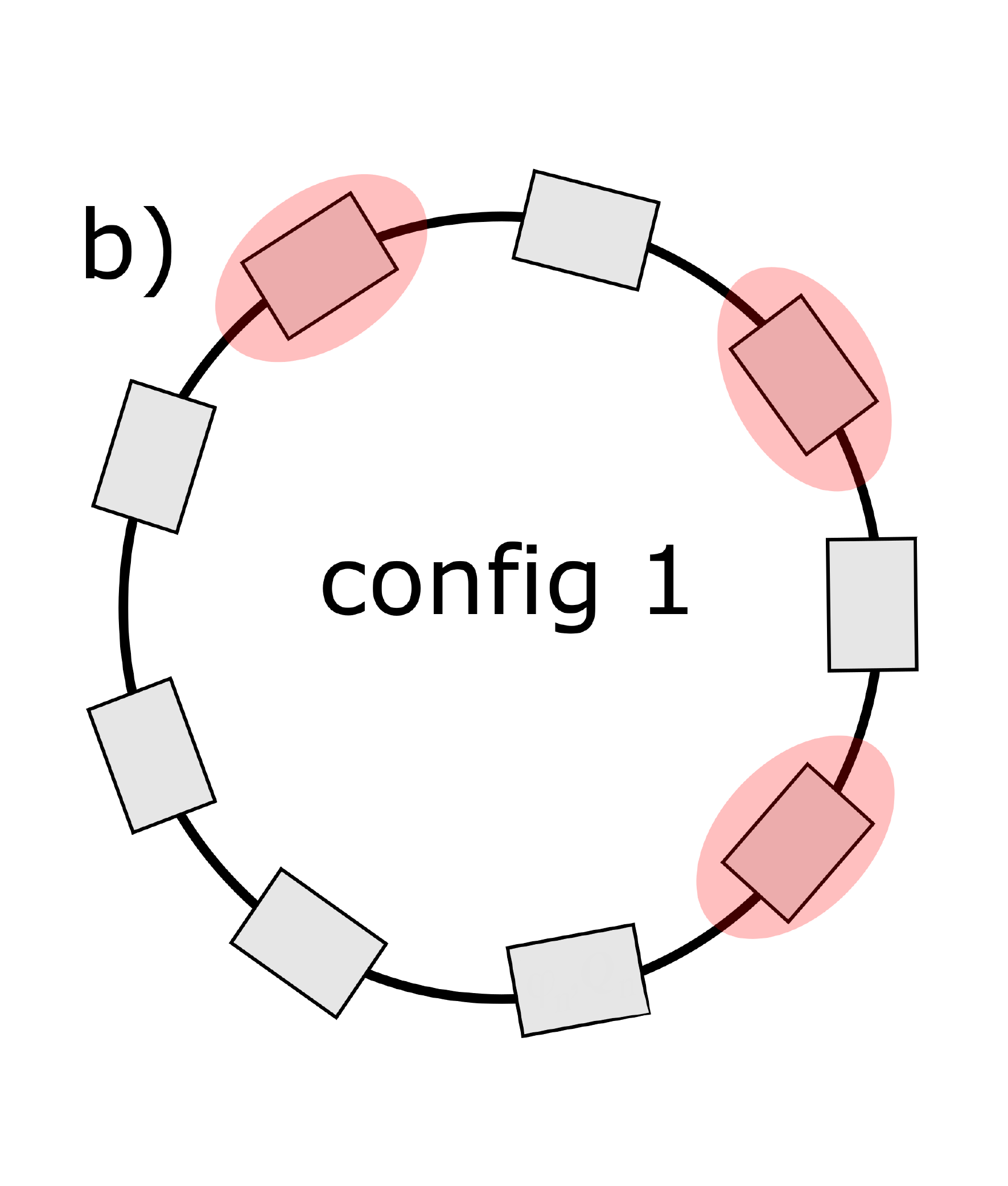}\hfill\includegraphics[width=0.33\columnwidth]{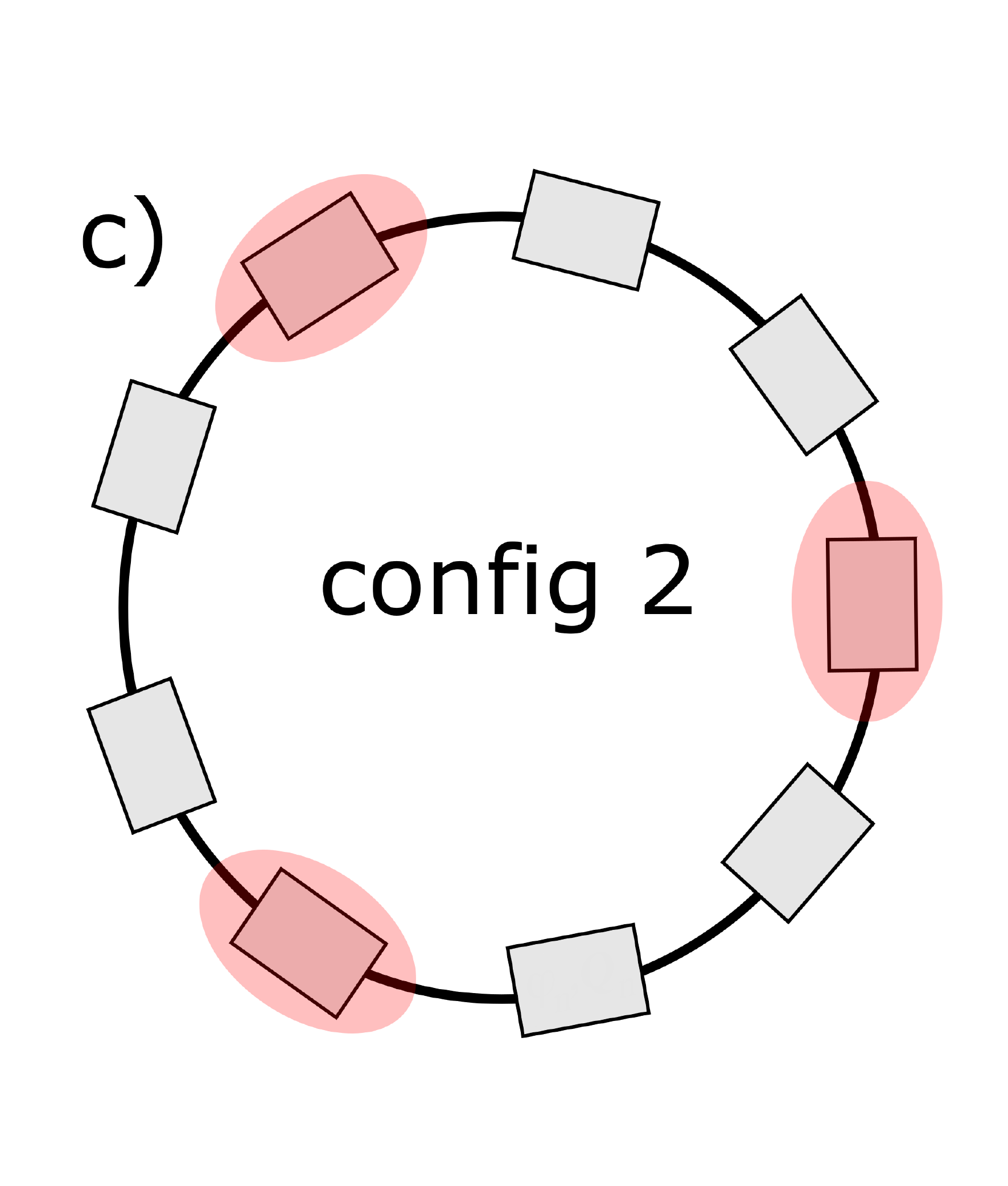}\\
	\caption{Example of the  ways to split the whole chain formed by $N$ sites in a 
		bipartite system formed by a subsystem $A$ and $B$ with $N_A$ and $N_B$ sites, respectively. 
		The red region denotes the subsystem $A$.
		The case in a) corresponds to the partition discussed in the main text whereas 
		the case b) and c) show configurations discussed only in this appendix.}
	\label{FIG:12}
\end{figure}
%
%
%
%

In this appendix we report the logarithmic negativity $E_{\mathcal{N}}$ of the system for different configurations. 
We focus on the size $N=9$ with frustrated dissipation. Here we only deal with the coupling $\alpha=0.2$ and the ratio $\widetilde{\alpha}/\alpha=0.3$.

The logarithmic negativity $E_{\mathcal{N}}$ is an entanglement measure defined for bipartite systems.
To quantify the entanglement in our single chain, we have to divide it in two parts and consider the whole chain 
as formed by two subsystems $A$ and $B$.
A priori, there are many possible choices for a such division. 
Few examples of different configurations are reported in  Fig.~\ref{FIG:12}.
In the first partition Fig.~\ref{FIG:12}(a), discussed in the main text,
the two subsystem are formed by neighboring islands.  
In the other two examples,  Figs.~\ref{FIG:12}(b) and \ref{FIG:12}(c), 
the internal sites forming the subsystem $A$ are equally spaced by one or two sites of the subsystem $B$, respectively.
%

%
%
%
%
%
\begin{figure}[t!]
	\includegraphics[width=0.49\columnwidth]{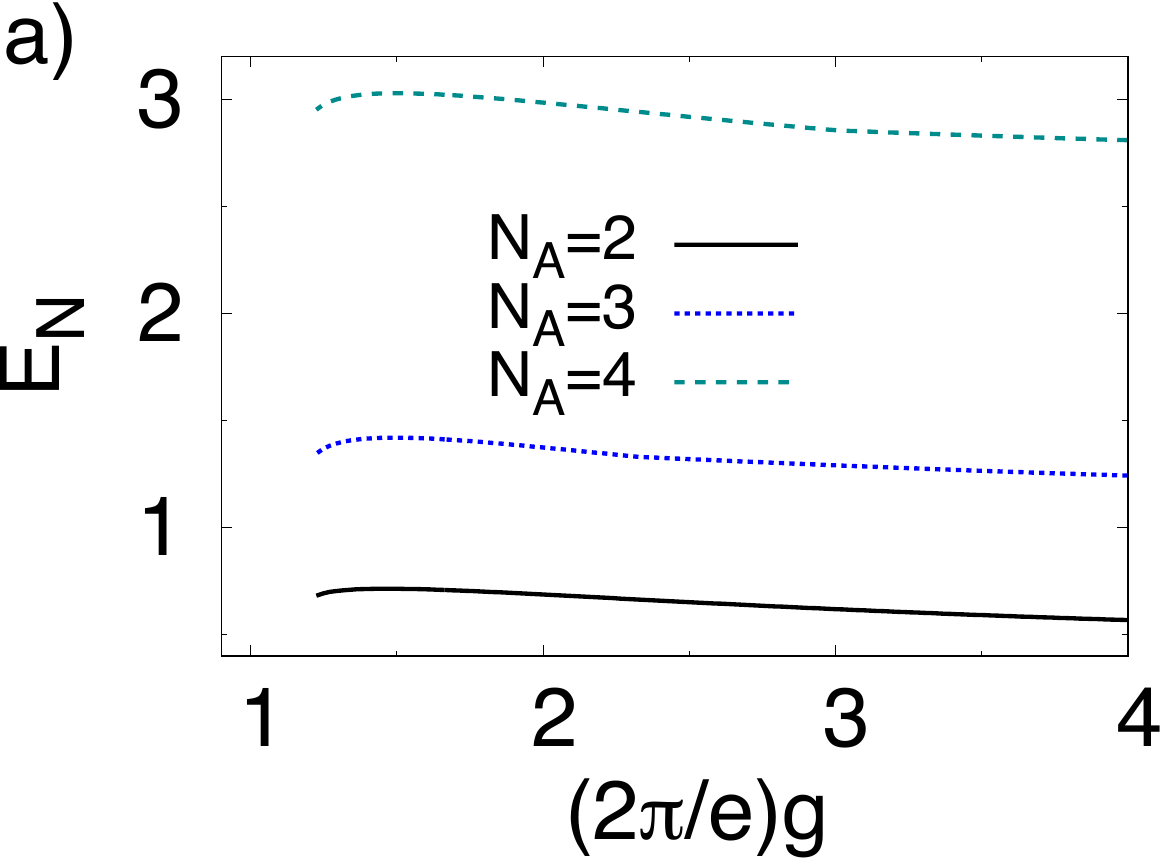}\hfill
	\includegraphics[width=0.49\columnwidth]{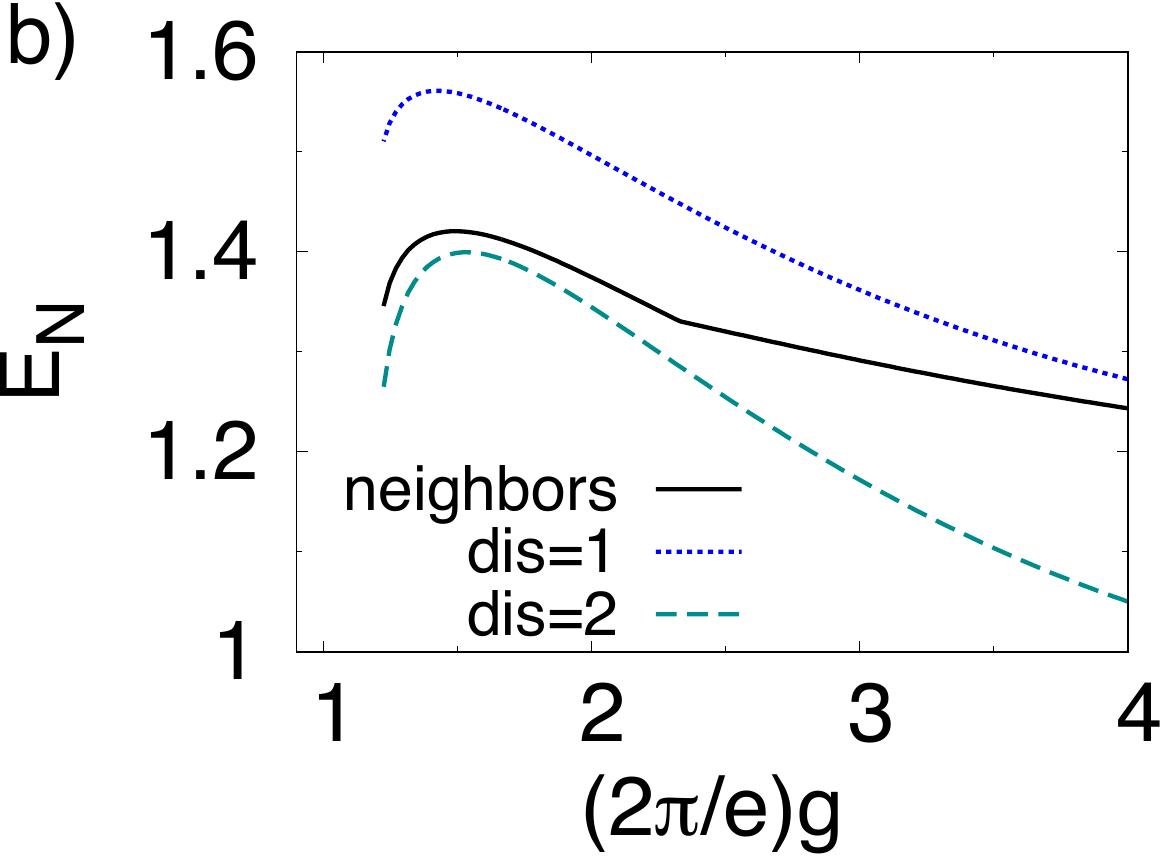}
	\caption{Logarithmic negativity for a system with $N=9$ as a function of $g$.
	a) The partition is fixed and corresponds to Fig.~\ref{FIG:12}(a) whereas the subsystem has different size $N_A=2,3,4$.
	b) The size of they subsystem is fixed $N_A =3$  whereas the different configurations are  reported as discussed in Fig.~\ref{FIG:12}.
	}
	\label{FIG:13}
\end{figure}
%
%
%
%

At a fixed configuration, corresponding to the one of Fig.~\ref{FIG:12}(a), we show the result for various partition sizes ($N_A,N-N_A$) with $N_A= 2,3,4$  in Fig.~\ref{FIG:13}(a). 
The logarithmic negativity grows with $N_A$ and the non-monotonic behavior is more pronounced in the latter case.
In Fig.~\ref{FIG:13}(b), we fix the size of the subsystem to $N_A=3$ and we  show the results 
for the different partitions of the chain.

We  conclude that, even if the specific slope depends on the configuration and size of the subsystem, 
the non-monotonic behavior still appears as a characteristic feature in the system affected by dissipative frustration. \\
\vspace{1cm}

\bibliography{bibliography}

\end{document}